\documentclass[traditabstract, amsmath]{aa}
\RequirePackage{aas_macros}
\usepackage[dvips]{graphicx}
\usepackage{txfonts}
\usepackage{natbib}
\bibpunct{(}{)}{;}{a}{}{,}
\usepackage{amsfonts,amssymb}
\usepackage{url}

\newcommand{\dif}{\ensuremath{\mathrm{d}}}

\begin{document}

\title{The evolution of \ion{H}{i} and \ion{C}{iv} quasar absorption
line systems at $1.9 < z < 3.2$\thanks{The 
data used in this study are taken from the ESO archive for the
UVES at the VLT, ESO, Paranal, Chile.}}

\author{T.-S. Kim\inst{1, 2} \and Adrian M. Partl\inst{1} \and R. F. Carswell\inst{3} \and Volker
  M\"{u}ller\inst{1}}

\institute{Leibniz-Institut f\"ur Astrophysik Potsdam, An der Sternwarte 16, 
D-14482 Potsdam, Germany \and Department of Astronomy, University of 
Wisconsin-Madison, 475 N. Charter St., Madison, WI 53706, USA \and Institute of 
Astronomy, Madingley Road, Cambridge CB3 0HA, UK}

\offprints{T.-S. Kim: kim@astro.wisc.edu}

\date{Received ; Accepted}

\titlerunning{The high-redshift \ion{H}{i} and \ion{C}{iv} forest}

\abstract{
We have investigated the distribution and evolution of
$\sim$3100 intergalactic neutral hydrogen
(\ion{H}{i}) absorbers with \ion{H}{i} column densities $\log N_{\mathrm{\ion{H}{i}}}\!=\![12.75, 17.0]$
at $1.9\!<\!z\!<\!3.2$, using 18 high resolution, high
signal-to-noise quasar spectra obtained from the ESO VLT/UVES archive.
We used two sets of Voigt profile fitting analysis, one including
all the available high-order Lyman lines to obtain reliable \ion{H}{i} column densities
of saturated lines, and another using only the Ly$\alpha$ transition.
There is no significant difference between the Ly$\alpha$-only fit 
and the high-order Lyman fit results. Combining our Ly$\alpha$-only fit results
at $1.7\!<\!z\!<\!3.6$ with high-quality  
literature data, the {\it mean} number density at $0 < z < 4$
is not well described by a single power law and strongly suggests 
that its evolution slows down at $z \le 1.5$ at the high and low column density ranges. 
We also divided our entire \ion{H}{i} absorbers at $1.9\!<\!z\!<\!3.2$ 
into two samples, the unenriched forest
and the \ion{C}{iv}-enriched forest, depending on whether \ion{H}{i} lines are associated
with \ion{C}{iv} at $\log N_{\mathrm{\ion{C}{iv}}}\!\ge\!12.2$ within a given velocity range. 
The entire \ion{H}{i} column density distribution function (CDDF) can be described as the combination
of these two well-characterised populations which overlap at $\log N_{\mathrm{\ion{H}{i}}} \sim 15$.
At $\log N_{\mathrm{\ion{H}{i}}} \le 15$,
the unenriched forest dominates, 
showing a similar power-law distribution to the entire forest.
The \ion{C}{iv}-enriched forest dominates at $\log N_{\mathrm{\ion{H}{i}}} \ge 15$, 
with its distribution function as $\propto N_{\mathrm{\ion{H}{i}}}^{\sim -1.45}$.
However, it
starts to flatten out at lower $N_{\mathrm{\ion{H}{i}}}$, since
the enriched forest fraction
decreases with decreasing $N_{\mathrm{\ion{H}{i}}}$.
The deviation from the power law
at $\log N_{\mathrm{\ion{H}{i}}} = [14, 17]$ shown
in the CDDF for the entire \ion{H}{i} sample
is a result of combining two
different \ion{H}{i} populations with a different CDDF shape.
The total \ion{H}{i} mass density relative to the critical density
is $\Omega_{\mathrm{\ion{H}{i}}} \sim 1.6\times10^{-6}h^{-1}$,
where the enriched forest accounts for $\sim 40$\% of $\Omega_{\mathrm{\ion{H}{i}}}$.}


\maketitle

\section{Introduction}

The resonant Ly$\alpha$ absorption by neutral hydrogen (\ion{H}{i}) in the
warm ($\sim 10^{4}$ K) photoionised intergalactic
medium (IGM)
produces rich absorption features
blueward of the Ly$\alpha$ emission line in high-redshift quasar spectra known as the
Ly$\alpha$ forest. The Ly$\alpha$ forest contains $\sim\!90$\% of the baryonic matter
at $z\!\sim\!3$ and can be observed in a wide range of
redshifts up to $z\!\sim\!6$.
Gas-dynamical simulations and semi-analytic models have been
very successful at explaining the observed properties of the Ly$\alpha$ forest
mainly at low \ion{H}{i} column densities $N_\mathrm{\ion{H}{i}} \le 10^{16}$ cm$^{-2}$.
These models 
have shown that the Ly$\alpha$ forest arises by mildly non-linear
density fluctuations
in the low-density \ion{H}{i} gas, which follows the
underlying dark matter distribution on large scales. This interpretation also predicts that
the Ly$\alpha$ forest provides
powerful observational constraints on the distribution and evolution
of the baryonic matter in the Universe, hence the evolution of galaxies
and the large-scale structure \citep{Cen:1994jl, Rauch:1997dz, Theuns:1998ud, Dave:1999kk, Schaye:2000kx,
Schaye:2001yk, Kim:2002fr}. In addition,
the discovery of triply ionised carbon (\ion{C}{iv}) associated with some of the
forest absorbers suggests that the forest metal abundances can be
utilised to probe early generations of star formation and
the feedback between high-redshift galaxies and the surrounding IGM
from which galaxies formed
\citep{Cowie:1995mb, Dave:1998mw, Aguirre:2001ud, Schaye:2003fc, Oppenheimer:2006lo}.


The physics of the Ly$\alpha$ forest is mainly governed by three competing processes,
the Hubble expansion, the gravitational growth and the ionizing UV background radiation.
The Hubble expansion which causes the gas to cool adiabatically and the gravitational 
growth 
are fairly well-constrained by the cosmological parameters
and the primordial power spectrum from the latest WMAP observations \citep{Jarosik:2010ye}.
On the other hand, the ionizing UV background radiation
controls the photoionisation heating and the gas ionisation fraction, thus determining the
fraction of the observable \ion{H}{i} gas compared to the unobservable \ion{H}{ii} gas.
The UV background is
assumed to be provided primarily by quasars and in some degree also by star-forming
galaxies \citep{Shapley:2006ay, Siana:2010bc} and Ly$\alpha$ emitters \citep{Iwata:2009zw}.
However, the intensity/spectral shape of the UV background and the relative contribution
from quasars and galaxies as a function of redshift are not well constrained
\citep{Bolton:2005yq, Faucher-Giguere:2008ux}.
One of the common methods to measure the UV background and its
evolution is the quasar proximity effect \citep{Dallaglio:2008ab}.
Unfortunately, measurements of the UV background through the proximity effect 
are biased by the large scale density distribution around the quasars which cannot
be easily quantified observationally
\citep{Partl:2010, Partl:2011}.

Two commonly explored quantities to constrain the properties of the Ly$\alpha$ forest 
are the
number of absorbers for a given \ion{H}{i}
column density range per unit redshift, $\dif n / \dif z$, and
the differential column density distribution function (CDDF, the number of absorbers
per unit absorption path length and per unit column density, an analogue 
to the galaxy luminosity function).
Compared with simulations, detailed structures seen in an overall 
power-law-like CDDF ($\propto N_{\mathrm{\ion{H}{i}}}^{\beta}$) such as a flattening or a steepening at 
different column density ranges constrain various forest physical and galactic feedback processes
\citep{Altay:2010ab, Dave:2010ab}. The CDDF is also one of the main observables required in calculating  
the mass density relative to the critical density contributed 
by the forest \citep{Schaye:2001yk}. The shape of the CDDF 
at lower $N_{\mathrm{\ion{H}{i}}} \le 10^{12.5-12.7}$ cm s$^{-2}$ (a typical detection limit for
most available high-quality data) is of particular importance, since the 
lower $N_{\mathrm{\ion{H}{i}}}$ absorbers are much more
numerous than higher $N_{\mathrm{\ion{H}{i}}}$ absorbers, thus 
they can trace a significant fraction of
baryons, depending on the steepness of the CCDF at the low $N_{\mathrm{\ion{H}{i}}}$ limit.    

On the other hand, $\dif n / \dif z$ provides an additional way to study  
the UV background radiation and its evolution. 
The gas density decreases with decreasing redshift due to the Hubble expansion. 
A lower gas density
results in a strong reduction of the recombination rate, allowing the gas to settle in to a
photoionisation equilibrium with a higher ionisation fraction.
With the non-decreasing background radiation,
this causes a steep number density evolution. However, the decrease of the quasar
number density at $z\!<\!2.5$ also decreases the available ionising photons \citep{Silverman:2005ab}. This
changes the ionisation fraction in the gas and also counteracts the gas density decrease,
and hence slows down the number density evolution
\citep{Theuns:1998ud, Dave:1999kk, Bianchi:2001ys}.

The result from the {\it HST}/FOS Quasar Absorption Line Key project
shows such a slow change in the $\dif n / \dif z$ evolution at $z\!<\!1.5$ \citep{Weymann:1998gf},
compared to a much steeper $\dif n / \dif z$ evolution shown 
at $z\!>\!2$ \citep{Kim:1997rt, Kim:2001fk, Kim:2002fr}.
Cosmic variance also seems to increase at lower $z$ \citep{Kim:2002fr}. Unfortunately,
recent work based on better-quality {\it HST} data at $z\!<\!1.5$ (or the observed
\ion{H}{i} Ly$\alpha$ at $<\!3050$ \AA\/)
have shown 
rather ambiguous $\dif n / \dif z$ results with a large scatter
along different sightlines \citep{Janknecht:2006eu, Lehner:2007ai, Williger:2010qe}. 
The only certain observational fact is that
all of these newer studies show a factor of $\sim 2-3$ lower number densities than the Weymann
et al. values at $z\!<\!1.5$. Considering a lack of results from good-quality data at
$1\!<\!z\!<\!1.5$ in the literature, the redshift evolution of $\dif n / \dif z$ 
can be considered as a single power
law without any abrupt change in $\dif n / \dif z$ at $0\!<\!z\!<\!3.5$. 

Here we present an in-depth Voigt profile fitting analysis of 18 high
resolution ($R\!\sim\!45\,000$), high signal-to-noise ($\sim\!35$--50 per pixel) quasar spectra
obtained with the UVES (Ultra-violet Visible Echelle Spectrograph) on the VLT, 
covering the Ly$\alpha$ forest at $1.9\!<\!z_{\mathrm{forest}}\!<\!3.2$.
Our main scientific aims are to derive
the redshift evolution of the absorber number density and the 
column density distribution function from a large and homogeneous set of data
available at $z\!>\!2$, since most previous high-quality forest studies at $z\!>\!2$
have been based on less than 5 sightlines. 
Even with few sightlines,
the statistics for the weak forest lines is 
robust due to the large number of weak 
absorbers with $N_{\mathrm{\ion{H}{i}}}\!=\!10^{13-15}$ cm$^{-2}$ 
(about 150 absorbers at $z\!\sim\!2.5$
per sightline, i.e. in the wavelength range between the quasar's Ly$\alpha$ and Ly$\beta$ emission lines). 
However, for the stronger forest systems with
$N_{\mathrm{\ion{H}{i}}}\!\ge\!10^{15}$ cm$^{-2}$, more sightlines are required 
since there are only about 10 absorbers per sightline at $z\!\sim\!2.5$.
Cosmic variance also plays an important role at lower redshifts, especially for
stronger absorbers \citep{Kim:2002fr}. Therefore, increasing the sample size at $z\!\sim\!2$ is
critical in addressing the $N_{\mathrm{\ion{H}{i}}}$ evolution for the Ly$\alpha$ forest. 

In addition to the increased sample size, we have improved previous results in two ways.
First, most previous studies on the forest from ground-based observations at $z>1.7$ have been
based on the Ly$\alpha$-only profile fitting analysis. This approach does not provide a
reliable $N_{\mathrm{\ion{H}{i}}}$ for saturated lines, 
$N_{\mathrm{\ion{H}{i}}}\!\ge\!10^{14.5}$ cm$^{-2}$ for the present UVES data. 
To derive a more reliable $N_{\mathrm{\ion{H}{i}}}$ of saturated lines, we have
included all the available high-order Lyman series in this study.

Second, strong evidence have been accumulated in recent studies that   
metals associated with the high-redshift Ly$\alpha$ forest are 
within $\sim\!100$~kpc of galaxies as in the circum-galactic medium rather
than in the intergalactic space far away from galaxies \citep{Adelberger:2005wa, Steidel:2010pd, 
Rudie:2012fk}.
This implies that the \ion{H}{i} absorbers containing metals might show different properties than
the ones without detectable metals. Taking \ion{C}{iv} as a metal proxy, 
we have divided our data into two samples, one with \ion{C}{iv}          
(the \ion{C}{iv}-enriched forest)
and another without \ion{C}{iv} (the unenriched forest), in order to
test this scenario of the circum-galactic medium.
Since our study lacks the imaging survey around the quasar targets,
we cannot claim that the \ion{C}{iv}-enriched forest is indeed located within $\sim\!100$~kpc
from a nearby galaxy. However, this study provides complementary results to
galaxy-absorber connection studies at high redshifts \citep{Steidel:2010pd, Rudie:2012fk}.

This study is also very timely since the Cosmic Origins Spectrograph (COS), a high-sensitivity 
FUV spectrograph onboard {\it HST} has started to produce many high-quality quasar spectra
at $z < 1$ \citep{Green:2012ab, Savage:2012ab}.
These COS quasar observations have opened a new tool to study the
low-$z$ Ly$\alpha$ forest. Combined with results at high redshifts such as our study,
COS observations will make it possible to characterise the $\dif n / \dif z$ evolution at
$0\!<\!z\!<\!3.5$ in a more robust way, thus a stringent constraint on the UV background
evolution. 

This paper is organised as follows. Section~\ref{sec:DataAndFitting} describes the analysed
data and two Voigt profile fitting methods. Comparisons with previous studies based
on the Ly$\alpha$-only fit are shown in Section~\ref{Sec3:Comparison}. The analysis based on
the high-order Lyman fit is presented in Section~\ref{Sec:Analysis}. Column density
distribution and evolution of the
Ly$\alpha$ forest containing \ion{C}{iv} are presented in Section~\ref{Sec:CIV}.
Finally, we discuss and summarise the main results in Section~\ref{Sec:Conclusions}.
All the results on the number density and the differential column density
distribution from our analysis are tabulated in Appendix A.
Throughout this study, the cosmological parameters are assumed to be
the matter density $\Omega_{m}=0.3$,
the cosmological constant $\Omega_{\Lambda}=0.7$ and
the current Hubble constant  $H_{\mathrm{0}} = 100 h$
km s$^{-1}$ Mpc$^{-1}$ with $h=0.7$, which is in concordance with latest WMAP measurements
\citep{Jarosik:2010ye}. The logarithm $N_{\mathrm{\ion{H}{i}}}$ is defined
as $\log N_{\mathrm{\ion{H}{i}}} = \log (N_{\mathrm{\ion{H}{i}}}/ 1 \, \, \mathrm{cm}^{-2})$.

\begin{table*}
\caption{Analysed quasars}
\label{tab1}
{\scriptsize{
\begin{tabular}{llccclcrl}
\hline
\noalign{\smallskip}
Quasar & $z_{\mathrm{em}}$ & $z_{\mathrm{Ly\alpha}}^{\mathrm{a}}$ &
   $z_{\mathrm{Ly\alpha\beta}}^{\mathrm{b}}$ &
   $z_{\mathrm{Ly,high-order}}^{\mathrm{c}}$ & Excluded $z_{\mathrm{\ion{C}{iv}}}^{\mathrm{d}}$ &
   S/N per pixel$^{\mathrm{e}}$ & LL (\AA\/) &  notes \\
\noalign{\smallskip}
\hline
\noalign{\smallskip}
Q0055--269          & 3.655$^{\mathrm{f}}$ & 2.936--3.605 &              & 2.936--3.205 & not used     
                    & 50--80 [..., ...] & 2288 &    \\
PKS2126--158        & 3.279       & 2.815--3.205 &              & 2.815--3.205 & 
                    & 50--200 [..., 170] & 3457 & two sub-DLAs at $z=$2.768 \& 2.638 \\  
Q0420--388          & 3.116$^{\mathrm{f}}$ & 2.480--3.038 & 2.665--3.038 & 2.665--3.038 & 2.607--2.670
                    & 100--140 [..., 120] & 3754 & a sub-DLA at $z=$3.087\\
HE0940--1050        & 3.078       & 2.452--3.006 &              &              & 2.714--2.778
                    & 50--130 [..., 115] &  $\le 3200$ & \\
HE2347--4342        & 2.874$^{\mathrm{f}}$ & 2.336--2.819 &              &              & 2.708--2.773 
                    & 100--160 [105, 120] & $\le 1160$ & multiple associated systems \\
Q0002--422          & 2.767       & 2.209--2.705 &              &              &   
                    &  60--70 [140, 170]  & 3025 &   \\
PKS0329--255        & 2.704$^{\mathrm{f}}$ & 2.138--2.651 &              &              & 
                    & 40--60 [90, 90]  & 3157 & an associated system at 4513.7 \AA \\
Q0453--423          & 2.658$^{\mathrm{f}}$ & 2.359--2.588 &              &              &  
                    & 90--100 [130, 120] & 3022 &  a sub-DLA at $z=$2.305 \\
                    &             & 2.091--2.217 &              &              &  & & &  \\
HE1347--2457        & 2.609$^{\mathrm{f}}$ & 2.048--2.553 &              &              & & 85--100 [115, 120] & 2237 &  \\
Q0329--385          & 2.434       & 1.902--2.377 &              &              & & 50--55 [(50, 85), ...]  & $\le 3050$ & \\

HE2217--2818        & 2.413       & 1.886--2.365 & 1.971--2.365 & 1.970--2.365 & & 65--120 [125, ...] & 2471 & \\
Q0109--3518         & 2.405       & 1.905--2.348 & 1.974--2.348 & 1.968--2.348 & & 60--80 [(80, 140), ...] & 2163 & \\
HE1122--1648        & 2.404       & 1.891--2.358 & 1.974--2.358 & 1.900--2.358 & & 70--170 [240, ...] & $\le 1629$ & \\
J2233--606          & 2.250       & 1.756--2.197 & 1.980--2.201 & 1.970--2.201 & not used & 30--50 [..., ...] & $\le 1750$ & \\
PKS0237--23         & 2.223$^{\mathrm{f}}$ & 1.765--2.179 & 1.974--2.179 & 1.961--2.179 & & 75--110 [(130, 200), ...] & $\le 3050$ & a sub-DLA at $z=$1.673 \\
PKS1448--232$^{\mathrm{g}}$  & 2.219       & 1.719--2.175 & 1.974--2.175 & 1.953--2.175 & & 30--90 [(70, 120), ...] & $\le 3050$ &  \\
Q0122--380          & 2.193       & 1.700--2.141 & 1.977--2.141 & 1.977--2.141 & & 30--80 [(85, 130), ...] & $\le 3052$ &  \\
Q1101--264          & 2.141       & 1.882--2.097 & 1.970--2.097 & 1.900--2.097 & & 80--110 [90, ...] & 2597 & a sub-DLA at $z=$1.839 \\
\noalign{\smallskip}
\hline
\end{tabular}
}}
\begin{list}{}{}
\item[$^{\mathrm{a}}$]
The redshift range of the Ly$\alpha$ forest region analysed for the number density
evolution in the Ly$\alpha$-only fit. For the differential column density evolution,
we used the redshift range listed in Column 5.
\item[$^{\mathrm{b}}$]
The redshift range for which the high-order Lyman fit {\it can be performed} is
listed only when it is different from the Ly$\alpha$-only fit region.
\item[$^{\mathrm{c}}$]
The redshift range of the Ly$\alpha$ forest region analysed for the high-order
Lyman fit is listed only when it is different from the Ly$\alpha$-only fit region.
\item[$^{\mathrm{d}}$]
The redshift range excluded for the \ion{C}{iv}-enriched \ion{H}{i} forest analysis in Section 5
due to the lack of the coverage.
Q0055$-$269 and J2233$-$606 are excluded
due to their lower S/N in the \ion{C}{iv} region. No 
entries mean that the analyzed $z_{\mathrm{\ion{C}{iv}}}$ is the same as
the one in Column 5.
\item[$^{\mathrm{e}}$]
The number outside the bracket is a S/N of the \ion{H}{i} forest region. The
first number inside the bracket is a typical S/N of the \ion{C}{iv} region at
$1.9 < z < 2.4$, while the second is for $2.4 < z < 3.2$. 
The dotted entries indicate that no \ion{C}{iv} forest
region is available for a given redshift range.
Two numbers inside the parentheses indicate the S/N of the \ion{C}{iv}
region at $1.9 < z < 2.1$ and $2.1 < z < 2.4$, respectively, due to the
dichroic setting toward some sightlines.
\item[$^{\mathrm{f}}$]
Due to the absorption systems at the peak of the Ly$\alpha$ emission line
or to the non-single-peak emission line, the measurement is uncertain.
\item[$^{\mathrm{g}}$]
Part of the continuum uncertainties toward shorter wavelengths
is due to the local, non-smooth
continuum shape, partly due to a lower S/N ($\sim$30--35).
\end{list}
\end{table*}

\section{Data and Voigt profile fitting}
\label{sec:DataAndFitting}

Table~\ref{tab1} lists the properties of the 18 high-redshift quasars analysed
in this study. 
The redshift quoted in Column 2 is measured from the                        
observed Ly$\alpha$ emission line of the quasars.
Note that the redshift based on the emission lines is known to be under-estimated
compared to the one measured from other quasar emission lines such as \ion{C}{iv}
\citep{Tytler:1992eb, VandenBerk:2001ab}.
The spectrum of Q1101$-$264 is the same one as analysed in
\cite{Kim:2002fr}, while the rest of spectra are from \cite{Kim:2007gf}.
The raw spectra were obtained from the ESO VLT/UVES archive and were
reduced with the UVES pipeline.
All of these spectra have
a resolution of $R\!\sim\!45\,000$ and heliocentric, vacuum-corrected
wavelengths. The spectrum is sampled at 0.05\AA\/.
A typical signal-to-noise ratio (S/N) in the Ly$\alpha$ forest
region is 35--50 per pixel (hereafter all the S/N ratios are given as per pixel).
Readers can refer to \cite{Kim:2004mz} and \cite{Kim:2007gf} for the details
of the data reduction.
To avoid the proximity effect,
the region of 4,000 $\textrm{km s}^{-1}$ blueward of
the quasar's Ly$\alpha$ emission was excluded.

In order to obtain the absorption line parameters (the redshift $z$,
the column density $N$ in cm$^{-2}$
and the Doppler parameter $b$ in $\textrm{km s}^{-1}$), we have performed a Voigt
profile fitting analysis using VPFIT{\footnote{Carswell et al.:
http://www.ast.cam.ac.uk/$\sim$rfc/vpfit.html}}. Details
can be found in the documentation provided with the software, Carswell,
Schaye \& Kim (2002) and \cite{Kim:2007gf}. Here, we only give a brief description of
the fitting procedure.

First, a localised initial continuum of each spectrum was defined
using the {\tt CONTINUUM/ECHELLE} command in {\tt IRAF}. Second,
we searched for metal lines in the entire spectrum, starting
from the longest wavelengths toward the shorter wavelengths.
We first fitted all the identified
metal lines. 
When metal lines were embedded in the \ion{H}{i}
forest regions, the \ion{H}{i} absorption lines blended with metals
are also included in the fit.
Sometimes the simultaneous fitting of different
transitions by the same ion reveals that the initial
continuum needs to be adjusted to obtain acceptable ion ratios.
In this case, we adjusted the initial continuum accordingly.
The rest of the absorption features were assumed to be \ion{H}{i}.

After fitting metal lines, we have fitted the entire spectrum
including all the available higher-order Lyman series in the UVES spectra.
This is absolutely necessary to obtain reliable $N_{\mathrm{\ion{H}{i}}}$
for saturated lines, as our study deals with saturated lines and
relies on line counting.
A typical $z\!\sim\!3$ IGM absorption feature having $b\!\sim\!30$ km~s$^{-1}$
starts to saturate around $N_\mathrm{\ion{H}{i}} \ge
10^{14.5} {\mathrm{cm}^{-2}}$
at the UVES resolution and S/N. Unfortunately,
$N_{\mathrm{\ion{H}{i}}}$ and $b$ values of saturated lines
are not well constrained. In order to derive reliable
$N_{\mathrm{\ion{H}{i}}}$ and $b$ values, higher-order Lyman
series, such as Ly$\beta$ and Ly$\gamma$, have to be included
in the fit, as higher-order Lyman series have smaller oscillator
strengths and start to saturate at much larger $N_{\mathrm{\ion{H}{i}}}$.

During this process, another small amount of continuum re-adjustment
was often required to achieve a satisfactory fit, i.e. a
reduced $\chi^{2}$ value of $\sim 1.2$. With
this re-adjusted continuum, we re-fitted the
entire spectrum. This iteration process of
continuum re-adjustments and re-fitting
was then repeated several times until satisfactory fitting parameters
were obtained. This produces
the final set of fitted parameters for each component of
the {\it high-order Lyman fit} analysis.

In addition to this high-order fit, we have also performed the
same analysis using only the Ly$\alpha$ transition region, i.e. the wavelength
range above the rest-frame Ly$\beta$ and
below the proximity effect zone.
This additional fitting analysis was done, since most
previous studies on the IGM $N_{\ion{H}{i}}$ analysis based on Voigt profile
fitting utilised only the Ly$\alpha$ region.
For the Ly$\alpha$-only fit, we kept
the same continuum used in the high-order Lyman fitting process.
In principle,
the difference between two sets of fitted parameters occurs
only in the regions where saturated absorption features are included.
In both fitting analyses, we did not tie the fitting parameters for different ions.

\begin{figure*}
\centering
\includegraphics[width=0.8\textwidth]{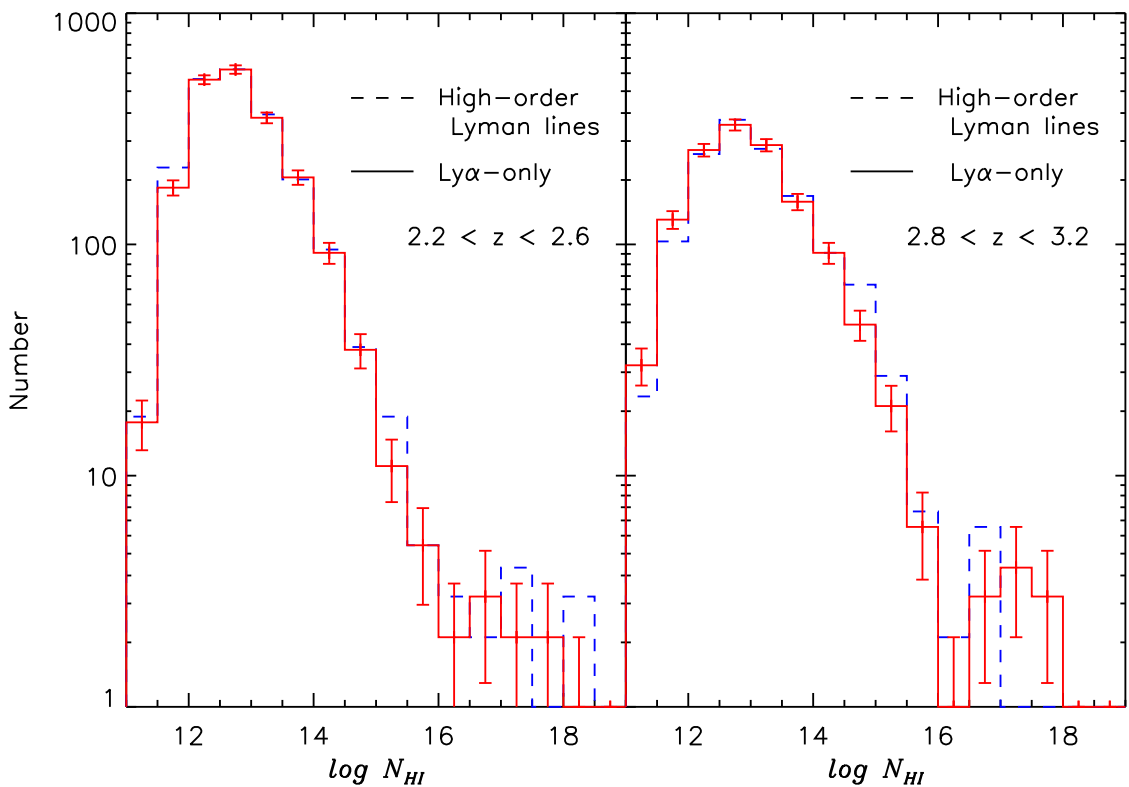}
\caption{Numbers of absorption lines as a function of
$N_{\mathrm{\ion{H}{i}}}$ at $2.2\! <\! z \! < \!2.6$ and $2.8\! <\! z\! <\! 3.2$.
The Ly$\alpha$-only
fits are shown as solid lines, while the high-order Lyman fits are marked as
dashed lines. Solid errors indicate the 1$\sigma$ Poisson errors
of the Ly$\alpha$-only fits.}
\label{fig:compare_all}
\end{figure*}

Fig.~\ref{fig:compare_all} shows the numbers of absorption lines as a function of
$N_{\mathrm{\ion{H}{i}}}$ for both fitting analyses at the two redshift ranges,
$2.2\!<\!z\!<\!2.6$ and $2.8\!<\!z\!<\!3.2$, in order to illustrate the differences
at high and low redshifts.
The differences between the two samples occurs mostly at
$N_{\ion{H}{i}} \ge 10^{14.5}$ cm$^{-2}$ and
at $N_{\ion{H}{i}} \le 10^{12}$ cm$^{-2}$.
This difference in the line numbers at $N_{\ion{H}{i}} \ge 10^{14.5}$ cm$^{-2}$
seems to be stronger at $2.8 < z < 3.2$, although it is still within 2$\sigma$
Poisson errors. The line
numbers at $N_{\ion{H}{i}} \le 10^{12}$ cm$^{-2}$ are more
susceptible to the incompleteness which depends on the local S/N 
than the difference
between the two fitting methods. 
The difference at other
column density ranges is smaller, which in turn leads us to expect
that there is no significant difference between
the Ly$\alpha$-only fit and the high-order Lyman fit.

We restrict our present analysis to
$\log N_{\ion{H}{i}}\!=\![12.75, 17]$ at all redshifts.
As clearly seen in Fig.~\ref{fig:compare_all}, the incompleteness
becomes quite severe for 
$\log N_{\ion{H}{i}}\!\le\!12.5$ and redshifts 
$z\!>\!3$ \citep{Kim:1997rt, Kim:2002fr}. Therefore, the lower $N_{\ion{H}{i}}$
limit was chosen to be $\log N_{\ion{H}{i}}\!=\!12.75$. 
We chose $\log N_{\ion{H}{i}}\!=\!17$ as 
the upper $N_{\ion{H}{i}}$ limit since  
we wanted to analyze only the Ly$\alpha$
forest whose traditional definition is an absorber with $\log N_{\ion{H}{i}}\!<\!17.2$
(above which it is referred to as a Lyman limit system \citep{Tytler:1982}).
Additionally,
absorbers at $\log N_{\ion{H}{i}}\!>\!17$ are very rare (Fig.~\ref{fig:compare_all}).

Note that the availability of the high-order Lyman series depends on
the redshift of the quasar and whether the sightline contains a Lyman limit system.
In addition, the amount of blending affects
whether a reliable column density can be measured. At high redshifts
$z_{\mathrm{em}}\,>\,3$, line blending becomes severe. However, most
UVES spectra also covers down to 3050 \AA\, where Lyman lines higher than
Ly$\delta$ are available. On the other hand, at $z_{\mathrm{em}} < 2.5$
the available high-order Lyman lines are rather limited, with mostly
Ly$\beta$ and Ly$\gamma$ available. However, line blending is less problematic than at higher
redshifts. We have generated tens of saturated artificial absorption lines and
fitted them including and excluding high-order Lyman lines. These simulations show
that {\it unblended} absorption features
at $N_{\ion{H}{i}} \le 10^{17}$ cm$^{-2}$ can be reasonably well constrained
with Ly$\alpha$ and Ly$\beta$ only. This indicates that our $N_{\ion{H}{i}}$ can
also be considered reliable even at
$z < 2.5$ with Ly$\alpha$ and Ly$\beta$ only.

\begin{figure}
\centering
\includegraphics[width=90mm]{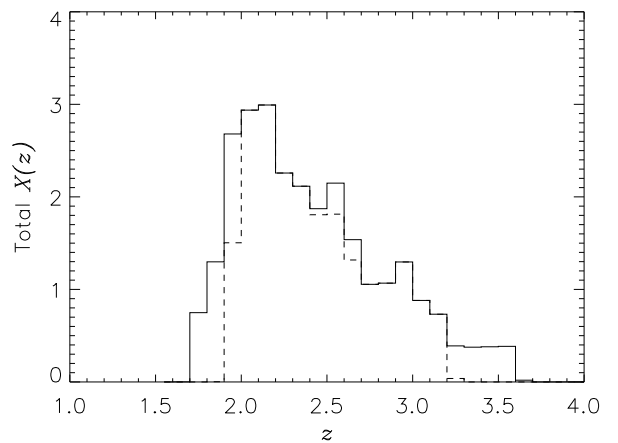}

\caption{Total absorption distance $X(z)$ covered with our sample of 18 high-redshift quasars.
The solid line is for the Ly$\alpha$-only fit, while the dashed one is for
the high-order fit.
}
\label{fig:totX_z}
\end{figure}

The absorption distance is obtained by
integrating the Friedmann equation for a $\Omega_m = 0.3$
and $\Omega_\Lambda = 0.7$ universe, and is given by
\begin{equation}
\label{eq:civRedPath}
X(z) = \int \frac{H_0}{H(z)}\left(1+z \right)^2 \dif z
\end{equation}
\citep{Bahcall:1969gd}, where $H_0$ is the Hubble constant at $z=0$.
The total absorption distance $X(z)$ covered by the spectra for both Ly$\alpha$-only
and high-order fits is shown in  Fig.~\ref{fig:totX_z}. 
The redshift coverage
of our sample steadily increases with decreasing redshift until it
reaches its maximum at $z\!\sim\!2.1$.
For redshifts below $z\!<\!1.9$ the coverage decreases rapidly and
our sample ends at $z=1.7$.
Note that the lowest redshift possible for the high-order Lyman line
analysis is
$z\!\sim\!1.97$, while the Ly$\alpha$-only fit analysis is 
possible down to $z \sim 1.7$. 
Due to the reduced redshift coverage in the high-order Lyman range 
of individual sight lines caused by intervening Lyman limit systems, the sample coverage 
of the high-order fit analyses is reduced between $2.4\!<\!z\!<\!2.7$. 
At the high redshift end $z\!>3.22$, the number of available forest lines 
decreases and the sample consists of only one line of sight. The low redshift
limit for the high-order fit was set to be the lowest redshift without
any saturated lines when no Ly$\beta$ is available for each quasar. This criterion
restricts our high-order Lyman fit analysis to $1.9\!<\!z\!<\!3.2$. Since the
redshift coverage of low-z quasars for the high-order fit is shorter than the one
for the Ly$\alpha$-only fit and the high-$N_{\mathrm{\ion{H}{i}}}$ forest 
clusters stronger at lower $z$
\citep{Kim:2002fr}, the quasar-by-quasar $\dif n / \dif z$ at $z\!\sim\!2$ from
the high-order fit analysis is expected to suffer from the low number statistics.

In Table~\ref{tab1}, Columns 3--6 summarise the redshift range used for the
different analysis. Column 3 lists the redshift range of the Ly$\alpha$ forest 
region analysed for the number density
evolution in the Ly$\alpha$-only fit.
For the differential column density evolution
of the Ly$\alpha$-only fit, we used the redshift range listed in Column 5.
Column 4 and Column 5 list the redshift range for which the high-order Lyman fit 
{\it can be performed} and the one for which the high-order Lyman fit was done,
respectively. The region is listed only when it is different from the Ly$\alpha$-only 
fit region in Column 3. 
Since there are no
strongly saturated Ly$\alpha$ lines at $1.90\!<\!z\!<\!1.98$ for some low-$z$ quasar
sightlines, we used a lower redshift range than the one listed in Column 4  
for the high-order Lyman fit analysis for these sightlines. Column 6 shows the redshift range excluded
for the \ion{C}{iv}-enriched \ion{H}{i} study in Section 5. 
Due to the wavelength gaps caused by the UVES dichroic setup, the covered \ion{C}{iv} redshift
ranges are smaller than the Ly$\alpha$ forest coverage listed in Column 3.
The region of $\pm 200$ km s$^{-1}$ from the gap was excluded, and only the redshift range covering
both \ion{C}{iv} doublets was included in the analysis.
The blank entries mean that the analyzed $z_{\mathrm{\ion{C}{iv}}}$ is the same as
the forest $z_{\mathrm{Ly, high-order}}$.
Q0055$-$269 and J2233$-$606 are excluded
in the Ly$\alpha$--\ion{C}{iv} forest study due to their lower S/N in the \ion{C}{iv} region.

In the HE2347$-$4342 Ly$\alpha$ forest region, 
there are very strong \ion{O}{vi} absorptions mixed with the two
saturated Ly$\alpha$
absorption systems at 4012--4052 \AA\/ \citep{Fechner:2004bc}.
Since the fitted line parameters
for these Ly$\alpha$ systems cannot be well constrained (their
corresponding Ly$\beta$ is below the partial Lyman limit
produced by the $z\!\sim\!2.738$ systems),
we excluded this forest region toward HE2347$-$4342.
In the J2233$-$606 sightline, there are two partial Lyman limit systems at 3489~\AA\/ ($z\!\sim\!1.870$) and
3558~\AA\/ ($z\!\sim\!1.926$)
and several high column density forest absorbers at 3400--3650\AA\/. To derive
a robust $N_\ion{H}{i}$, we included the {\it HST}/STIS echelle spectrum of 
J2233$-$606{\footnote{
The STIS spectrum is taken from {\scriptsize{\tt http://www.stsci.edu/ftp/observing/hdf/hdfsouth/hdfs.html}}
\citep{Savaglio:1999rr}.}} at 2280--3150~\AA\/. 
The resolution in this wavelength region is $\sim\!10$ km s$^{-1}$ and 
its S/N is $\sim\!8$ per pixel (0.05~\AA\/).

Table~\ref{tab1} also lists the S/N of each quasar spectrum in Column 7.
The number outside the bracket is a S/N of the \ion{H}{i} forest region. The
first number inside the bracket is a typical S/N of the \ion{C}{iv} region at
$1.9 < z < 2.4$, while the second is for $2.4 < z < 3.2$.
The dotted entries inside the bracket indicate that no \ion{C}{iv} forest
region is available for a given redshift range. The low redshift bin of
the \ion{C}{iv} forest covers the wavelength region where the different CCDs from two dichroic settings
were used at $\sim\!4780$ \AA (or $z \sim 2.1$). This leads to a much lower S/N at $\le\!4780$ \AA\/
($z < 2.1$).
When the lower S/N region is larger than 20\% of the whole \ion{C}{iv}
forest range, two numbers were listed inside the parentheses. The first number corresponds to
the lower S/N at $1.9 < z < 2.1$, while the second number is for the higher S/N
at $2.1 < z < 2.4$.

\begin{figure}
\includegraphics[width=90mm]{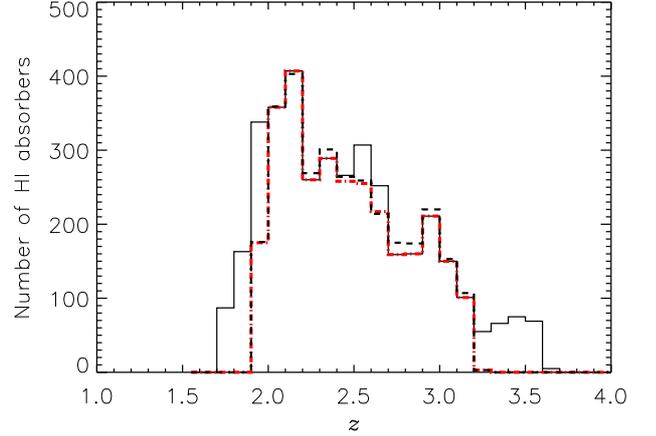}

\caption{Number of \ion{H}{i} absorbers with
$\log N_{\ion{H}{i}}\!=\![12.75, 17]$ as function of redshift
in our our sample of 18 high-redshift quasars.
The solid line is for the Ly$\alpha$-only fit. The dashed line is for
the high-order fit, while the red heavy dot-dashed line is for
the Ly$\alpha$-only fit for the redshift range used for the high-order fit.}
\label{fig:totNHI_z}
\end{figure}

In addition, regions within $\pm 50$ \AA\/ to the center of
a sub-damped Ly$\alpha$ system ($N_{\mathrm{\ion{H}{i}}} \ge
10^{19}$ cm$^{-2}$) are
excluded, since they are associated directly with
intervening high-$z$ galactic disks/halos and could have a possible influence on 
the apparent line densities in the forest. 
The sightline toward Q0453$-$423 includes a sub-DLA, which introduces a gap in the
Ly$\alpha$ redshift range. All the calculations toward Q0453$-$423
account for this redshift gap correctly. However, they are plotted as a single
data point and their plotted redshift range is the whole Ly$\alpha$ redshift range
without showing a gap. The sightlines toward PKS2126$-$158 and Q0420$-$388
also contain an intervening sub-DLA, which shortens the continuously available redshift coverage 
for the high-order fit. Column 8 of Table~\ref{tab1} lists the
observed wavelength of a Lyman limit (LL, 912~\AA\/ in the rest-frame wavelength) 
of each quasar, which is defined as the wavelength 
below which the observed flux becomes 0. The values are taken from \cite{Kim:2004mz}.
When a Lyman limit is not detected within available data, it is denoted to be less
than the lowest available wavelength. Column 9 of Table~\ref{tab1} notes information
on sub-DLAs along the sightline. When a sub-damped Ly$\alpha$ system exists
along the sightline, we discarded 50 \AA\/ centred at the sub-DLA
each side to exclude its influence on the forest, such as a higher frequency or
lack of higher-column density forest.
The total number of \ion{H}{i} lines for $\log N_{\ion{H}{i}}\!=\![12.75, 17]$
at $1.9 < z < 3.2$ is 3077 for the high-order Lyman fit sample. 
The Ly$\alpha$-only fit sample has 3778 \ion{H}{i} lines at the total redshift
range listed in the 3rd column of Table~\ref{tab1}.

In Fig.~\ref{fig:totNHI_z} the number of \ion{H}{i} absorbers 
with $\log N_{\mathrm{\ion{H}{i}}} = [12.75, 17]$ from both fitting
methods is shown as a function of redshift.
The number of absorbers obtained from each fitting analysis is roughly proportional to the
absorption distance coverage. Therefore, our sample shows the highest \ion{H}{i} absorber numbers
around redshift $z \sim 2$ for each fitting analysis,
where the sample absorption distance coverage also reaches its maximum. Sometimes the high-order fit
analysis (dashed line) reveals a slightly higher number of absorbers between $2 < z < 3$. This is because 
what appear to be single saturated Ly$\alpha$ lines may have more than one component
present in the corresponding higher order Lyman lines.
At $z \sim 2$, the number of the Ly$\alpha$-only-fit absorbers (heavy dot-dashed line) is
slightly larger than the high-order-fit absorbers. This is caused by the fact that
some simple saturated lines with $\log N_{\mathrm{\ion{H}{i}}}\!<\!17$ in the Ly$\alpha$-only 
fit analysis are actually absorbers with $\log N_{\mathrm{\ion{H}{i}}}\!>\!17$ in the 
high-order fit analysis. Since the Ly$\alpha$-only fit gives a lower $N_{\mathrm{\ion{H}{i}}}$
limit for a saturated line, these lines are included in the Ly$\alpha$-only fit
sample, but excluded in the high-order fit sample in Fig.~\ref{fig:totNHI_z}.

\section{Comparison with previous studies using Ly$\alpha$ only}
\label{Sec3:Comparison}

In Sect. \ref{sec:DataAndFitting} we have shown that including higher order
transitions in the fitting process slightly alters the column density statistics at 
$\log N_\ion{H}{i}\!>\!15.0$.
In order to compare our quasar sample with previous studies based only on the Ly$\alpha$ transition,
we briefly present the column
density distribution and evolution derived from the Ly$\alpha$-only fit in this section.
A large redshift coverage is very important in the study of the absorber number density.
Therefore we used all Ly$\alpha$ lines found in the whole available Ly$\alpha$ redshift ranges 
listed in Column 3 of Table~\ref{tab1} in this section. On the other hand, the
differential density distribution function is not sensitive to a large redshift coverage.
Thus, only the Ly$\alpha$ lines at $1.9\!<\!z\!<\!3.2$ are analysed for the
distribution function study.
A detailed analysis 
using the high-order fit is presented in Section \ref{Sec:Analysis}.
All the results from this section are tabulated in Appendix A.

\subsection{Absorber number density evolution $\mathrm{d} n / \mathrm{d} z$}
\label{Sec3:1}

The absorber number density $n(z)$ is measured by counting the number of \ion{H}{i}
absorption lines for a given column density range for each line of sight.
The line
count $n$ is then divided by the covered redshift range $\Delta z$ to
obtain $\mathrm{d} n / \mathrm{d} z$.
If forest absorbers have a constant size and a constant comoving number density, its number density
evolution due to the Hubble expansion can be described as
\begin{equation}
\label{eq1}
\frac{\dif n}{\dif z} = \pi R^{2} N_{0} c H(z)^{-1} (1+z)^{2}, 
\end{equation}
where $R$ is the size of an absorber, $N_{0}$ is the local comoving number density and $c$
is the speed of light \citep{Bahcall:1969gd}. For our assumed cosmology, Eq.~\ref{eq1} becomes
\begin{equation}
\label{eq2}
\frac{\dif n}{dz} \propto \frac{(1+z)^{2}}{\sqrt{0.3 (1+z)^{3} + 0.7}}.
\end{equation}
At $1 < z < 4.5$, Eq.~\ref{eq2} has an asymptotic behaviour of $\dif n / \dif z \propto (1+z)^{\sim 0.6}$, 
while at $z < 1$ it becomes $\dif n / \dif z \propto (1+z)^{\sim 1.15}$. For higher redshifts the
asymptotic behaviour becomes $\dif n / \dif z \propto (1+z)^{0.5}$. Any differences in the observed
exponent from what is expected from Eq.~\ref{eq2} indicate that the absorber size or/and
the comoving density are not constant.

Empirically, $\dif n / \dif z$ is
described as $\dif n / \dif z  = A (1+z)^{\gamma}$. It has been known that $\dif n / \dif z$ evolves
more rapidly at higher column densities. At $z\!>\!1.5$, a $\gamma\!\sim\!2.9$ is found
for $N_\mathrm{\ion{H}{i}} = 10^{14-17}$ cm$^{-2}$, and $\gamma\!\sim\!1.4$ for
$N_\mathrm{\ion{H}{i}} = 10^{13.1-14}$ cm$^{-2}$ \citep{Kim:2002fr}. At $z\!<\!1.5$,
\citet{Weymann:1998gf} found $\gamma\!\sim\!0.16$ and $A\!\sim\!35$ for absorbers with a
rest-frame equivalent
width greater than 0.24~\AA\/ from {\it HST}/FOS data. Later studies on $\dif n / \dif z$ based on
the profile fitting or curve of growth analysis using 
better-quality data from {\it HST}/STIS and {\it HST}/GHRS
show a factor of $\sim$2--3 lower
$\dif n / \dif z$ than the one found by \citet{Weymann:1998gf}. 
These studies
also show a larger scatter in $\dif n / \dif z$ at $z < 0.2$ with $A\!\sim\!5$--22
\citep{Lehner:2007ai, Williger:2010qe}. 
Part of this scatter is thought to be caused by inhomogeneous data
quality, analysis methods, and cosmic variance. 
Unfortunately high-quality data lack a complete $z$ coverage at $z\!<\!1.5$,
missing mostly at $0.4\!<\!z\!<\!1.0$. 
Keep in mind that the FOS
result and most available ground-based results at $z > 1.5$ in the literature are 
based on the Ly$\alpha$ lines
only, while most space-based results at $z < 1.5$ are using the available high-order
Lyman series. Therefore,
it is not possible to derive a robust power-law slope $\gamma$ of $\dif n / \dif z$ 
at $0\!<\!z\!<\!3.5$. Strictly speaking, a fair comparison should be made on
the data with similar qualities and uniform analyses.

\begin{figure}
\includegraphics[width=90mm]{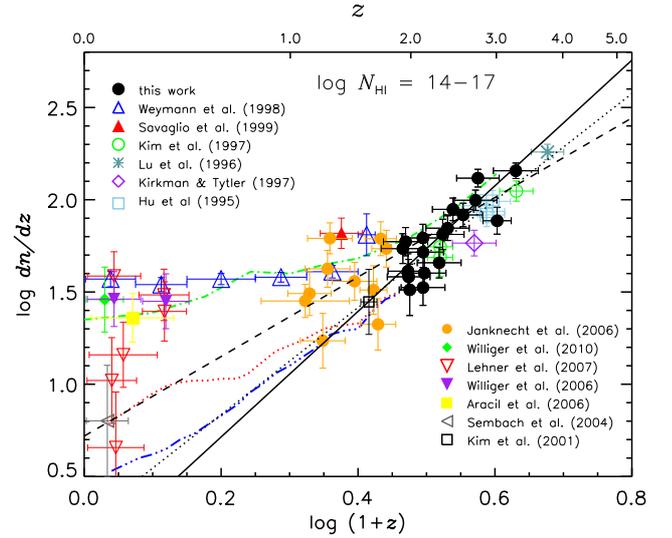}

\caption{The number density evolution of the Ly$\alpha$ forest in the
column density range $\log N_{\mathrm{\ion{H}{i}}} = [14,17]$ of the Ly$\alpha$-only
fits. Black filled circles show results
from our data set, which is tabulated in
Table~\ref{tbl:dndz_ap1}.
Other data points indicate various results obtained from the
literature. The vertical error bars give the $1\sigma$ Poisson error, while
the x-axis error bars show the redshift range covered by each sightline.
The solid line shows the fit to our data only. Dashed line is
the result including the literature data for $z>1$ ($\log(1+z) > 0.3$).
The dotted line gives the fit given in
\citet{Kim:2002fr}.
The green dot-dashed curve shows the predicted
$\dif n / \dif z$ evolution based on a quasar-only UV background by \citet{Dave:1999kk}.
The red dotted and the blue dot-dot-dot-dashed curves at $z < 2$ illustrate the predicted
$\dif n / \dif z$ based on momentum-driven wind and no-wind models
with a UV background by quasars and galaxies, respectively \citep{Dave:2010ab}.}
\label{fig:litComp_KimFig7}
\end{figure}

The number density evolution is illustrated in Figs.~\ref{fig:litComp_KimFig7},
\ref{fig:litComp_KimFig9} and \ref{fig:litComp_KimFig8} for two different column 
density ranges: $\log N_{\ion{H}{i}} = [14, 17]$,
and $[13.1, 14]$.
Data compiled from the literature are indicated in the figures:
\citet{Hu:1995lq}, \citet{Lu:1996ul}, \citet{Kim:1997rt},  \citet{Kirkman:1997pd},
\citet{Weymann:1998gf}, \citet{Savaglio:1999rr}, \citet{Kim:2001fk}, \citet{Sembach:2004fy}, 
\citet{Williger:2006dp},
\citet{Aracil:2006sp}{\footnote{The revised line list was used.}, 
\citet{Janknecht:2006eu}{\footnote{The fitted line parameters
by \citet{Janknecht:2006eu} show many \ion{H}{i} lines with $b < 20$ km~s$^{-1}$, about 25\%
of all lines. They attribute this to their low signal-to-noise data of less than 10
per resolution element. Although there are 9 sightlines analysed, one sightline has a
long wavelength coverage from VLT/UVES and {\it HST}/STIS. This sightline was split
into two data points in Figs.~\ref{fig:litComp_KimFig7} and \ref{fig:litComp_KimFig9}.}}, 
\citet{Lehner:2007ai}
and \citet{Williger:2010qe}. 
To be consistent with our definition of the
proximity effect zone, we applied the same $4\,000$ km s$^{-1}$ exclusion within
the quasar's Ly$\alpha$ emission line for all the literature data, whenever
the line lists from the literature include all the Ly$\alpha$ lines below
the Ly$\alpha$ emission line of the quasar. When the published line lists are only
for the shorter wavelength region than the entire, available forest region outside
the $4\,000$ km s$^{-1}$ proximity zone, such as the ones by \citet{Hu:1995lq}, no such an exclusion
is required.  We used all the reported \ion{H}{i} lines in
the literature mentioned above, without any pre-selection imposed on $N_{\mathrm{\ion{H}{i}}}$
or $b$ parameters. The latest study on the low-redshift IGM by  \citet{Williger:2010qe}
found that the number density from the {\it HST}/STIS results is a factor of 2--3 lower than 
the {\it HST}/FOS results by \citet{Weymann:1998gf}. They applied the same selection criteria 
on \ion{H}{i} absorbers used by \citet{Lehner:2007ai},
i.e. measurement errors less than 40\% and 
$b < 40$ km s$^{-1}$. As \ion{H}{i} absorbers
tend to have a larger $b$ parameter at lower redshift \citep{Lehner:2007ai} and larger
measurement errors in general, selecting 
\ion{H}{i} absorbers at $b > 40$ km s$^{-1}$ has a larger impact on $\dif n / \dif z$ at
lower redshift. In addition, as the {\it HST}/FOS results are based on the \ion{H}{i}
sample without any imposed selection criteria, using the full \ion{H}{i} lines provides
a more straightforward comparison to the {\it HST}/FOS result.

We have performed a linear regression to our data in logarithmic space for the various column density bins, 
using the 
maximum likelihood method described in \citet{Ripley:1987dk}. This method  
accounts for the uncertainties in the number density and incorporates the weighting using the
uncertainties. Errors of the fit parameters 
were obtained using the maximum likelihood method.
Linear regressions were once obtained from our data including the literature data and once without
them. Since for redshifts $z \lesssim 1$ (or $\log(1+z) \lesssim 0.3$)
the number density evolution {\it could} remain constant with
redshift, cf. \citet{Weymann:1998gf}, only the literature data with redshift $z\!>\!1$ was used for the fit. 
The resulting parameters are given in Tables~\ref{tbl:linRegDnDzLyAlpha}.

Fig.~\ref{fig:litComp_KimFig7} shows the $\dif n / \dif z$ evolution for the
column density interval of $\log N_{\ion{H}{i}} = [14,17]$.
Our results (filled circles) agree well with previous findings 
at $z > 1.5$ ($\log(1+z) > 0.4$), confirming that
there is a real sightline variation in $\dif n / \dif z$.
\citet{Kim:2002fr} notes that the scatter between different sightlines increases as $z$ decreases
down to $z\!\sim\!2$.
In fact, the data of \citet{Janknecht:2006eu} at redshifts below
$z\!\sim\!2$ ($\log(1+z) \sim 0.45$) indicate that the scatter might well increase at lower
$z$, although the errors are still very large to draw any firm conclusions.
Considering that the FOS result is based on the equivalent width measurement,
and the conversion from the equivalent width to the column density requires  
the $b$ parameters of individual absorbers, which are ill-constrained 
at the FOS resolution, the full {\it HST}/STIS \ion{H}{i} sample 
toward some sightlines is in good agreement with the {\it HST}/FOS result
(blue open triangles),
although there still is a large sightline variation. The full \ion{H}{i} sample at $z < 0.4$ strongly 
supports the previous
conclusion obtained by the {\it HST}/FOS result, that $\dif n / \dif z$ flattens out at $z \le 1.5$.

\begin{table}
\centering
\small
\caption{Linear regression results for $\dif n / \dif z$
\label{tbl:linRegDnDzLyAlpha}}
\begin{tabular}{@{}c@{~~~}c@{~~~}c@{~~~}c@{~~~}c@{}}
\hline
\noalign{\smallskip}

\multicolumn{5}{c}{a) Using the Ly$\alpha$-only fits and fits including literature data for $z\!>\!1$} \\
\noalign{\smallskip}
\noalign{\smallskip}
\noalign{\smallskip}

 & \multicolumn{2}{c}{UVES quasar by quasar} & \multicolumn{2}{c}{quasar by quasar with lit.}  \tabularnewline
$\Delta \log N_\ion{H}{i}$ & $\log A$ & $\gamma$ & $\log A$ & $\gamma$ \tabularnewline
\hline
\noalign{\smallskip}
$13.64-17.0$ & $0.78\pm0.14$ & $2.55\pm0.27$ & $1.13\pm0.06$ & $1.90\pm0.11$ \tabularnewline
$14.0-17.0$ & $0.03\pm0.20$ & $3.40\pm0.36$ & $0.72\pm0.08$ & $2.16\pm0.14$ \tabularnewline
$14.5-17.0$ & $-0.14\pm0.26$ & $3.02\pm0.48$ & $0.46\pm0.11$ & $1.94\pm0.20$ \tabularnewline
$13.1-14.0$ & $1.46\pm0.11$ & $1.67\pm0.21$ & $1.52\pm0.05$ & $1.51\pm0.09$ \tabularnewline
$12.75-14.0$ & $1.98\pm0.08$ & $1.13\pm0.16$ &  & \tabularnewline
\hline
\hline

\noalign{\smallskip}
\noalign{\smallskip}

\multicolumn{5}{c}{b) The {\it mean} $\dif n / \dif z$ using the Ly$\alpha$-only fits at $0 < z < 4$} \\

\noalign{\smallskip}
\noalign{\smallskip}

$\Delta \log N_\ion{H}{i}$ & $\log A$ & $\gamma$ &  &  \tabularnewline
\hline
\noalign{\smallskip}

$13.1-14.0$ & $1.88\pm0.03$ & $0.89\pm0.06$ &  & \tabularnewline
$14.0-17.0$ & $1.03\pm0.07$ & $1.61\pm0.12$ &  & \tabularnewline

\hline
\hline

\noalign{\smallskip}
\noalign{\smallskip}

\multicolumn{5}{c}{c) The UVES high-order fits at $1.9 < z < 3.2$} \\
\noalign{\smallskip}
\noalign{\smallskip}

 & \multicolumn{2}{c}{quasar by quasar} & \multicolumn{2}{c}{Mean sample}\tabularnewline
$\Delta \log N_\ion{H}{i}$ & $\log A$ & $\gamma$ & $\log A$ & $\gamma$ \tabularnewline
\hline
\noalign{\smallskip}

$12.75-14.0$ & $1.85\pm0.12$ & $1.38\pm0.22$ & $1.89\pm0.13$ & $1.28\pm0.24$ \tabularnewline
$14.0-17.0$ & $-0.76\pm0.29$ & $4.91\pm0.53$ & $-0.65\pm0.36$ & $4.65\pm0.66$ \tabularnewline

\hline
\hline
\end{tabular}
\end{table}

The linear regression to 
our results only (the solid line) with $\gamma = 3.40\pm0.36$
is different at 3$\sigma$ from the fit
to all the available data at $z > 1$ ($\log(1+z) > 0.3$)
which yields $\gamma = 2.16\pm0.14$ (the dashed line). 
This discrepancy is mainly due to the sparse
data of our sample at higher redshift $z > 3.5$ ($\log (1+z) > 0.65$) and the missing
constraints at $z<2.0$. 
The discrepancy is also in part caused by how the power-law fit is performed.
Our maximum likelihood fit does the weighted fit. This gives a higher
weight on higher-$z$ data points where the 1$\sigma$ Poisson error is usually smaller.
The non-weighted fit for our UVES data only results in 
a steeper power-law slope,  
$(-0.84 \pm 0.24) \times (1+z)^{5.02\pm0.76}$. The non-weighted fit for all the data at $z > 1$
is $(0.76 \pm 0.18) \times (1+z)^{2.00 \pm 0.25}$.


Interestingly, some earlier numerical simulations and theories with a quasar-only UV background
have shown that there should be
a break in the $\dif n / \dif z$ evolution at $z\!\sim\!2$ due to the decrease in the quasar number density,
thus less available \ion{H}{i} ionising photons \citep{Theuns:1998ud, Dave:1999kk, Bianchi:2001ys}.
The green dot-dashed curve in Fig.~\ref{fig:litComp_KimFig7} shows one of such predicted 
$\dif n / \dif z$ evolutions by \citet{Dave:1999kk}, which outlines the Weymann et al. $\dif n / \dif z$
reasonably well. However, more recent simulations by \citet{Dave:2010ab} predict
different $\dif n / \dif z$ evolutions. These simulations are based on the various galactic 
wind models and the UV background contributed both by quasars and galaxies.  
The red dotted and the blue dot-dot-dot-dashed curves at $z < 2$ illustrate their predicted 
$\dif n / \dif z$ based on momentum-driven wind and no-wind models,
respectively. These newer simulations predict that $\dif n / \dif z$ continuously decreases
with decreasing redshift. Their momentum-driven wind model agrees reasonably
well with the observations by {\it HST}/STIS with the \ion{H}{i} absorber selection imposed
(measurement errors less than 40\% and $b < 40$ km s$^{-1}$), 
but not with the Weymann et al. data. A better, uniform dataset from {\it HST}/COS observations
should resolve this discrepancy at $z < 0.5$.

For the column density interval for stronger
absorbers $\log N_{\mathrm{\ion{H}{i}}} = [14.5,17.0]$, 
our data shows that the evolution continues to follow the 
empirical power-law with $\gamma = 3.02\pm0.48$ (see Table \ref{tbl:linRegDnDzLyAlpha}). 
However, the scatter between different sightlines is large as stronger absorbers are rare at all redshifts
\citep{Dave:2010ab}. There are more than 3$\sigma$ difference
between the lowest $\dif n / \dif z$ sightline and the highest $\dif n / \dif z$ sightline at $z\!\sim\!2$.
\citet{Kim:2002fr}
discuss the possibility on whether the column density evolution flattens out at $z\!<\!2.5$
($\log (1+z)\!<\!0.55$)
for this column density interval. Even though more data points are available in this study,
this question cannot be conclusively answered and more data covering lower redshifts are required.

\begin{figure}
\includegraphics[width=90mm]{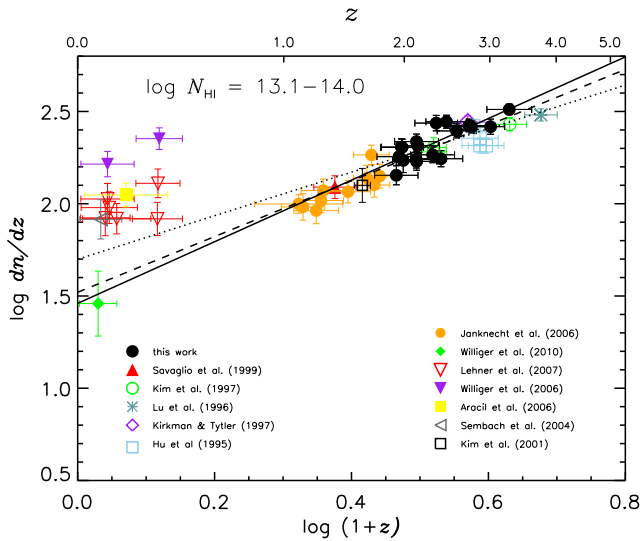}

\caption{The number density evolution of the Ly$\alpha$ forest in the
column density range $\log N_{\mathrm{\ion{H}{i}}} = [13.1,14.0]$. All the symbols have the same
meaning as in Fig.~\ref{fig:litComp_KimFig7}.}
\label{fig:litComp_KimFig9}
\end{figure}

The line number density evolution for low column density systems in the range of 
$\log N_{\mathrm{\ion{H}{i}}} = [13.1, 14.0]$ is presented in Fig. \ref{fig:litComp_KimFig9}.
Similar to Fig.~\ref{fig:litComp_KimFig7}, it suggests that the flattening of $\dif n / \dif z$ at $z < 1.5$
might continue at the lower column density range. However,  the sightline variation at $z < 0.4$ is 
larger at this column density range. This is in part caused by different analysis
methods and different S/N STIS data used by different studies. For example, the 
number density measured in the STIS spectrum toward
PKS0405$-$123 is different between the Williger et al. (2006) work (filled purple upside-down
triangles) and the Lehner et al. (2007) work (two of open red upside-down triangles).
Again the results from our data agree well with previous results found in the literature
at $z\!>\!1.5$.
The linear regression to our data at $z\!>1.5$ 
gives $\gamma\approx 1.67$, comparable to the fit including all available
literature data points at $z\! >\! 1.0$. 
However, these results do not compare well with the linear regression obtained by
\citet{Kim:2002fr} with $\gamma = 1.18\pm0.14$ (the dotted line),
a shallower $\dif n / \dif z$ evolution.
This discrepancy
arises due to their rather small sample size at $z\!<\!2.5$ and
more severe line blending at higher redshifts. Given a larger cosmic variance
at lower redshifts, the sample size becomes more important. At the same time,
line blending at high redshifts makes the detection of weak absorbers difficult.
This incompleteness effect has been shown to underestimate the line number density 
of low column density systems
at $\log N_{\mathrm{\ion{H}{i}}} = [13.3, 13.6]$ by $\sim\!17\%$ at $z\!\sim\!3$ 
($\log(1+z)\!\sim\!0.6$) and by $\sim\!35\%$ at $z\!\sim\!4$ \citep{Giallongo:1996lr}.
Both effects tend to flatten the evolution observationally from its true value.
In addition, the robust estimate of the exponent $\gamma$ requires a large $z$ leverage.  


\begin{figure}
\includegraphics[width=90mm]{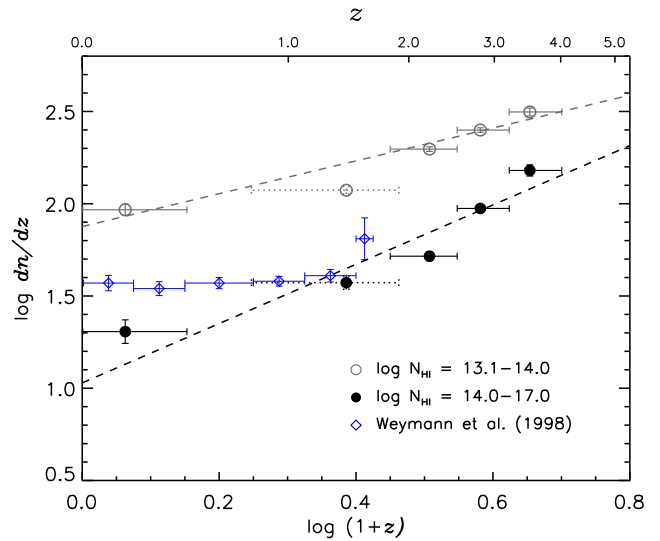}
\caption{The {\it mean} number density evolution of the Ly$\alpha$ forest.
The vertical error bars give the $1\sigma$ Poisson error,  while
the x-axis error bars show the redshift range covered by each data point.
The data point with the
dotted error bar indicate the \citet{Janknecht:2006eu} data. The dashed straight
lines mark the fit to the mean data excluding the \citet{Janknecht:2006eu} data.}
\label{fig:litComp_KimFig8}
\end{figure}

Even though there are not many sightlines covering $0.5\!<\!z\!<\!1.5$, we calculated
the {\it mean} $\dif n / \dif z$ from all the combined \ion{H}{i} fitted 
line lists including the literature data
in Fig. \ref{fig:litComp_KimFig8}. This {\it mean} $\dif n / \dif z$ is not
an averaged value of the individual sightlines.  
The literature data used in the combined line list include all the quasar sightlines
shown in Figs.~\ref{fig:litComp_KimFig7} and \ref{fig:litComp_KimFig9}, except
the {\it HST}/FOS Weymann et al. (1998) data, the Williger et al. (2006) data and the
Savaglio et al. (1999) data. The {\it HST}/FOS data was excluded since they were
based on the equivalent width measurements, while the Williger et al. (2006) data 
suffered from noise features. The Savaglio et al. (1999) result is from a single sightline and
provides the only data point besides the Janknecht et al. (2006) data at $z \sim 1$. 
Although the Janknecht et al. (2006) data also suffer from noise, they were from 9 sightlines.
We opted to use a result based on the analysis of multiple
sightlines from a single study. This helps to reduce any systematics caused by combining
results from different studies at $z \sim 1$. For $z < 0.4$, the systematic uncertainty is
larger since the line lists used are produced by different studies.

At $\log N_\mathrm{\ion{H}{i}}\!=\![13.1, 14.0]$, there might occur a flattening at $z\!\sim\!1$,
if the Janknecht data were included.
At $\log N_\mathrm{\ion{H}{i}}\!=\![14, 17]$, 
a single power law with $\gamma = 1.61\pm0.12$ 
does not give a good fit at $0\!<\!z\!<\!4$, regardless of the inclusion of the Janknecht et al. data.
It remains to be seen whether a single
power law fits the $\dif n / \dif z$ evolution for both high and low column 
density ranges at $z=0.4$.
It should be noted that the $\dif n / \dif z$ of
Lyman limit systems with a                                        
column density of $\log N_{\ion{H}{i}} = [17.2, 19.0]$ does not fit to a single
power law. It shows a slower evolution at $z < 2$ and evolves rapidly at $z > 2$
\citep{Prochaska:2010ab}, while the $\dif n / \dif z$ of damped Ly$\alpha$ systems with
$\log N_{\ion{H}{i}} = [20.3, 22.0]$ shows a single power-law evolution with a slope
$\gamma = 1.27\pm0.11$ at $0 < z < 4.5$
\citep{Rao:2006ab} .

Our results indicate that higher column density forest systems evolve more rapidly
than low column density systems and the number density of high column density systems
decreases faster with decreasing redshift. The increase in the scatter
at redshifts $z\!<\!2.5$ {\it might indicate} the transition point
where the evolving number density changes into
a non-evolving one, as is predicted in
earlier numerical simulations by \citet{Theuns:1998ud} and \citet{Dave:1999kk}.

\subsection{The differential column density distribution function}
\label{sec:diffDensLyAlpha}

The differential column
density distribution function (CDDF) is defined as the number of absorbers per
unit absorption distance $X(z)$ and per unit column density $N_\ion{H}{i}$. 
The absorption distance is calculated using Eq. \ref{eq:civRedPath}.
Empirically, the differential distribution function is reasonably well described by
a single power law at $z \sim 3$ at $\log N_{\ion{H}{i}} = [13, 22]$ as 
\begin{equation}
\frac{\dif n}{\dif N_\ion{H}{i} \, \dif X} = \left( \frac{\dif n}{\dif N_\ion{H}{i} \, \dif X} \right)_0
N_\ion{H}{i}^{\beta}, 
\end{equation}
where $\left( \dif n / (\dif N_\ion{H}{i} \, \dif X) \right)_0$ gives the normalisation
point of the distribution function and $\beta$ denotes its slope.
However, the detailed shape of the differential column density distribution
function is dependent on the $N_{\ion{H}{i}}$ column density range
\citep{Prochaska:2010ab, Altay:2010ab}. It shows a flattening around the transition
from the forest to the Lyman limit systems at $N_\ion{H}{i}$ at $\log N_\ion{H}{i}\!\sim\!17$.
Then it shows a steepening at $\log N_\ion{H}{i}\!\sim\!20$ where a transition occurs from
the sub-damped Ly$\alpha$ systems to the damped Ly$\alpha$ systems.

\begin{figure}
\includegraphics[width=1\columnwidth]{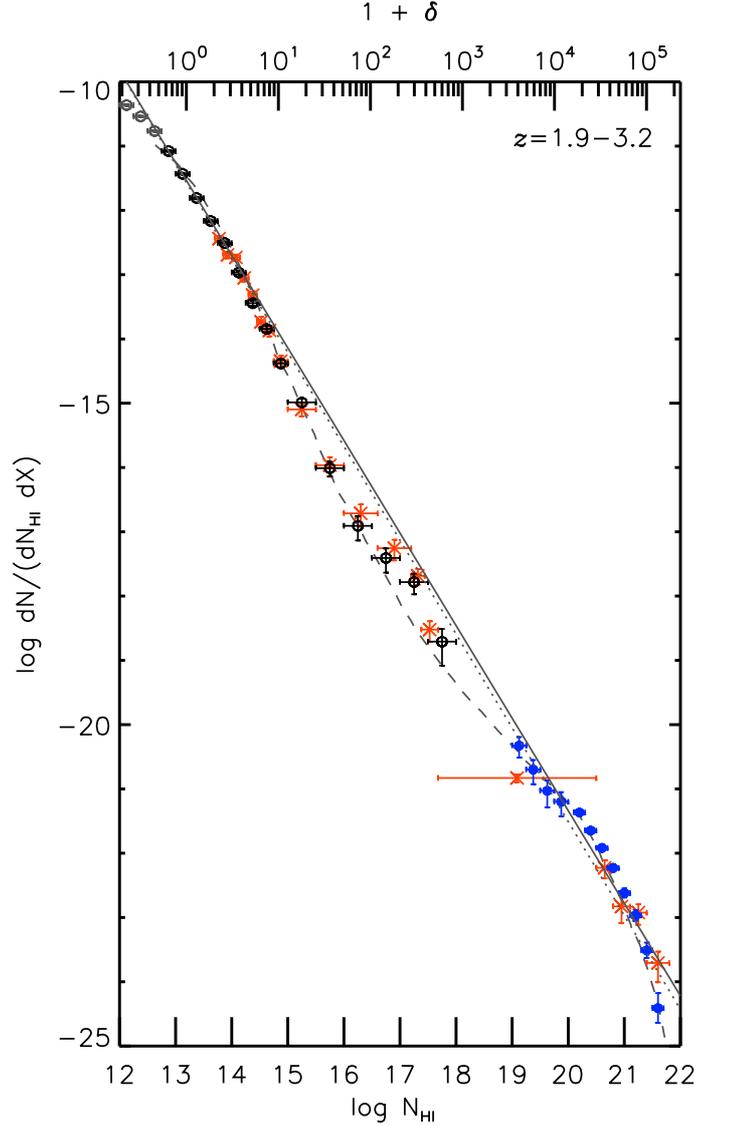}
\caption{
 The differential column density distribution at $1.9 < z < 3.2$
 using the Ly$\alpha$-only fits. Both black ($\log N_\ion{H}{i}\!\ge\!12.75$)
 and grey ($\log N_\ion{H}{i}\!<\!12.75$) data points
 show the results from our quasar sample. The grey data points
 mark the column
 densities that are affected by incompleteness. The stars are the data points obtained
 by \citet{Petitjean:1993bq}. The filled circles at $\log N_\ion{H}{i}\!>\! 20.1$ 
 and at $19\!<\!\log N_\ion{H}{i}\!<\! 20.1$ are from the 
 SDSS II DR7 at $<z>\, = 3.02$ by \citet{Noterdaeme:2009ab} and from \citet{OMeara:2007td}
 at $<z>\, = 3.1$, respectively. The
 solid line gives the power law fit to our data for $\log N_\ion{H}{i}\!=\![12.75, 14]$.
 The dotted
 line represent the fit obtained by \citet{Hu:1995lq}, while the dashed line represents
 a theoretical prediction at $z \sim 3$ by \citet{Altay:2010ab}.
 The vertical error bars indicate $1\sigma$ Poisson errors, while
the x-axis error bars show the $N_{\mathrm{\ion{H}{i}}}$ range covered by each data point.
 Gas overdensities on the top x-axis are computed using Eq. 10 from
 \citet{Schaye:2001yk} at $z\!=\!2.55$ (see text for details).
 }

 \label{fig:figure7}
\end{figure}

In Fig.~\ref{fig:figure7} we present the results using the Ly$\alpha$-only fits at
$1.9 < z < 3.2$. Note that the redshift range used for the CDDF analysis
is different from the one used for the $dn/dz$ analysis in Section~\ref{Sec3:1}.
The total absorption  
distance $X(z)$ at $1.9\! <\! z\! <\! 3.2$ is 21.8165. 
The binsize of $\log N_{\mathrm{\ion{H}{i}}}\!=\!0.25$ is used
at $\log N_{\mathrm{\ion{H}{i}}}\!= [12.0, 15.0]$, then the binsize of 0.5 at 
$\log N_{\mathrm{\ion{H}{i}}}\!= [15.0, 18.0]$. 
To increase the column density coverage, we include
results from \citet{Noterdaeme:2009ab} and \citet{OMeara:2007td} for
$\log N_{\mathrm{\ion{H}{i}}}>19$ at $z\sim 3$ from the SDSS II DR7 data{\footnote{The 
plotted data points from the literature are the reported ones in each study.
Both \citet{OMeara:2007td} and \citet{Noterdaeme:2009ab} used the same
cosmology as ours, while \citet{Petitjean:1993bq} used the $q_{0} = 0$ cosmology.
The absorption distance $X(z)$ in our cosmology is about 6\% smaller than theirs.
Since the CDDF uses the logarithm value of $X(z)$, the difference in the CDDF
is negligible even without converting their CDDF to our cosmology.}.
The top x-axis is in units of the gas overdensity $\delta$ which was computed according to Eq. 10
by \citet{Schaye:2001yk}
\label{eq:dcdf_columnDensToOverdens}
\begin{eqnarray}
N_\ion{H}{i} & \sim &  2.7 \times 10^{13} \, \left( 1+\delta \right)^{3/2} \, \times  \nonumber \\
			&&       \times  \,T_4^{-0.26} \, \Gamma_{12}^{-1}
                                \left( \frac{1+z}{4} \right)^{9/2} 
                        \left( \frac{\Omega_b h^2}{0.02} \right)^{3/2}
                                \left( \frac{f_g}{0.16} \right)^{1/2} \textrm{ cm}^{-2}.
\end{eqnarray}
Here, the gas temperature $T$ is assumed to be $T = T_4 \times 10^4 \textrm{ K}$, the photoionisation rate
$\Gamma = \Gamma_{12} \times 10^{-12} \textrm{ sec}^{-1}$. The parameter $f_g$ denotes the fraction of mass
in gas. The IGM gas temperature is assumed to be governed by the effective equation of state
$T = T_0 (1+\delta)^{\gamma-1}$, where $T_0$ is the temperature at the cosmic density \citep{Hui:1997uq}.
For $\Gamma_{12}$ and $\gamma$, we interpolated results obtained by \citet{Bolton:2008kx}.
We assumed that $T_0$ is $2 \times 10^4 \textrm{ K}$, $f_g = 0.16$, and
$\Omega_b h^2 = 0.0227$ \citep{Schaye:2001yk}. As the same overdensity corresponds to a
different $N_\ion{H}{i}$ at different $z$, the overdensity plotted in Fig.~\ref{fig:figure7}
is at the mean redshift, $z\!=\!2.55$.

We compare our results with the observations by \citet{Petitjean:1993bq}
and \citet{Hu:1995lq}.
Our results are in good agreement with the \citet{Petitjean:1993bq} data
over the whole column density range down to $\log N_\ion{H}{i} \sim 13.5$,
following a power law at $\log N_\ion{H}{i} = [12.75, 14]$. 
At smaller column densities $\log N_\ion{H}{i} < 12.75$, the CDDF 
starts to deviate from a power law due to the sample
incompleteness for weak absorbers \citep{Kim:1997rt}.
From the linear regression, we
find $\log \left( \dif n / (\dif N_{\ion{H}{i}} \, \dif X) \right)_0
= 7.34 \pm 0.42$ and a slope of $\beta = -1.43 \pm 0.03$ for the
$\log N_\ion{H}{i} = [12.75, 14]$ range  (the solid line).
This result is slightly lower than
$\beta = -1.46$ (no errors given) by \citet{Hu:1995lq} (the dotted line) or
$\beta = -1.49\pm0.02$ by \citet{Petitjean:1993bq}.

The distribution function
becomes steeper at $\log N_{\ion{H}{i}}\!>\!15$, then becomes shallower at
higher $\log N_{\ion{H}{i}}$, as previously observed
\citep{Petitjean:1993bq, Kim:1997rt, Prochaska:2010ab}.
This result agrees well with the theoretical prediction at $z \sim 3$ (the dashed line)
by \citet{Altay:2010ab} at $\log N_\ion{H}{i} \le 16$, but starts to show a noticeable disagreement
at the 1--3$\sigma$ level at $\log N_\ion{H}{i} = [16, 18]$, in part due to the lack
of enough high-column density systems in our small sample.
We will address the shape of the CDDF in more detail
in the next section
using results from the high-order fit sample. 

\section{Analysis using higher-order Lyman lines} 
\label{Sec:Analysis}

In the last section, we checked the Ly$\alpha$ absorber number density 
evolution and the differential column density distribution obtained from the Ly$\alpha$-only fits
for consistency with previous studies. The analysis is now revisited with the
results from the Voigt profile analysis including the higher order transitions
at $1.9\!<\!z\!<\!3.2$,
hence a sample with a more reliable $N_{\mathrm{\ion{H}{i}}}$.
Therefore, it can be established whether the dip 
seen in the differential column density distribution 
at $\log N_{\mathrm{\ion{H}{i}}}$ between $14.5$
and $18$ is a physical feature or 
just an imprint of uncertainties
in $N_\ion{H}{i}$. 
All the results from this section are tabulated in Appendix A.

\subsection{The mean number density evolution \label{sec:combNumDens}}

We now revisit the line number density evolution using the high-order Lyman sample, as described
in the previous section. On a quasar by quasar analysis we
determine $\dif n / \dif z$ for a low column density range of $\log N_\ion{H}{i} = [12.75, 14.0]$
and for high column densities of $\log N_\ion{H}{i} = [14, 17]$. The lower column
density range is chosen in such a way that the part of the differential column density distribution 
function which follows a power-law is covered,
whereas the $\log N_\ion{H}{i} = [14, 17]$ interval covers those systems responsible
for the dip in the column density distribution function.

The results are presented in Fig.~\ref{fig:dNdz_QSOonly}.
Linear regressions from the data are obtained and the resulting parameters are summarised
in Table \ref{tbl:linRegDnDzLyAlpha}.
Similar to the previous analysis, the line
number density shows a decrease with decreasing redshift. No significant
differences between the two different fits are present, even though
the total redshift coverage used for the high-order fit is about 20\% smaller.
In the case of the $\log N_\ion{H}{i} = [14, 17]$ interval,
the slope of the power law steepens from the Ly$\alpha$-only slope of
$\gamma = 3.40\pm0.36$ to $\gamma = 4.91\pm0.53$ for the high-order fit.
This is in part caused by that the number of high column density absorbers
is larger in the high-order fit sample.
However, the slopes of the two samples are still in the $2\sigma$ uncertainty range,
rendering the two results consistent to each other. Similar results are obtained
for the $\log N_\ion{H}{i} = [12.75, 14.0]$ range. The slope for the high-order
fit increases from $\gamma = 1.13\pm0.16$ for the Ly$\alpha$-only fit to 
$\gamma = 1.38\pm0.22$. Again, 
the results from the two samples agree within the $1\sigma$ 
uncertainty range.

\begin{figure}
 \includegraphics[width=90mm]{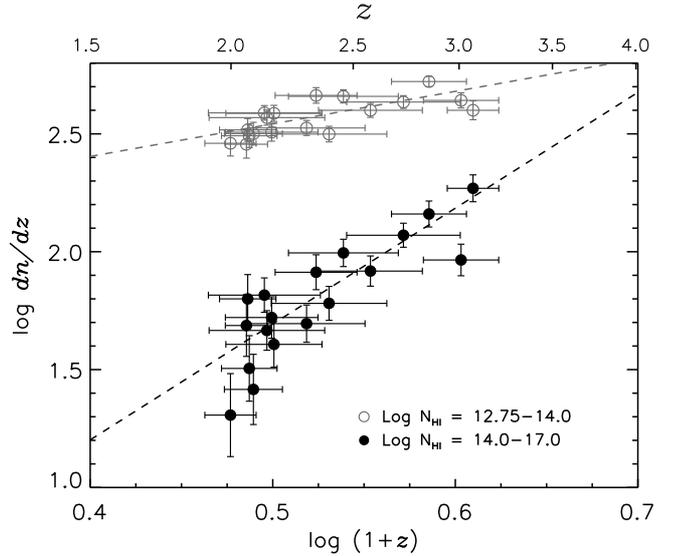}

 \caption{
 The line number density evolution derived on a quasar by quasar analysis using the high-order Lyman
 sample for column density intervals of $\log N_\ion{H}{i} = [12.75, 14.0]$ and
 $\log N_\ion{H}{i} = [14, 17]$. The vertical error bars indicate 
 $1\sigma$ Poisson errors, while
the x-axis error bars show the redshift range covered by each sightline.
The straight lines denote results from a linear regression
 to the data with parameters given in Table \ref{tbl:linRegDnDzLyAlpha}.
 The data are tabulated in
 Table~\ref{tbl:dndz_ap1}. 
 }
 \label{fig:dNdz_QSOonly}
\end{figure}

\begin{figure}
 \includegraphics[width=90mm]{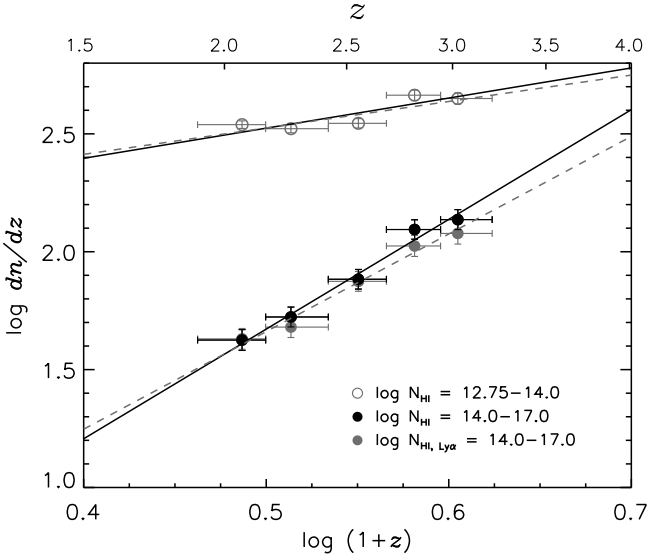}

 \caption{
 The {\it mean} line number density evolution of the combined sample as a function of redshift using the 
 high-order Lyman-series sample for column density intervals of $\log N_\ion{H}{i} = [12.75, 14.0]$ 
 and $\log N_\ion{H}{i} = [14, 17]$. The sample is binned in redshift with $\Delta z = 0.26$,
 starting from $z = 1.90$.
 The vertical error bars indicate 
 $1\sigma$ Poisson errors, while
the x-axis error bars show the redshift range covered by each data point.
 For comparison, the results of the Ly$\alpha$-only fits (grey open circles) are shown for the 
 $\log N_\ion{H}{i} = [14, 17]$ interval. 
 The straight solid lines denote results from a 
 linear regression to the binned data. Two dashed lines represent the  
 {\it mean} number density evolution of the Ly$\alpha$-only fit sample for 
 $\log N_\ion{H}{i} = [14, 17]$ ($\log \dif n / \dif z = (-0.41\pm0.35) + (4.14 \pm 0.63) \times \log (1+z)$)
 and for $\log N_\ion{H}{i} = [12.75, 14]$
 ($\log \dif n / \dif z = (1.96\pm0.13) + (1.12 \pm 0.24) \times \log (1+z)$), respectively.
 Exactly same redshift range was used for both fit samples. The data are tabulated in
 Table~\ref{tbl:dndz_ap2}. 
 The parameters of the fits are given in Table~\ref{tbl:linRegDnDzLyAlpha}.
 }
 \label{fig:dNdz}
\end{figure}

In previous studies the number density evolution has been usually 
derived on a quasar
by quasar analysis. Previous studies did not have enough quasar sight lines available
to sample the number density evolution at smaller redshift interval $\Delta z$,
without suffering from small number statistics. 
Our sample of 18 high-redshift quasars is characterised by a large 
redshift distance coverage in the redshift range of $1.9 < z < 3.2$ (see Fig. \ref{fig:totX_z}). 
As a result, a large number of absorption lines is available for small 
redshift intervals to combine
the individual quasar line lists into one big sample. However, due to the larger
cosmic variance at low redshifts from the structure formation, 
the redshift bin size should not be too small.
From this combined sample, 
the evolution of the {\it mean} number
density is derived in redshift bins of $\Delta z = 0.26$, starting from $z = 1.90$.

Results of the combined line number density evolution are shown in Fig. \ref{fig:dNdz} for identical
column density ranges as used in the quasar by quasar analysis. Error bars have been determined 
using the bootstrap technique. For comparison, results using the Ly$\alpha$-only fits
are overplotted as grey open circles for the high column density bin.

The high column density results are similar to the ones obtained
from the Ly$\alpha$-only fits. 
The number density itself is higher in the high-order
fits,
since some strongly saturated systems break up into multiple, strong components in the high-order
Lyman transition. 
In addition, three absorbers (two toward HE0940$-$1050 and one toward Q0420$-$388) 
were found to be a Lyman limit system with $\log N_\ion{H}{i}\!>\!17$ in the Ly$\alpha$-only fit. Therefore,
these systems were not included in the Ly$\alpha$-only results. However,
these Lyman limit systems break up into multiple weaker components in the high-order fit
and contribute to the number count in the high order fit analysis. 
However, the differences between the
two samples are smaller than the statistical uncertainties. 

At low column densities, no noticeable differences between the
two samples are observed, as expected. 

Again, linear regressions have been determined 
and their parameters are given in
Table \ref{tbl:linRegDnDzLyAlpha}.
At $\log N_\mathrm{\ion{H}{i}} = [12.75, 14.0]$,
the slope of our combined sample is $\gamma =
1.28\pm0.24$, similar to $\gamma = 1.38\pm0.22$ from the quasar by quasar analysis.
At $\log N_\ion{H}{i} = [14, 17]$, 
the slope of the combined sample $\gamma = 4.65\pm0.66$ is 
also similar to $\gamma = 4.91\pm0.53$ obtained from the quasar by quasar analysis.
The slopes from both analyses of our high-order fit sample at $\log N_\ion{H}{i} = [14, 17]$  
are steeper than the ones obtained from the Ly$\alpha$-only fit sample.
In particular, the ones from the combined sample differ more than 3$\sigma$. 
This difference is
mainly caused by that the redshift range used for the combined sample 
is different for two analyses. For the Ly$\alpha$-only fit, the {\it mean} $\dif n / \dif z$
is derived for $0 < z < 4$, while for the high-order fit it is restricted to 
$1.9 < z < 3.2$.

\begin{figure}
 \centering \includegraphics[width=1\columnwidth]{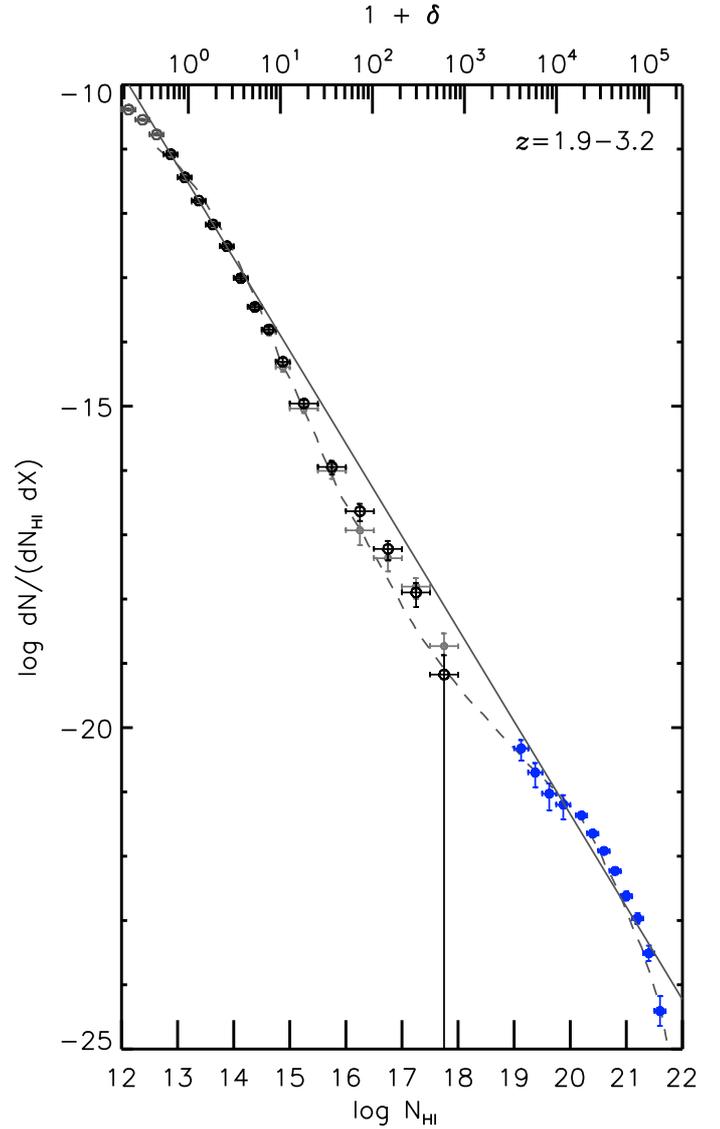}
 \caption{The differential column density distribution at $1.9 < z < 3.2$
 of our quasar sample using the high-order Lyman fit. Black and grey data points
 show the results from our quasar sample. The grey data points below
 $\log N_\ion{H}{i} < 12.75$ mark the column densities that are affected by incompleteness.
 The grey data points
 above $\log N_\ion{H}{i}\!>\!12.75$
 represent the results from the Ly$\alpha$-only fit. The dashed line represents
 a theoretical prediction at $z\!\sim\!3$ by \citet{Altay:2010ab}. The vertical error bars indicate
 $1\sigma$ Poisson errors, while
the x-axis error bars show the $N_{\mathrm{\ion{H}{i}}}$ range covered by each data point.
The filled circles at $\log N_{\mathrm{\ion{H}{i}}} \!>\! 20.1$ and
 at $19.0\!<\!\log N_{\mathrm{\ion{H}{i}}}\!<\! 20.1$ are from the SDSS II DR7 at $<\!z\!>\, = 3.02$
 by \citet{Noterdaeme:2009ab} and from \citet{OMeara:2007td} at $<\!z\!>\, = 3.1$,
 respectively. The solid line gives the power law fit to our data for 
 $\log N_{\mathrm{\ion{H}{i}}} = [12.75, 14.0]$.
 The overdensity plotted on the top x-axis is calculated at $z=2.55$.}
 \label{fig:dNdNdX}
\end{figure}

\subsection{The differential column density distribution function \label{sec:highOrderDCDF}}

Using the high-order fits, we have derived the differential column density distribution
function (CDDF) for $1.9\!<\!z\!<\!3.2$, analogous
to Section \ref{sec:diffDensLyAlpha}. In Fig.~\ref{fig:dNdNdX} we show
the results for the entire redshift range. As in Fig.~\ref{fig:figure7},
the binsize of $\log N_{\mathrm{\ion{H}{i}}}\!=\!0.25$ is used
at $\log N_{\mathrm{\ion{H}{i}}}\!= [12.0, 15.0]$, then the binsize of 0.5 at
$\log N_{\mathrm{\ion{H}{i}}}\!= [15.0, 18.0]$.
The total absorption distance is $X(z) = 21.8165$, the same value used
for the Ly$\alpha$-only fit CDDF analysis.
As with the Ly$\alpha$-only fits,
we included observations by \citet{Noterdaeme:2009ab} and \citet{OMeara:2007td}.

The high-order fit results show a power law relation which is almost identical to the
results of the Ly$\alpha$-only fits. As with the
Ly$\alpha$-only fits, the differential column density distribution function shows a deviation from the
empirical power law at column densities between $14\!<\!\log N_\ion{H}{i}\!<\!19$.
Since the column density distribution deviates from a single power law at
$\log N_\ion{H}{i}\!\sim\!14$, we have individually fitted power laws to four
column density intervals of $\log N_\ion{H}{i} = [12.75, 14.0]$, $[14,15]$, $[15,18]$, and
$[12.75, 18.0]$ at $1.9 < z< 3.2$, characterising the shape of the distribution function.
The resulting parameters are listed in Table \ref{tbl:dNdNdX_diffN}.

At $\log N_\ion{H}{i} = [12.75, 14.0]$, the
linear regression yields a normalisation point of $\log \left( \dif N / (\dif N_\ion{H}{i} \, \dif X) \right)_0
= 7.41 \pm 0.42$ and a slope of $\beta = -1.44 \pm 0.03$. 
This result is almost identical to the Ly$\alpha$-only
fit, since differences between the Ly$\alpha$-only and the high-order fits start to be
significant at
$\log N_\ion{H}{i}\!>\!15$ (see Fig. \ref{fig:dNdNdX}).
The high-order fits show a larger number of absorbers at
$15\!<\!\log N_\ion{H}{i}\!<\!17$ than the Ly$\alpha$-only fits. However, at higher column densities, the number
of absorbers is lower for the high-order fits than for the Ly$\alpha$-only fits. This again
indicates the breaking up of high column density systems into multiple lower-$N_\ion{H}{i}$ ones when
including higher
transitions than Ly$\alpha$.
For the entire redshift sample,
the slope becomes steeper from $\sim\!-1.44$
to $\sim\!-1.67$ at $\log N_\ion{H}{i} = [14, 15]$.
Then at the higher column density range $\log N_\ion{H}{i} = [15, 18]$,
the slope becomes shallower to $\sim\!-1.55$, a trend shown in the
numerical simulation (the dashed line) by \citet{Altay:2010ab} in
Fig. \ref{fig:dNdNdX}.

\begin{table*}
\centering
\begin{minipage}{\textwidth}
\centering
\caption{Linear regression results for the differential column density distribution as a function
of redshift and column density using the high-order fit. Here the normalisation
point $\log \left( \dif n / (\dif N_\ion{H}{i} \, \dif X) \right)_0$ is denoted by $B$. \label{tbl:dNdNdX_diffN}}
\begin{tabular}{ccccccccc}
\hline
\noalign{\smallskip}

 & \multicolumn{2}{c}{$\log N_\ion{H}{i}=12.75-14.0$} & \multicolumn{2}{c}{$\log N_\ion{H}{i}=14.0-15.0$} &
\multicolumn{2}{c}{$\log N_\ion{H}{i}=15.0-18.0$} & \multicolumn{2}{c}{$\log N_\ion{H}{i}=12.75-18.0$}\tabularnewline
$z$ & $B$ & $\beta$ & $B$ & $\beta$ & $B$ & $\beta$ & $B$ & $\beta$\tabularnewline
\hline
\noalign{\smallskip}

$1.9-3.2$ & $7.41\pm0.42$ & $-1.44\pm0.03$ &  $10.63\pm1.32$ & $-1.67\pm0.09$  &
    $8.70\pm1.33$ & $-1.55\pm0.08$ & $9.57\pm0.20$ & $-1.60\pm0.01$ \tabularnewline
$1.9-2.4$ & $8.08\pm0.55$ & $-1.49\pm0.04$ & $11.16\pm2.07$ & $-1.72\pm0.14$  &
    $7.39\pm3.00$ & $-1.48\pm0.19$ & $10.32\pm0.31$ & $-1.66\pm0.02$ \tabularnewline
$2.4-3.2$ & $6.72\pm0.55$ & $-1.38\pm0.04$ & $9.96\pm1.70$ & $-1.62\pm0.12$ &
    $8.64\pm1.59$ & $-1.54\pm0.10$  & $8.59\pm0.25$ & $-1.52\pm0.02$ \tabularnewline

\hline
\end{tabular}
\end{minipage}
\end{table*}

\begin{figure*}
 \includegraphics[width=170mm]{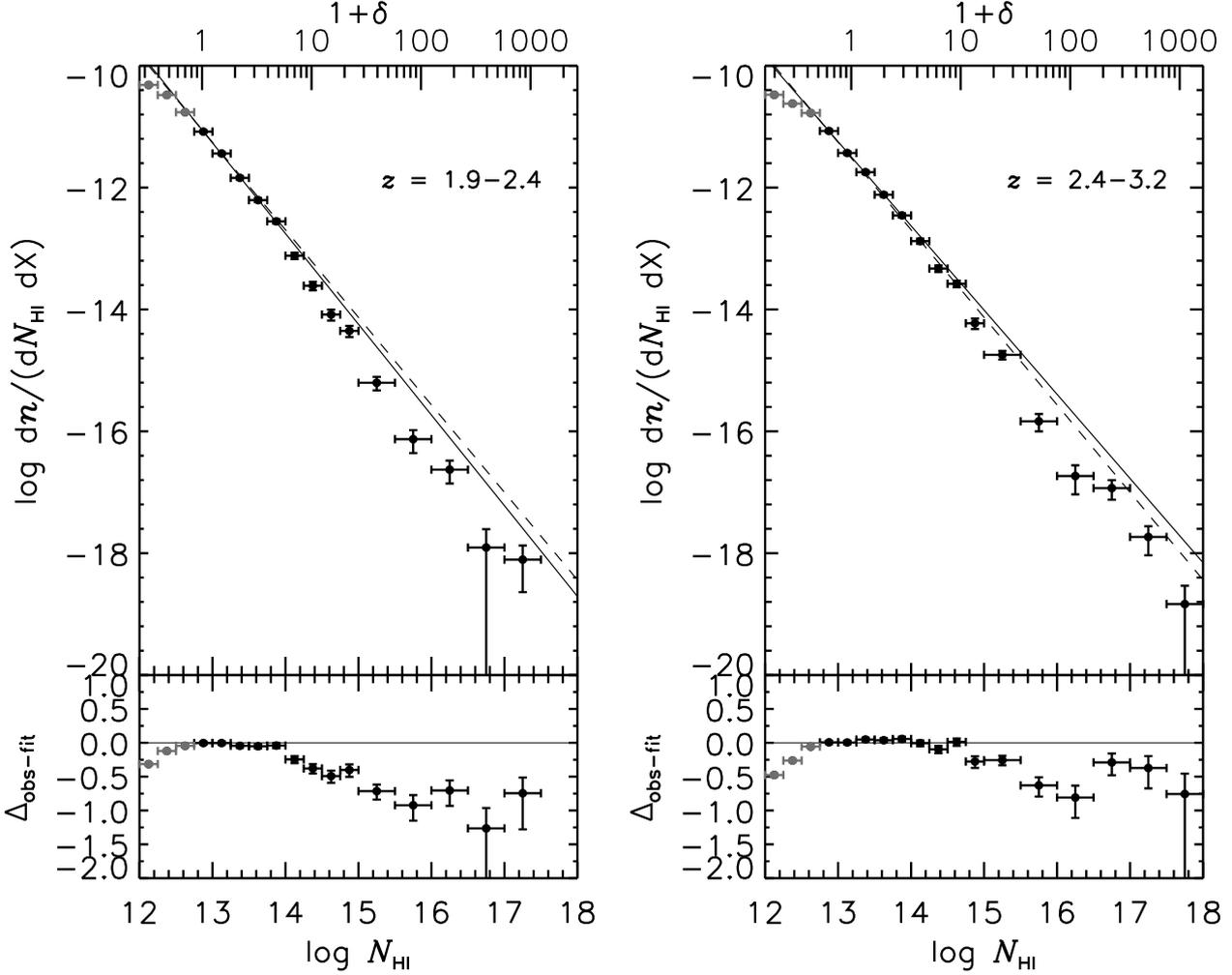}
 \vspace{-0.5cm}
 \caption{Upper panels: The differential column density distribution as a function of redshift.
Both black and grey data points show the results from the high-order Lyman sample.
The grey data points at $\log N_{\mathrm{\ion{H}{i}}}\!<\!12.75$
mark the column densities that are affected by incompleteness.
The vertical error bars indicate $1\sigma$ Poisson errors, while
the x-axis error bars show the $N_{\mathrm{\ion{H}{i}}}$ range covered by each data point.
The black solid line gives the power law fit for $\log N_\ion{H}{i} = [12.75, 14.0]$
at each redshift bin,
whereas the dashed line is the fit to the $z=[1.9, 3.2]$ redshift range (see Fig.
\ref{fig:dNdNdX}). The overdensity plotted on the top x-axis is calculated at the
mean $z$ for each redshift bin.
Lower panels: the residuals from the power law fit from the entire redshift range at
$\log N_\ion{H}{i} = [12.75, 14.0]$.}
 \label{fig:dNdNdXfuncZ}
\end{figure*}

In order to determine the redshift evolution of the differential column density distribution, 
we split the sample into two redshift bins: $z=[1.9, 2.4]$ and $[2.4, 3.2]$.
Fig. \ref{fig:dNdNdXfuncZ} shows the CDDF at two different
redshift bins, where we overplot the power-law fit at $\log N_\ion{H}{i} = [12.75, 14.0]$
for each redshift bin as the solid line.
We also overplot the results of the power-law fit to the entire
redshift range $1.9\!<\!z\!<\!3.2$ at the same column density range as the dashed line.
For the redshift intervals of
$z = [1.9, 2.4]$ and [2.4, 3.2], the absorption distance is
$X(z) = $ 11.8049 and 10.0116, respectively.

Unfortunately, the uncertainties in the power-law fit parameters at each redshift bin
do not allow us to constrain the shape of the
distribution as a function of redshift reliably. 
Comparing the slope of the linear relations shows that
the CDDF becomes slightly steeper at low redshift for 
$\log N_{\mathrm{\ion{H}{i}}} = [12.75, 14.0]$,
from $\gamma = -1.38\pm0.04$ at high $z$ to $\gamma = -1.49\pm0.04$ at low $z$.
However, the slopes are still
consistent within 2$\sigma$, i.e. no significant CDDF evolution, cf. \cite{Williger:2010qe}. 
They are also consistent with the result from the entire redshift range within $1\sigma$.


Let us now focus on column densities above $\log N_{\mathrm{\ion{H}{i}}}\!>\!14.0$. From  
Fig.~\ref{fig:dNdNdX} we have seen that the differential column density distribution 
deviates from the power 
law form for column densities $\log N_{\mathrm{\ion{H}{i}}}\!>\!14.0$. 
The lower
panels of Fig.~\ref{fig:dNdNdXfuncZ} show the difference between the observed CDDF
and the power-law fit to the CDDF for the entire redshift range (the dashed lines).
The entire redshift fit was used since the comparison requires an absolute reference.
From the lower panels, it is clear that 
the deviation from the power law is
stronger for the low redshift bin. At the same time, the deviation column density above which
the deviation starts to be noticeable is lower at low redshift, from 
$\log N_{\mathrm{\ion{H}{i}}}\! \sim \! 14.7$ at $2.4\!<\!z\!<\!3.2$ to 
$\log N_{\mathrm{\ion{H}{i}}}\! \sim \! 14.0$ at $1.9\!<\!z\!<\!2.4$.

Note that no such break in the CDDF has been seen in the 
$\log N_{\mathrm{\ion{H}{i}}} = [12.5, 16]$ at $z < 2$, cf. Fig. 5 of \citet{Williger:2010qe}
at $z\!\sim\!0.08$ and Fig. 9 of \citet{Ribaudo:2011ab} at $z\!<\!2$.
Both works also found a steeper CDDF slope of $\sim$1.75. Some of the discrepancy
is caused by the different fitting methods, 
the \ion{H}{i} selection criterion discussed in Section~\ref{Sec3:1} and 
the column density range over which the power law was performed.
On the other hand, \citet{Prochaska:2010ab} found a more significant dip in the column density
distribution function at
$\log N_{\mathrm{\ion{H}{I}}} = [14, 19]$ at $z \sim 3.7$ (similar
to the Altay simulation at $z\!\sim\!3$ indicated by the dashed curve in Figs. \ref{fig:figure7} and
\ref{fig:dNdNdX}). However, the dip shown at $z\!\sim\!3$ in Fig.~\ref{fig:dNdNdXfuncZ} 
(the high redshift bin)
is not as strong as the one predicted by the Altay simulation, although both results
are still considered to be consistent within 2$\sigma$.
These differences could be simply
due to our small sample size, or due to the different analysis method or due to
the strong CDDF evolution between $z\!\sim\!4$ and
$z\!\sim\!0$.

Note that the dip shown in Fig. \ref{fig:dNdNdXfuncZ} is not caused by
self-shielding. Self-shielding causes the number density of absorbers to increase. 
Self-shielding
becomes important at $\log N_{\mathrm{\ion{H}{i}}} \ge 16$ and its effect becomes evident at 
$\log N_{\mathrm{\ion{H}{i}}}\!\ge\!17$ with a shallower slope than the extrapolated one 
at the lower $\log N_{\mathrm{\ion{H}{i}}}$ 
\citep{Altay:2010ab}.
However, the dip in discussion occurs at $\log N_{\mathrm{\ion{H}{i}}}\!=\![14.5, 17.0]$ 
compared to the extrapolated power-law slope at 
$\log N_{\mathrm{\ion{H}{i}}}\!=\![12.75, 14.5]$. In addition,
the deviation $N_{\mathrm{\ion{H}{i}}}$ from this single power law starts at
$\log N_{\mathrm{\ion{H}{i}}}\!\sim\!14.5$, where self-shielding has no effect.

\section{Characteristics of the metal enriched forest}
\label{Sec:CIV}

The discovery of metal lines which are associated with 
\ion{H}{i} absorber in the Ly$\alpha$ forest, 
such as \ion{C}{iv} or \ion{O}{vi} \citep{Cowie:1995mb, Songaila:1998kh, Schaye:2000bd}, 
have raised the question of
how the IGM has been metal enriched.
As the forest has a high temperature and a low gas density, it is not likely to form stars in-situ.
Metals should be transferred from galaxies by 
e.g. galactic outflows
\citep{Aguirre:2001ud, Schaye:2003fc, Oppenheimer:2006lo}.
In recent years, studies on galaxy-galaxy pairs at high redshift have revealed some evidence
that metals associated with the Ly$\alpha$ forest reside in the circum-galactic medium 
\citep{Adelberger:2005wa, Steidel:2010pd, Rudie:2012fk}. 
In this interpretation, the metal-enriched forest cannot be called
the IGM in the conventional sense and is likely to show a different evolutionary behaviour
compared to the metal-free forest.
In order to learn more about these enriched hydrogen absorbers, we characterise \ion{C}{iv} enriched 
\ion{H}{i} absorbers in this section by determining their number density evolution and differential column 
density distribution.
Note that we excluded Q0055$-$269 and J2233$-$606 for both the \ion{C}{iv} enriched
forest and the unenriched forest samples in this section, as
their \ion{C}{iv} region has a much lower S/N of $\sim \!40$ per pixel 
compared to the other 
16 quasar spectra whose S/N is greater than 100 per pixel in most \ion{C}{iv} regions.
Due to the wavelength gap caused by the UVES dichroic setup, the \ion{C}{iv} redshift coverage
is shorter than the \ion{H}{i} coverage for Q0420$-$388, HE0940$-$1050 and
HE2347$-$4342. We excluded the $\pm 200$ km s$^{-1}$ region from the wavelength
gap and included the \ion{C}{iv} region only when it covered both doublets. 
The excluded \ion{C}{iv} redshift range for these
three quasars is listed in Table~\ref{tab1}. 
In this section, we used the column density and $b$ parameter of \ion{H}{i} 
from the high-order Lyman fit, unless stated otherwise.
All the results from this section are tabulated in Appendix A.

\begin{figure}
\includegraphics[width=84mm]{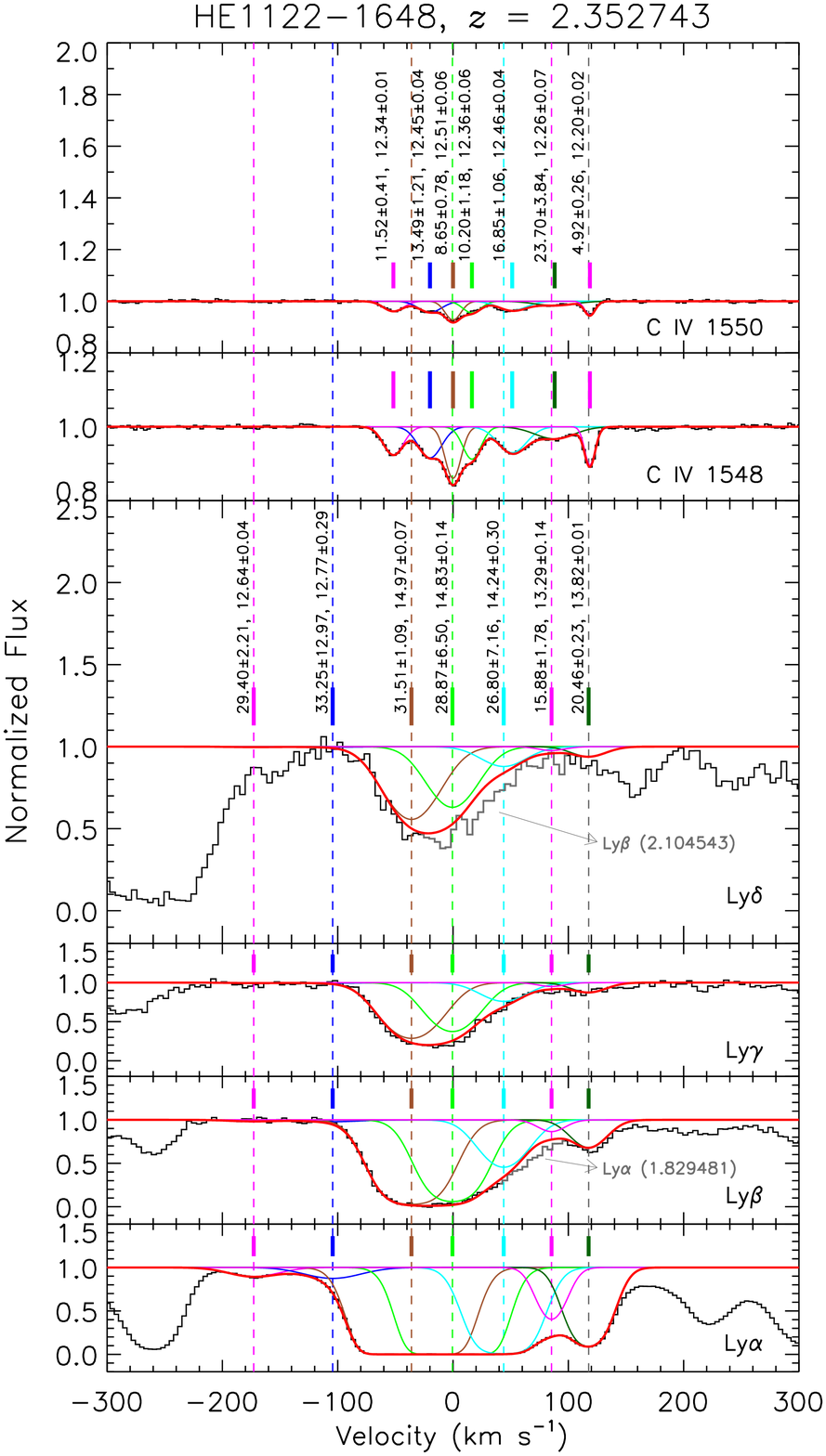}

\caption{Example of a velocity plot (a relative velocity vs normalised absorption
profile plot) of \ion{H}{i} and
associated \ion{C}{iv} detected in the $z = 2.352743$ absorber in the
spectrum of HE1122$-$1648.
The strongest \ion{C}{iv} component is set to be at the zero velocity.
The observed spectra are plotted as a histogram, while Voigt-profile
fits are as a smooth curve.
Thick red curves are the combined fit from
individual components.
The heavy tick marks
above the profiles indicate the velocity centroid of each component.
Non-negligible blends by other ions
are indicated in gray. The $b$ value (in
km s$^{-1}$)
and $\log N_{\ion{H}{i}}$ with the VPFIT fitting errors are displayed next a
tick mark indicating the center of the component.
}
\label{fig:velocityplot}
\end{figure}

\subsection{Method}
\label{sec:Method}

Unfortunately there is no one-to-one relation between \ion{H}{i} lines and \ion{C}{iv} lines.
Fig.~\ref{fig:velocityplot} shows a velocity plot 
(the relative velocity centered at the redshift of an absorber vs normalised flux)
of a typical \ion{C}{iv}-enriched \ion{H}{i} absorber in the spectrum of HE1122$-$1648. 
The vertical dashed lines indicate the velocity of individual \ion{H}{i} components. Not
all \ion{H}{i} lines can be directly assigned to one or only one 
\ion{C}{iv} component. For example,
the \ion{H}{i} component at $-36.02$ km s$^{-1}$ could be associated either with the first \ion{C}{iv}
component at $-51.72$ km s$^{-1}$ or with the second one at $-19.97$ km s$^{-1}$, or with both.
A general trend is that the associated \ion{C}{iv} features show an increased number
of velocity components as $N_\mathrm{\ion{H}{i}}$ increases.
The absorption line centers of \ion{H}{i} and \ion{C}{iv} lines often show velocity 
differences as well,
indicating that the \ion{H}{i}-absorbing gas might not be co-spatial with 
the \ion{C}{iv}-producing gas.
Therefore, we apply a simple assigning method to our fitted absorber line lists, 
in order to determine if
an \ion{H}{i} absorption line is associated with \ion{C}{iv}. 

We consider an
\ion{H}{i} absorber to be metal enriched if a   
\ion{C}{iv} line with $N_{\mathrm{\ion{C}{iv}}}$ greater than a threshold 
value exists within  
the velocity range $\pm \Delta v_{\mathrm{\ion{C}{iv}}}$
centered at each identified \ion{H}{i} line.
The threshold $N_{\mathrm{\ion{C}{iv}}}$ 
should be large enough not to be affected by the incompleteness 
of weak \ion{C}{iv} detection, but not too large so that there are enough
\ion{C}{iv} enriched absorbers to have a meaningful statistics.
This method can assign one \ion{H}{i} component with multiple \ion{C}{iv}
components and vice versa. As we are not concerned with the one-to-one relation between
$N_{\mathrm{\ion{C}{iv}}}$ and $N_{\mathrm{\ion{H}{i}}}$ of each \ion{H}{i} component,
but the existence of the \ion{C}{iv} line for a given search velocity range,
the multiple assigning of the same component does not affect the results.

Two arbitrary choices of $\Delta v_{\mathrm{\ion{C}{iv}}}$ are considered:
a conservative narrow range of $\pm 10 \textrm{ km s}^{-1}$ (a minimum $b$ value of 
a single Ly$\alpha$ absorption line is roughly 20 km s$^{-1}$) and a more
generous interval of $\pm 100 \textrm{ km s}^{-1}$.

\begin{figure}
\includegraphics[width=90mm]{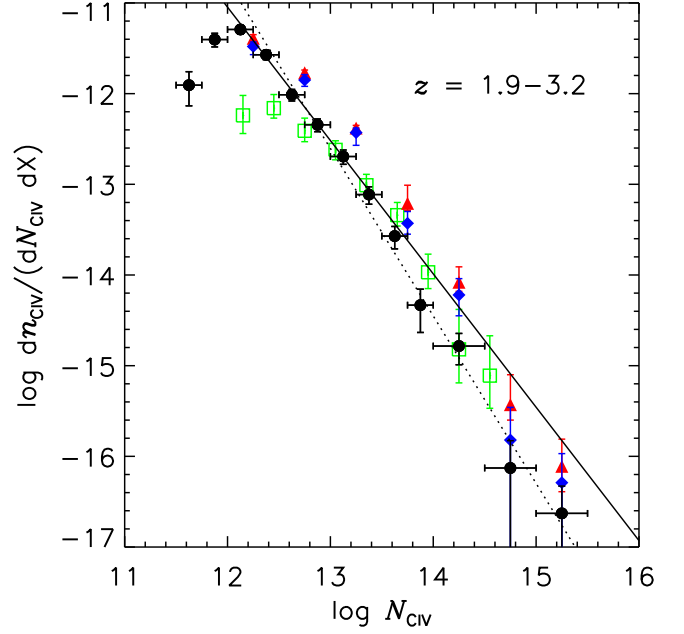}

\caption{The \ion{C}{iv} column density distribution. Filled circles
are our results at the redshift range used for the high-order fit \ion{C}{iv}-enriched
\ion{H}{i} sample
at $1.9\!<\!z\!<\!3.2$.
The CCD gap
in the \ion{C}{iv} region was accounted for.  
Red filled triangles and blue filled diamonds
are from \citet{Pichon:2003uq}
at $1.5\!<\!z\!<\!2.3$ and $2.3 < z < 3.1$, respectively.
Green open squares are taken from
\citet{Songaila:2001fk} at $2.90 < z < 3.54$. 
The vertical error bars indicate $1\sigma$ Poisson errors, while
the x-axis error bars show the $N_{\mathrm{\ion{H}{i}}}$ range covered by each data point.
The black dotted line shows the linear regression to filled circles at 
$\log N_{\mathrm{\ion{C}{iv}}} = [12.25, 15.5]:
\log \dif n_{\mathrm{\ion{C}{iv}}} / (\dif N_{\mathrm{\ion{C}{iv}}} \dif X) = (11.41\pm1.61)
+ (-1.85\pm0.13) \times \log N_{\mathrm{\ion{C}{iv}}}$. The solid line is the power
law fit at $\log N_{\mathrm{\ion{C}{iv}}} = [12.25, 13.5]:  
\log \dif n_{\mathrm{\ion{C}{iv}}}  / (\dif N_{\mathrm{\ion{C}{iv}}} \dif X) = (6.60\pm1.23)
+ (-1.47\pm0.10) \times \log N_{\mathrm{\ion{C}{iv}}}$.
The turn-over at $\log N_{\mathrm{\ion{C}{iv}}}\!\sim\!12.5$ shown in green data
is simply due to the incompleteness for weak \ion{C}{iv}.
Similarly, the turn-over seen at $\log N_{\mathrm{\ion{C}{iv}}}\!\sim\!12.1$
in our data is also due to the incompleteness.}
\label{fig:litComp_KimFigNew}
\end{figure}

Fig.~\ref{fig:litComp_KimFigNew} shows the \ion{C}{iv} column density distribution function
at $1.9\!<\!z\!<\!3.2$ from our sample (black filled circles). For comparison, 
other results from the literature are also included: red filled triangles and blue 
filled diamonds 
from \citet{Pichon:2003uq} at $1.5\!<\!z\!<\!2.3$ and $2.3\!<\!z\!<\!3.1$,
respectively, and green open squares from \citet{Songaila:2001fk}
at $2.90\!<\!z\!<\!3.54$. The turn-over seen in green open squares is due to
the incompleteness effect, i.e. not all weak \ion{C}{iv} can be detected due to noise.

Similar to the \ion{H}{i} density distribution, the \ion{C}{iv} 
CDDF does not fit with a single power law over a large $N_{\mathrm{\ion{C}{iv}}}$ range.
The Pichon et al. result even suggests
that the \ion{C}{iv} density distribution might have a non-linear functional form.  
At $\log N_{\mathrm{\ion{C}{iv}}} = [12.25, 13.5]$, a single power-law fit gives
$\log \dif N / (\dif N_{\mathrm{\ion{C}{iv}}} \dif X) = (6.60\pm1.23)
+ (-1.47\pm0.10) \times \log N_{\mathrm{\ion{C}{iv}}}$ (the solid line).  
At $\log {\mathrm{\ion{C}{iv}}} = [12.25, 15.5]$, a single power law is  
$\log \dif N / (\dif N_{\mathrm{\ion{C}{iv}}} \dif X) = (11.41\pm1.61)
+ (-1.85\pm0.13) \times \log N_{\mathrm{\ion{C}{iv}}}$ (the dotted line). 
If the solid line is taken as a reasonable CDDF since it fits the low-$N_{\mathrm{\ion{C}{iv}}}$
CDDF better, our \ion{C}{iv} detection can be considered
complete at $\log N_{\mathrm{\ion{C}{iv}}}\!\ge\!12.2$.

Another way to look at whether our $N_{\mathrm{\ion{C}{iv}}}$ completeness limit 
is reasonable is with the column density--$b$ value
diagram. As seen in
the 7th column of Table~\ref{tab1}, the S/N differs for different sightlines, and
changes even along a single spectrum. This makes it extremely difficult to quantify the 
correct 3$\sigma$ detection limit for a dataset containing spectra with different S/N.

\begin{figure}
\includegraphics[width=90mm]{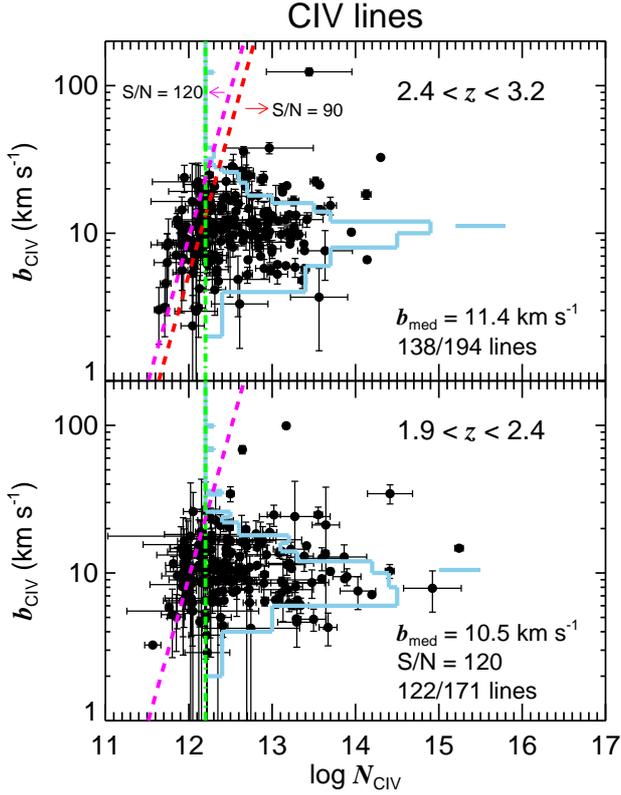}

\caption{Line width vs. column density for the \ion{C}{iv} absorption lines
along 16 sightlines excluding Q0055$-$269 and J2233$-$606 at the two
redshift bins. Error bars are fitting errors from the VPFIT profile fitting.  
In the upper panel, two heavy dashed lines delineate a 3$\sigma$ detection limit
for a spectrum with S/N $ = 120$ and S/N $= 90$
per pixel. At $2.4 < z < 3.2$, most spectra show S/N greater than 90.
In the lower panel, the heavy dashed line shows a 3$\sigma$ detection limit
for S/N $= 120$.
Broader and weaker absorption lines at the left of the detection limit are missed
in lower S/N spectra. The vertical dotted line
indicates the adopted low $N_{\mathrm{\ion{C}{iv}}}$ bound of
$\log N_{\mathrm{\ion{C}{iv}}}\!=\!12.2$ above which the
incompleteness does not affect the \ion{C}{iv} detection significantly.
The histogram shown with the base at
$\log N_{\mathrm{\ion{C}{iv}}}\!=\!12.2$ is the number of \ion{C}{iv}
lines as a function of $b_{\mathrm{\ion{C}{iv}}}$ with the $b_{\mathrm{\ion{C}{iv}}}$ binsize
of 2 km s$^{-1}$. Thick ticks above the
number distribution mark the
median $b_{\mathrm{\ion{C}{iv}}}$ for
$\log N_{\mathrm{\ion{C}{iv}}}\!\ge\!12.2$. The total number of \ion{C}{iv}
lines is 194 and 171 at $2.4 < z < 3.2$ and
$1.9 < z < 2.4$, respectively. Among them, 138 and 122 lines have
$\log N_{\mathrm{\ion{C}{iv}}}\!\ge\!12.2$ at the same redshift range.}
\label{fig:civblogn}
\end{figure}

Fig.~\ref{fig:civblogn} shows the $\log N_{\mathrm{\ion{C}{iv}}}$--$b_{\mathrm{\ion{C}{iv}}}$ 
diagram at the two redshift bins. The vertical heavy dot-dashed lines mark 
$\log N_{\mathrm{\ion{C}{iv}}} = 12.2$.
In the upper panel, two heavy dashed lines show a 3$\sigma$ detection limit for 
a spectrum with S/N $= 120$ (the left side) and 90 (the right side,
an approximate lowest S/N) per pixel, respectively. 
In the lower panel, the heavy dashed line is a 3$\sigma$ detection limit for S/N $= 120$.
Absorption lines at the left-side of the detection limit, i.e. broader and weaker lines, 
can be only detected for S/N greater than the given S/N. Overlaid as a histogram is
the distribution of the number of \ion{C}{iv} lines with 
$\log N_{\mathrm{\ion{C}{iv}}}\!\ge\!12.2$ as a function of $b_{\mathrm{\ion{C}{iv}}}$.
For the distribution, the zero base is set to be $\log N_{\mathrm{\ion{C}{iv}}} = 12.2$.
Thick ticks above the distribution mark the median $b_{\mathrm{\ion{C}{iv}}}$. 
There is no correlation between $N_{\mathrm{\ion{C}{iv}}}$ and $b_{\mathrm{\ion{C}{iv}}}$
above the S/N detection limit at all of the reasonable expected $b_{\mathrm{\ion{C}{iv}}}$ values.

At $2.4 < z < 3.2$ (the upper panel), the 3$\sigma$ $b_{\mathrm{\ion{C}{iv}}}$ detection limit
is 23.6 (13.4) km s$^{-1}$ for S/N $= 120$ (90) at $\log N_{\mathrm{\ion{C}{iv}}}\!\sim\!12.2$.        
The total wavelength coverage of \ion{C}{iv}
at the high redshift bin is $\sim\!3192$ \AA\/. For about half of the spectra 
there is contamination from weak telluric lines
in $\le\!10$\% of the \ion{C}{iv} region. This contamination prevents isolated weak
\ion{C}{iv} lines from being detected, however, can be treated as a lower-S/N region.
Including the telluric-contaminated region, the wavelength coverage with S/N $\le\!120$
is about 1018 \AA\/.
In the \ion{C}{iv} wavelength region with S/N $\ge\!120$,
the total number of \ion{C}{iv} lines with
$\log N_{\mathrm{\ion{C}{iv}}} = [12.2, 12.3]$ is 8. Out of those 8, none has 
$b_{\mathrm{\ion{C}{iv}}}\!\ge\!23.6$ km s$^{-1}$.
It is possible that
a large fraction of \ion{C}{iv} has a $b_{\mathrm{\ion{C}{iv}}}$
value greater than 23.6 km s$^{-1}$, and therefore, would be 
completely missed even in the high-S/N
spectra analysed here. However, as clearly seen in the upper panel of Fig.~\ref{fig:civblogn},
the $b_{\mathrm{\ion{C}{iv}}}$ distribution
at $\log N_{\mathrm{\ion{C}{iv}}}\!\ge\!12.2$ shows that only 9\% of \ion{C}{iv}
has $b_{\mathrm{\ion{C}{iv}}}\!\ge\!23.6$ km s$^{-1}$.
If a large fraction of \ion{C}{iv}
lines were broader regardless of $N_{\mathrm{\ion{C}{iv}}}$, the region around
$\log N_{\mathrm{\ion{C}{iv}}}\!\sim\!12.4$ and $b_{\mathrm{\ion{C}{iv}}}\!\sim\!25$ km s$^{-1}$
in Fig.~\ref{fig:civblogn} should have been more crowded.
Therefore, it is not likely that many weak \ion{C}{iv} lines with 
$b_{\mathrm{\ion{C}{iv}}}\!\ge\!23.6$ km s$^{-1}$ have been missed for S/N $\ge\!120$.

Only 2 out of 8 have 
$b_{\mathrm{\ion{C}{iv}}}\!\ge\!13.4$ km s$^{-1}$ at
$\log N_{\mathrm{\ion{C}{iv}}} = [12.2, 12.3]$. In other words, these 2 \ion{C}{iv} lines
would have been missed in the S/N $\le\!120$ region. One is a single isolated line, while
the other is part of a multi-component \ion{C}{iv} complex. 
We assumed that the number of \ion{C}{iv} lines 
with $\log N_{\mathrm{\ion{C}{iv}}}\!\sim\!12.2$ and $b_{\mathrm{\ion{C}{iv}}}\!\ge\!13.4$ km s$^{-1}$   
is 2 in the wavelength range of 2174 \AA\/,
i.e. the total wavelength range with S/N $\ge\!120$.
If we assume that weak \ion{C}{iv} lines
have a negligible clustering, about 1 (or $2 \times 1018/(3192-1018) = 0.9$) \ion{C}{iv}
line with $\log N_{\mathrm{\ion{C}{iv}}}\!\sim\!12.2$ and
$b_{\mathrm{\ion{C}{iv}}}\!\ge\!13.4$ km s$^{-1}$ could have been missed in the
\ion{C}{iv} forest region with S/N $\le\!120$.

A total of 5 \ion{H}{i} lines with
$\log N_{\mathrm{\ion{H}{i}}}\!=\![12.8, 17.5]$ is found within
$\pm$100 km s$^{-1}$ centered at these two \ion{C}{iv} lines.
The total number of
high-order-fit \ion{H}{i} lines in the \ion{H}{i} forest region corresponding to 
the S/N $\!\ge\!120$ \ion{C}{iv} forest region is
[265, 363, 233, 120, 50, 27, 5] for $\log N_{\mathrm{\ion{H}{i}}}\!=\!$ [12.75--13.00, 13.0--13.5,
13.5--14.0, 14.0--14.5, 14.5--15.0, 15.0--16.0, 16.0--17.0], respectively. Among them, a negligible number
of \ion{H}{i} lines, [0, 0, 2, 0, 1, 0, 1], is associated with these two \ion{C}{iv}
lines for the same $N_{\mathrm{\ion{H}{i}}}$ range, or less than 2\%. 
The remaining one \ion{H}{i} line
has $\log N_{\mathrm{\ion{H}{i}}}\!\ge\!17.0$ as the associated \ion{C}{iv} line belongs to
a \ion{C}{iv} complex of a partial Lyman limit system.
Although the number of undetected   
weak and broad \ion{C}{iv} lines in the S/N $\le\!120$ region is a very rough estimate,
less than 2\% of the \ion{H}{i} lines would be mis-classified as the unenriched forest due to 
the incompleteness at $\log N_{\mathrm{\ion{C}{iv}}}\!\sim\!12.2$.

The situation becomes more complicated in the low redshift bin, where the 
variance of the S/N limits of individual spectra is much higher than in the high redshift bin.
If a similar logic were applied to, the total
\ion{C}{iv} coverage is 5485 \AA\/, and the one with S/N $\!\le\!120$ is 2719 \AA\/. 
In the S/N $\!\ge\!120$ \ion{C}{iv} region, there is a total of 10 \ion{C}{iv} lines
with $\log N_{\mathrm{\ion{C}{iv}}} = [12.2, 12.3]$.
Out of 10, 6 lines have 
$b_{\mathrm{\ion{C}{iv}}}\!\ge\!13.4$ km s$^{-1}$, the maximum $b_{\mathrm{\ion{C}{iv}}}$
value to be detected for a line with $\log N_{\mathrm{\ion{C}{iv}}} = [12.2, 12.3]$
in a S/N $= 90$ spectrum. 
Among those 6 \ion{C}{iv} lines, two \ion{C}{iv} lines are part of a two-isolated-component
complex, with the rest being part of a multi-component complex.
Since stronger \ion{H}{i} lines tend to be associated with a \ion{C}{iv} complex,
using all these 6 \ion{C}{iv} lines to calculate the associated \ion{H}{i} fraction
leads to a biased result. Therefore, we used 4 \ion{C}{iv} lines which are part of
a \ion{C}{iv} complex with less than 3 components in order 
to estimate the missed enriched \ion{H}{i} fraction.

There is a total of 11 \ion{H}{i} lines at $\log N_{\mathrm{\ion{H}{i}}} = [12.75, 16.00]$  
within 100 km s$^{-1}$ centered at the 4 weak \ion{C}{iv} lines. 
The ratio of the \ion{C}{iv} enriched \ion{H}{i} lines
and the total \ion{H}{i} lines in the wavelength regions corresponding to the S/N $\ge\!120$
\ion{C}{iv} forest is
[2/231, 4/315, 3/174, 1/59, 1/28, 0/10] for
$\log N_{\mathrm{\ion{H}{i}}}\!=\!$ [12.75--13.00, 13.0--13.5,
13.5--14.0, 14.0--14.5, 14.5--15.0, 15.0--16.0], respectively, or $\le\!3$\%. 
Again the fraction of missed \ion{C}{iv} is negligible even at the low redshift bin. 

Note that our estimate on the true undetected \ion{C}{iv} fraction is  
uncertain.
However, from Fig.~\ref{fig:litComp_KimFigNew},
the incompleteness at $\log N_{\mathrm{\ion{C}{iv}}}=12.2$ is less than 10\% or within
the 1$\sigma$ Poisson error.

\begin{figure}
\includegraphics[width=90mm]{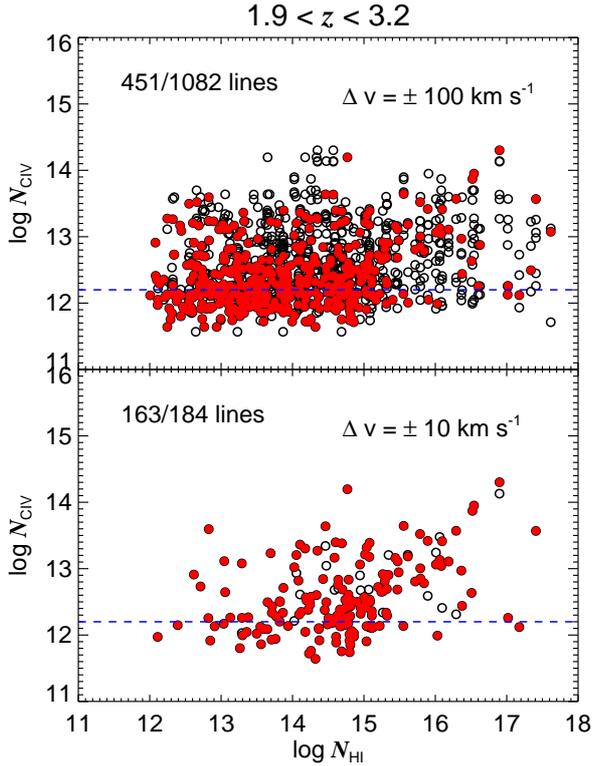}
\caption{
The $N_{\mathrm{\ion{H}{i}}}$--$ N_{\mathrm{\ion{C}{iv}}}$
diagram for $\log N_{\mathrm{\ion{H}{i}}} = [12.0, 17.8]$
from the high-order fit sample at $1.9\!<\!z\!<\!3.2$. The upper panel is for the
$\Delta v_{\mathrm{\ion{C}{iv}}} = \pm 100 \textrm{ km s}^{-1}$ sample,
while the lower panel for the
$\Delta v_{\mathrm{\ion{C}{iv}}} = \pm 10 \textrm{ km s}^{-1}$ sample.
Open circles represent \ion{H}{i} absorbers associated with all the possible
\ion{C}{iv} components, since a single \ion{H}{i} line could be assigned to several
\ion{C}{iv} lines. On the other hand, red filled circles indicate a \ion{H}{i} absorber
associated with only one closest \ion{C}{iv}.
The total number of open (filled) circles
is 1082 (451) and 184 (163) for the
$\Delta v_{\mathrm{\ion{C}{iv}}} = \pm 100 \textrm{ km s}^{-1}$ and
$\Delta v_{\mathrm{\ion{C}{iv}}} = \pm 10 \textrm{ km s}^{-1}$ sample, respectively.
}
\label{fig:nh1c4}
\end{figure}

While it is clear that the incompleteness does not play a significant role in the
\ion{H}{i} detection down to $\log N_{\mathrm{\ion{H}{i}}} = 12.75$ and
the \ion{C}{iv} detection down to $\log N_{\mathrm{\ion{C}{iv}}} = 12.2$,
the combination of the \ion{H}{i} and \ion{C}{iv} detection could introduce a bias
in the \ion{C}{iv} assigning method.
The pixel optical depth method which correlates the optical depth of \ion{H}{i} ($\tau_{\mathrm{\ion{H}{i}}}$) 
and \ion{C}{iv} ($\tau_{\mathrm{\ion{C}{iv}}}$) at the same redshift shows that at $z \sim 3$ there 
is a one-to-one positive correlation between the {\it median} $\tau_{\mathrm{\ion{H}{i}}}$
and the {\it median} $\tau_{\mathrm{\ion{C}{iv}}}$
down to $\log \tau_{\mathrm{\ion{H}{i}}}\!\sim\!0.15$ or $\log N_{\mathrm{\ion{H}{i}}}\!\sim\!13.73$ for
$b_{\mathrm{\ion{H}{i}}} = 28$ km s$^{-1}$ (a median $b_{\mathrm{\ion{H}{i}}}$ 
of the forest at $z\!\sim\!2.5$) 
\citep{Schaye:2003fc}. Below $\log \tau_{\mathrm{\ion{H}{i}}}\!\sim\!0.15$,
the $\tau_{\mathrm{\ion{H}{i}}}$ signal is blended with noise at $\log \tau_{\mathrm{\ion{C}{iv}}}\!\sim\!0.001$
or $\log N_{\mathrm{\ion{C}{iv}}}\!\le\!11.0$ for $b_{\mathrm{\ion{C}{iv}}} = 9.5$ km s$^{-1}$ (a median
$b_{\mathrm{\ion{C}{iv}}}$ of all the \ion{C}{iv} lines in our UVES sample). 

This result         
suggests that many low-$N_{\mathrm{\ion{H}{i}}}$ absorbers
might be mis-assigned as unenriched \ion{H}{i} absorber in our \ion{C}{iv} assigning method.
Unfortunately, the lower $\log N_{\mathrm{\ion{C}{iv}}}\!\sim\!11$ limit that a typical 
optical depth analysis explores
is an order of magnitude lower than our adopted low 
$N_{\mathrm{\ion{C}{iv}}}$ limit of $\log N_{\mathrm{\ion{C}{iv}}} = 12.2$.
This $\log N_{\mathrm{\ion{C}{iv}}}\!\sim\!11$ limit cannot be obtained even in 
the highest S/N \ion{C}{iv} region with S/N $\ge\!220$ in our UVES spectra. 
Therefore,
our \ion{C}{iv} analysis can not confirm, nor refute the results from the optical depth method.

Fig.~\ref{fig:nh1c4} shows the $N_{\mathrm{\ion{H}{i}}}$--$N_{\mathrm{\ion{C}{iv}}}$
diagram for the $\Delta v_{\mathrm{\ion{C}{iv}}} = \pm 100 \textrm{ km s}^{-1}$ sample
(the upper panel) and for the
$\Delta v_{\mathrm{\ion{C}{iv}}} = \pm 10 \textrm{ km s}^{-1}$ sample (the lower panel).
Since one \ion{H}{i} line can be associated with several \ion{C}{iv} lines,
data points at the same $N_{\mathrm{\ion{H}{i}}}$ represent the same \ion{H}{i} absorber.
Open circles show all the \ion{H}{i} absorbers associated with all the possible
\ion{C}{iv} lines. Red filled circles indicate
\ion{H}{i} absorbers
associated with only one closest \ion{C}{iv} within the search velocity range.
With a larger search velocity range, the
$\Delta v_{\mathrm{\ion{C}{iv}}} = \pm 100 \textrm{ km s}^{-1}$ sample has more
lines. The number of the red filled circles increases abruptly at
$\log N_{\mathrm{\ion{H}{i}}} \le 15.0$ at both redshift ranges, more prominently at
the high redshift bin. This is simply due to the
fact that the number of weaker \ion{H}{i} absorbers is larger than stronger \ion{H}{i}
absorbers.

\begin{figure*}
 \includegraphics[width=84mm]{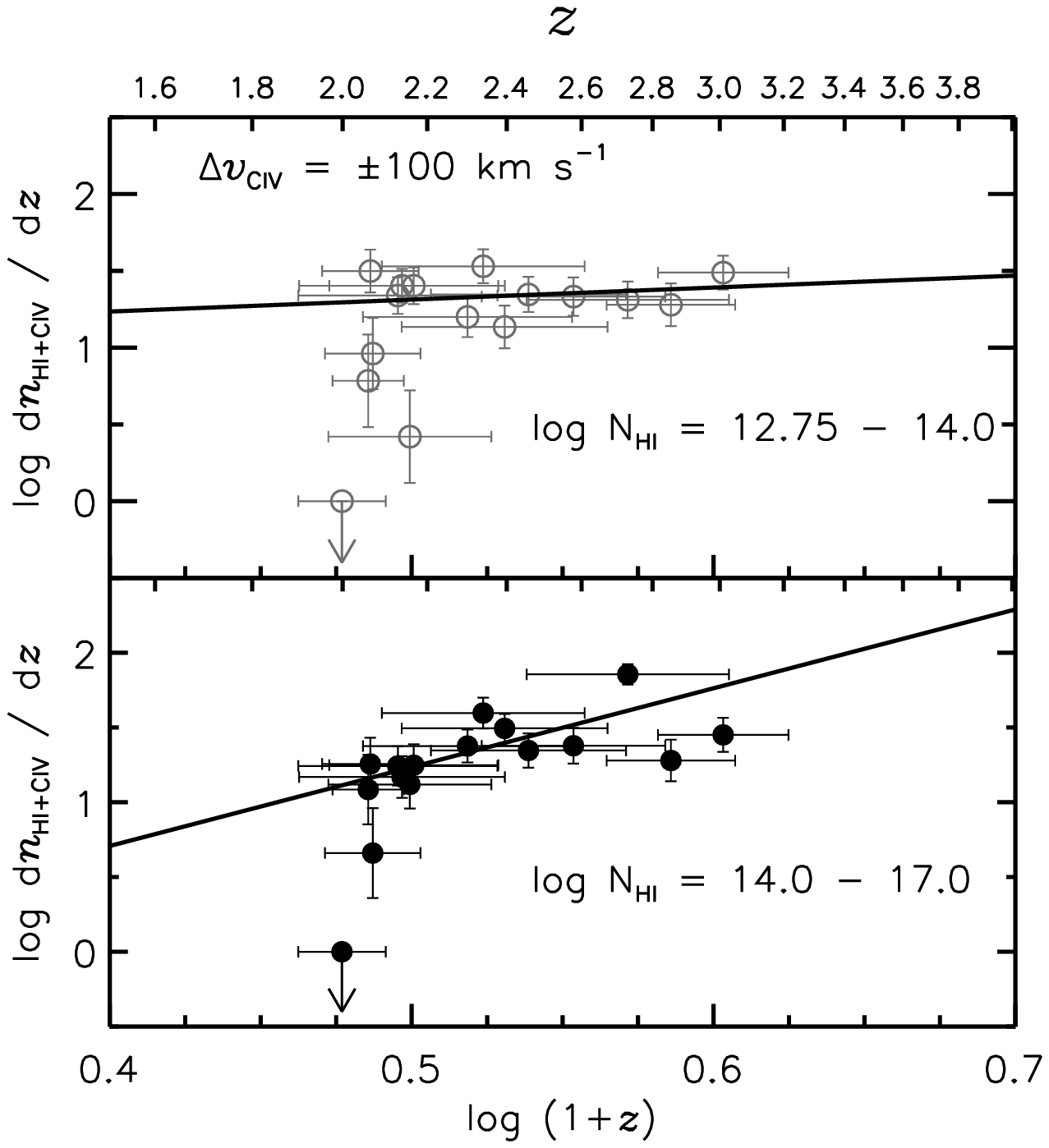}
 \hspace{5mm}
 \includegraphics[width=84mm]{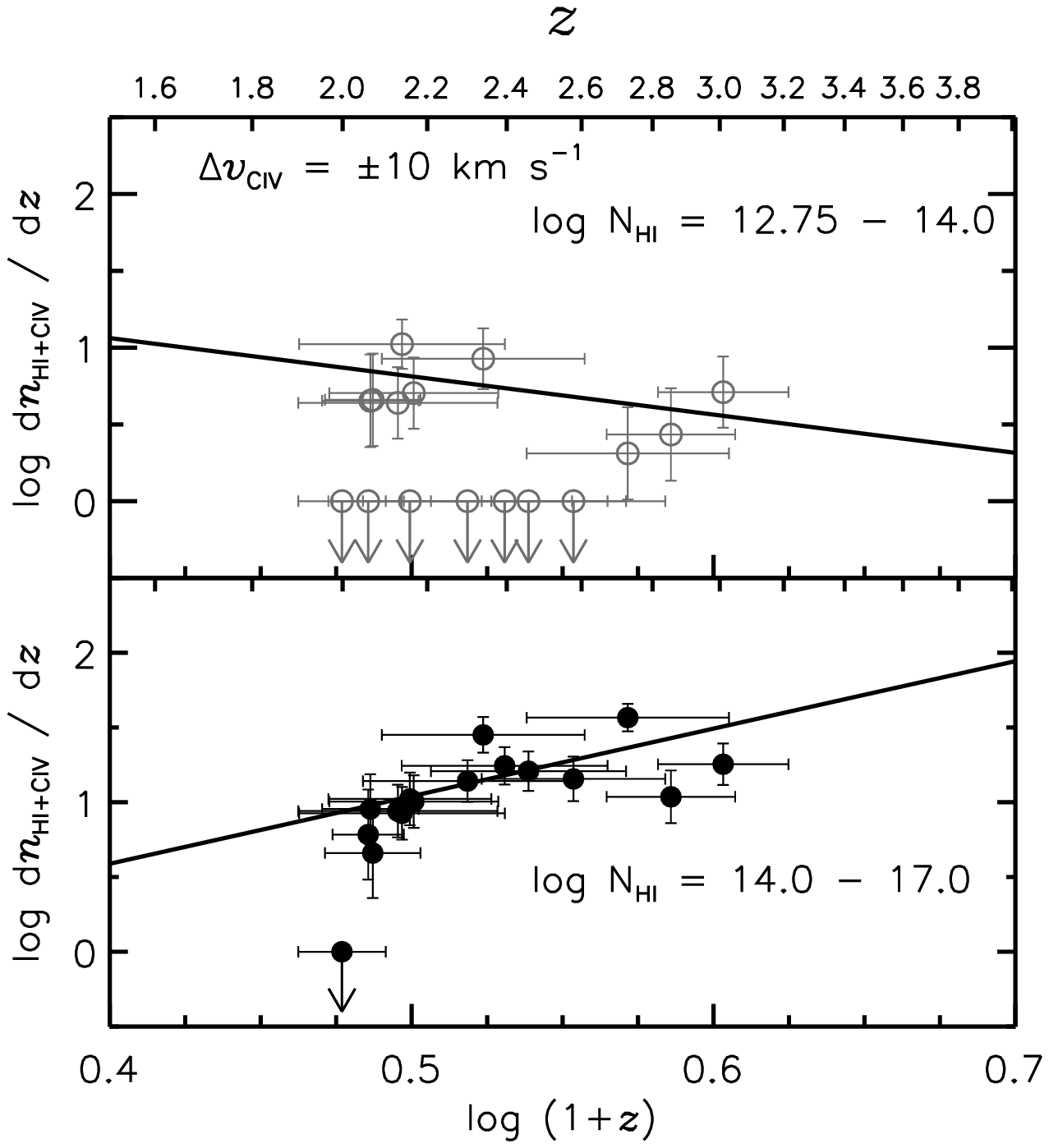}

 \caption{Quasar by quasar line number density evolution of
   \ion{C}{iv}-enriched \ion{H}{i} absorbers. The left panels are derived from the
 $\Delta v_{\mathrm{\ion{C}{iv}}} = \pm 100 \textrm{ km s}^{-1}$ sample,
 while the right panels represent the $\Delta v_{\mathrm{\ion{C}{iv}}} = \pm 10 \textrm{ km s}^{-1}$ one.
 The open circles
 in the upper panel represent \ion{C}{iv}-enriched absorbers having $\log N_\ion{H}{i} = [12.75, 14.0]$ and the
 filled circles in the lower panel represent $\log N_\ion{H}{i} = [14,17]$. 
 The vertical error bars mark $1\sigma$ Poisson errors, while
the x-axis error bars show the redshift range covered by each sightline. Sightlines having no
\ion{C}{iv}-enriched \ion{H}{i} absorbers for a given velocity range are plotted at 
$\log \dif dn_{\mathrm{\ion{C}{iv}}} / \dif d z = 0$.
 The solid lines represent linear regressions to the data, using the parameters summarised
 in Table \ref{tbl:linRegDnDzMetals}. Sightlines with no \ion{C}{iv}-enriched absorbers,
 $\log \dif n_{\ion{H}{i}+\ion{C}{iv}} / \dif z$ is plotted to be 0 with a downward arrow.}
 \label{fig:dNdzMetals}
\end{figure*}

If our \ion{C}{iv} assigning method were biased due to our failure to detect \ion{C}{iv} lines
toward lower $N_{\mathrm{\ion{H}{i}}}$ values,
there should be a correlation in $N_{\mathrm{\ion{H}{i}}}$
and $N_{\mathrm{\ion{C}{iv}}}$, such that a lower $N_{\mathrm{\ion{H}{i}}}$ line tends to be
associated with a lower $N_{\mathrm{\ion{C}{iv}}}$ line (cf. the relation between the 
median $\tau_{\mathrm{\ion{H}{i}}}$ and the median $\tau_{\mathrm{\ion{C}{iv}}}$) 
or the number of the \ion{C}{iv}-enriched 
\ion{H}{i} lines at lower $N_{\mathrm{\ion{H}{i}}}$ is smaller.
No such correlations are seen in Fig.~\ref{fig:nh1c4}. Note that our method deals with
the fitted individual lines, while the optical depth analysis works with {\it statistical,
median} values. The optical depth analysis is not sensitive to any minor \ion{C}{iv}
population, such as high-metallicity absorbers \citep{Schaye:2007ss}.

In reality, the detection of weak
\ion{C}{iv} is dependent on the local S/N as well as the
combination of $b_{\mathrm{\ion{C}{iv}}}$ and $N_{\mathrm{\ion{C}{iv}}}$. The S/N of a
spectrum does not change in a way to satisfy a higher S/N at strong \ion{H}{i} absorbers
and a lower S/N at weaker \ion{H}{i} absorbers or vice versa. 
Usually the S/N changes over a larger
wavelength interval than the wavelength interval between typical strong \ion{H}{i} lines. 
In addition, strong and weak \ion{H}{i} lines
do not occupy a portion of a spectrum separately, but exist mixed along the spectrum.
If a weak
\ion{C}{iv} were detected associated with a high-$N_{\mathrm{\ion{H}{i}}}$ line,
a similar strength of \ion{C}{iv}, if exists, 
should be detected for low-$N_{\mathrm{\ion{H}{i}}}$
lines nearby or in a similar S/N region. 
Therefore, unless a majority \ion{C}{iv} fraction at lower $N_{\mathrm{\ion{C}{iv}}}$
and/or lower $N_{\mathrm{\ion{H}{i}}}$ has a very large $b_{\mathrm{\ion{C}{iv}}}$ value, 
i.e. high gas temperature, our \ion{C}{iv} assigning
method does not introduce a serious selection bias within the adopted 
$N_{\mathrm{\ion{C}{iv}}}$ limit.

\subsection{Results}

\subsubsection{Number density evolution of the \ion{C}{iv}-enriched absorbers}
\label{sec:ts2_enrichedNumDens}

In a similar way to the analysis of all the \ion{H}{i} absorbers, we calculate the absorber number
density evolution $\dif n_{\ion{H}{i}+\ion{C}{iv}} / \dif z$ on a quasar by quasar analysis for all the 
\ion{C}{iv}-enriched \ion{H}{i} absorbers. 
The resulting
$\dif n_{\ion{H}{i}+\ion{C}{iv}} / \dif z$ evolution is shown in Fig. \ref{fig:dNdzMetals} for the 
$\Delta v_{\mathrm{\ion{C}{iv}}} = \pm 100 \textrm{ km s}^{-1}$ 
and $\Delta v_{\mathrm{\ion{C}{iv}}} = \pm 10 \textrm{ km s}^{-1}$
interval from the high-order Lyman fit samples. 
For the $\Delta v_{\mathrm{\ion{C}{iv}}} = \pm 100 \textrm{ km s}^{-1}$ sample,
the Q1101$-$264 sightline does not show any \ion{C}{iv} in the redshift range of interest
due to its short redshift coverage. For the 
$\Delta v_{\mathrm{\ion{C}{iv}}} = \pm 10 \textrm{ km s}^{-1}$ sample, 7 sightlines
(HE2347$-$4342, Q0002$-$422, PKS0329$-$255, HE1347$-$2457, Q0109$-$3518,
Q0122$-$380 and Q1101$-$264) out
of 16 have no \ion{C}{iv}-enriched \ion{H}{i} absorbers at $\log N_{\mathrm{\ion{H}{i}}}
= [12.75, 14.0]$, while only only sightline (Q1101$-$264) has no \ion{C}{iv}-enriched
\ion{H}{i} absorber at $\log N_{\mathrm{\ion{H}{i}}} = [14, 17]$. This is caused by the
combination of two facts that \ion{C}{iv} tends to be associated with strong \ion{H}{i}
absorbers and that the small search velocity is not adequate due to the velocity difference 
between \ion{H}{i} and \ion{C}{iv} observed in many enriched absorbers. For these
sightlines, 
$\dif n_{\ion{H}{i}+\ion{C}{iv}} / \dif z$ is 0. Therefore, 
their $\log n_{\ion{H}{i}+\ion{C}{iv}} / \dif z$ is
set to be 0 with a downward arrow in Fig.~\ref{fig:dNdzMetals}.

\begin{figure*}
 \includegraphics[width=84mm]{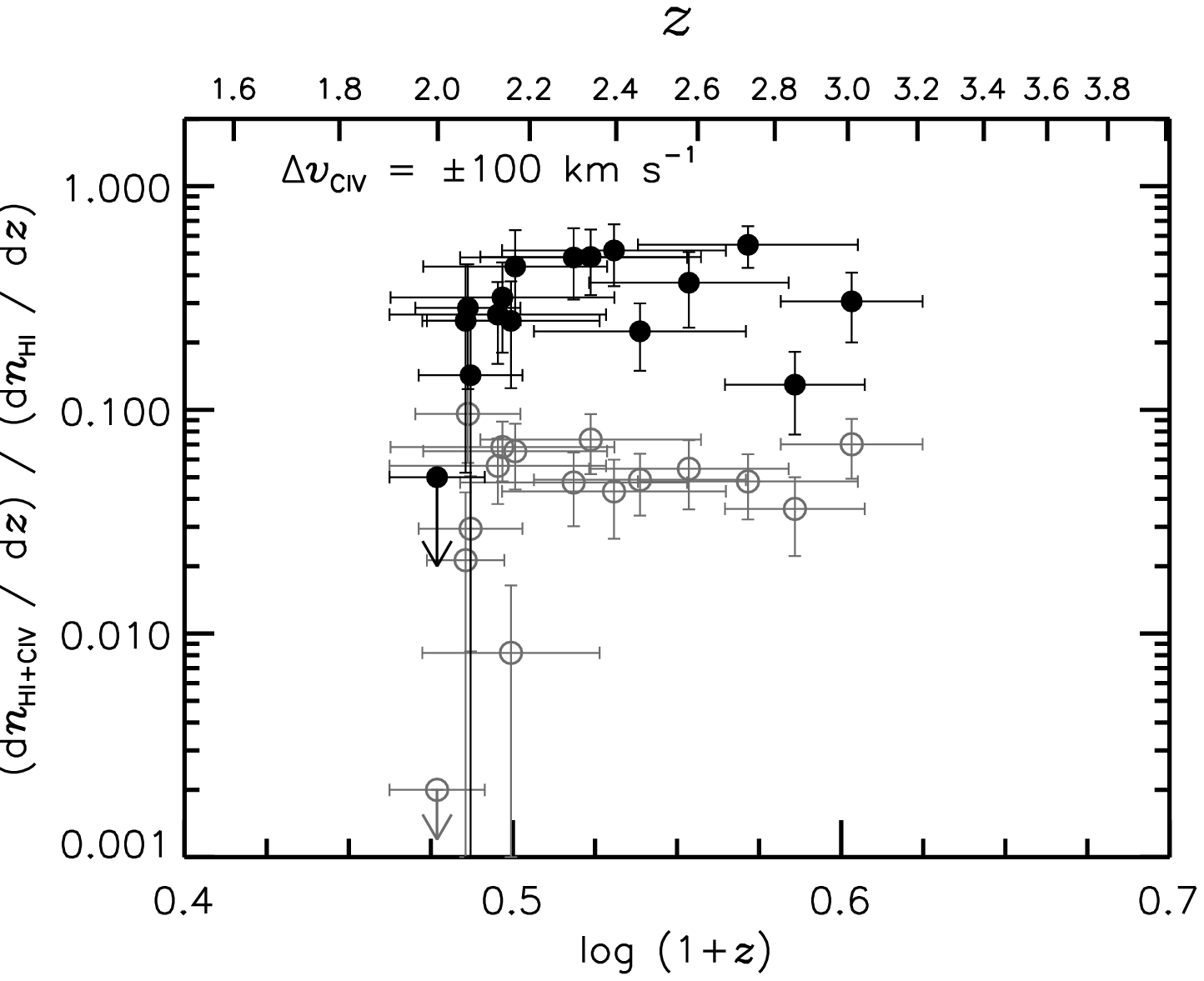} 
 \hspace{5mm}
 \includegraphics[width=84mm]{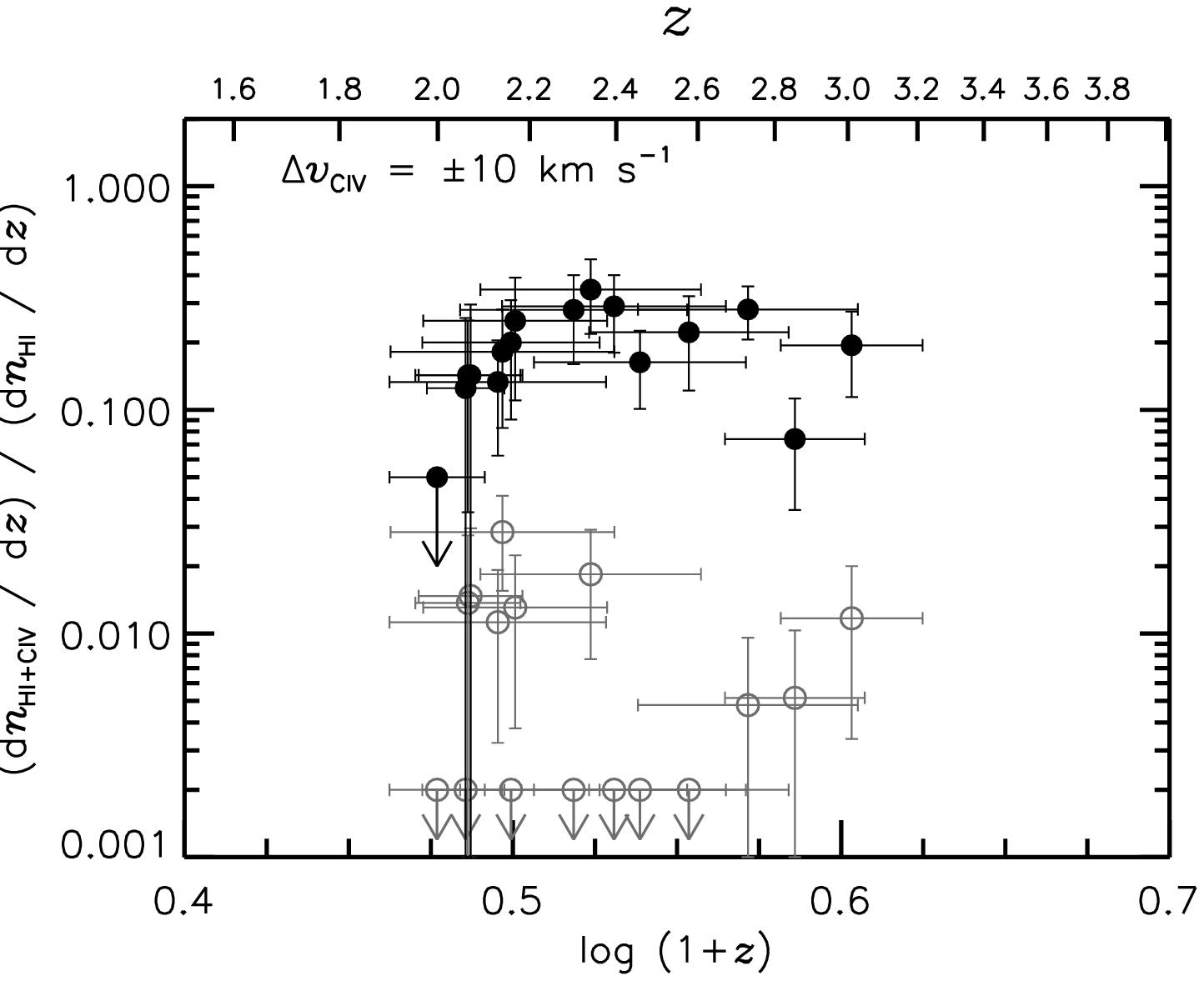} \\
 \vspace{-0.3cm}
 \caption{The fraction of the \ion{C}{iv}-enriched \ion{H}{i} absorber number density to the total absorber
 number density as a function of
 redshift. 
 The left panels are derived from the $\Delta v_{\mathrm{\ion{C}{iv}}} = \pm 100 \textrm{ km s}^{-1}$
 sample, while the right panels represent the $\Delta v_{\mathrm{\ion{C}{iv}}} = \pm 10 \textrm{ km s}^{-1}$
 sample.
 The black open circles represent a column density interval of $\log N_\ion{H}{i} = [12.75, 14.0]$ 
 and the filled circles represent $\log N_\ion{H}{i} = [14,17]$. The vertical error bars mark $1 \sigma$
 Poisson errors, while
the x-axis error bars show the redshift range covered by each sightline.
In the left panel for $\log N_\ion{H}{i} = [12.75, 14.0]$ (gray open circles), two lowest data points 
at $\log (1+z) \sim 0.5$ (or $z \sim 2$) including an upper limit are
from Q1101$-$264
 and Q0122$-$380. Both have a short redshift coverage, therefore become more susceptible to
 cosmic variance. Sightlines with no \ion{C}{iv}-enriched \ion{H}{i} absorbers are plotted as
 upper limits with an arbitrary value of 0.05 and 0.002 for $\log N_\ion{H}{i} = [14,17]$
 and $\log N_\ion{H}{i} = [12.75, 14.0]$, respectively.}
 \label{fig:dNdzMetalsFraction}
\end{figure*}

As for the entire Ly$\alpha$ forest analysis, the $\dif n_{\ion{H}{i}+\ion{C}{iv}} / \dif z$ evolution resembles a power law.
Therefore, linear regressions have been obtained from the data set and its results are summarised
in Table \ref{tbl:linRegDnDzMetals}. Sightlines showing no \ion{C}{iv}-enriched \ion{H}{i}
absorbers were not included in the regression. Similar to the entire high-order-fit \ion{H}{i} sample, the 
$\Delta v_{\mathrm{\ion{C}{iv}}} = \pm 100 \textrm{ km s}^{-1}$ sample shows a decline in the \ion{C}{iv}-enriched 
absorber number density with decreasing redshift. This behaviour is present in both column density
ranges. Comparing these results with the quasar-by-quasar $\dif n / \dif z$ 
of the entire high-order fit sample at $1.9 < z < 3.2$ shows that 
the \ion{C}{iv}-enriched absorbers at $\log N_\ion{H}{i} = [14,17]$
has a steeper slope ($5.27\pm0.99$), but completely consistent within 1$\sigma$.
The robust result on the 
$\dif n_{\ion{H}{i}+\ion{C}{iv}} / \dif z$ evolution
requires a large redshift coverage and more sightlines per redshift coverage,
especially at high column density range.
With a lack of more \ion{C}{iv} forest data at $z_{\mathrm{forest}} > 3$,  
$\dif n_{\ion{H}{i}+\ion{C}{iv}} / \dif z$ derived in this study 
at the high column density range should be considered less robust
compared to the entire \ion{H}{i} $\dif n / \dif z$.
Similarly, the $\dif n_{\ion{H}{i}+\ion{C}{iv}} / \dif z$ slope
($0.78\pm0.92$) at
$\log N_\ion{H}{i} = [12.75, 14.0]$ is also consistent with the one ($1.38 \pm 0.22$)
of the entire high-order-fit forest sample, given the rather large uncertainty. 
The actual number densities are lower at both column density ranges.

This becomes apparent in the left panel of Fig. \ref{fig:dNdzMetalsFraction}, 
where the 
ratios of the
number densities of the \ion{C}{iv}-enriched systems $\dif n_{\ion{H}{i}+\ion{C}{iv}} / \dif z$
and the number density of the entire sample $\dif n / \dif z$ are shown. The results for the 
$\Delta v_{\mathrm{\ion{C}{iv}}} = \pm 100 \textrm{ km s}^{-1}$ sample (filled circles) show 
that there is no significant evolution
of the \ion{C}{iv} enrichment fraction for $\log N_\ion{H}{i} = [12.75, 14.0]$.
For $\log N_\ion{H}{i} = [14, 17]$, the enrichment
fraction is consistent with no redshift evolution, considering a large scatter at
$z \sim 2$ and a lack of data at $z > 3$.
For the low column 
density $\log N_\ion{H}{i} = [12.75, 14.0]$ sample we find that around  5\% of all the \ion{H}{i} 
absorbers show \ion{C}{iv} enrichment.
The \ion{C}{iv} enrichment fraction is higher for larger column densities 
of $\log N_\ion{H}{i} = [14,17]$, where around 40\% of the
absorbers are \ion{C}{iv}-enriched.

\begin{table}
\centering
\small
\caption{\label{tbl:linRegDnDzMetals} Linear regression results for the number density evolution
$\dif n_{\ion{H}{i}+\ion{C}{iv}} / \dif z$ of the \ion{C}{iv}-enriched \ion{H}{i} forest absorbers in 
the quasar by quasar analysis.}
\begin{tabular}{@{}c@{~~~}c@{~~~}c@{~~~}c@{~~~}c@{}}
\hline
\noalign{\smallskip}

 & \multicolumn{2}{c}{$\Delta v_{\mathrm{\ion{C}{iv}}} = \pm 100 \textrm{ km s}^{-1}$} & \multicolumn{2}{c}{$\Delta v_{\mathrm{\ion{C}{iv}}} = \pm 10 \textrm{ km s}^{-1}$}  \tabularnewline
$\Delta \log N_\ion{H}{i}$ & $\log A$ & $\gamma$ & $\log A$ & $\gamma$ \tabularnewline
\hline
\noalign{\smallskip}

$12.75-14.0$ & $0.92\pm0.49$ & $0.78\pm0.92$ & $2.06\pm1.02$ & $-2.49\pm1.94$ \tabularnewline
$14.0-17.0$ & $-1.40\pm0.53$ & $5.27\pm0.99$ & $-1.22\pm0.64$ & $4.52\pm1.22$ \tabularnewline
\hline

\end{tabular}
\end{table}

This picture changes slightly for the 
$\Delta v_{\mathrm{\ion{C}{iv}}} = \pm\, 10 \textrm{ km s}^{-1}$ sample.
For the high column densities, the $\dif n_{\ion{H}{i}+\ion{C}{iv}} / \dif z$ evolution
is less strong compared to the one of  
the $\Delta v_{\mathrm{\ion{C}{iv}}} = \pm 100 \textrm{ km s}^{-1}$ sample. However,
both are still consistent within 1$\sigma$ due to a large uncertainty.
Only the number density itself decreases by a factor of 1.7. 
The enrichment fractions in the right panel of Fig.~\ref{fig:dNdzMetalsFraction} show
that now around 20\% to 30\% of the high column density \ion{H}{i} absorbers 
are \ion{C}{iv}-enriched.


On the other hand, $\dif n_{\ion{H}{i}+\ion{C}{iv}} / \dif z$ increases with decreasing redshift for the 
low column densities. Its negative slope of $\gamma = -2.49\pm1.94$ shows an opposite
behaviour from the one ($\gamma = 0.78 \pm 0.92$) of the 
$\Delta v_{\mathrm{\ion{C}{iv}}} = \pm 100 \textrm{ km s}^{-1}$ sample.
This negative slope is in part caused by the inadequacy in our \ion{C}{iv} assigning method
at the small search velocity,
and in part by the fact that the number of high-metallicity absorbers
increases at low redshift \citep{Schaye:2007ss}. However,
due to several sightlines containing no \ion{C}{iv}-enriched weak \ion{H}{i} absorbers
which are not included in the power-law fit,
the negative slope should not be taken literally. The fraction of enriched absorbers
increases from $\sim$0.5\% at $z \sim 3$ to $\sim$1.5\% at $z \sim 2.1$, as expected from 
$\dif n_{\ion{H}{i}+\ion{C}{iv}} / \dif z$ at the low \ion{H}{i} column density. However,
keep in mind that the cosmic variance is large as some sightlines show no enriched 
weak \ion{H}{i} absorbers.

There are two distinct groups of \ion{C}{iv} absorbers 
assigned to the low \ion{H}{i} column density. One group is
associated with strong, saturated high column density \ion{H}{i} absorbers. These absorbers are
sometimes accompanied by lower $N_{\mathrm{\ion{H}{i}}}$ absorbers
within a velocity range of $\Delta v < 200 
\textrm{ km s}^{-1}$. In these systems, the \ion{C}{iv} absorption is usually found within $20
\textrm{ km s}^{-1}$ to the strongest \ion{H}{i} lines (Kim et al. 2013, {\it in preparation}). Therefore,
these accompanied low \ion{H}{i} column density systems get associated with the \ion{C}{iv} absorbers 
if the velocity range $\Delta v_{\mathrm{\ion{C}{iv}}}$ is large. 
With a small velocity search interval, however, 
only \ion{H}{i} systems that have \ion{C}{iv} in their direct vicinity are flagged as \ion{C}{iv}-enriched. 
This means that the aforementioned low column density systems 
around strong absorbers are not considered
\ion{C}{iv}-enriched in a small velocity search interval.

Another \ion{C}{iv}-enriched group consists of usually 
isolated, low column density \ion{H}{i} absorbers associated with strong \ion{C}{iv} absorption,
i.e. the same high-metallicity forest population 
studied by \citet{Schaye:2007ss}. An example of such a system
toward HE1122$-$1648 is shown in Fig.~\ref{fig:he1122_highly}. In this velocity plot,
an \ion{H}{i} absorption feature
is hardly recognisable, while strong \ion{C}{iv} and \ion{N}{v} doublets are present. The existence
of both doublets makes the identification of this absorber secure. Due to the low 
$N_{\mathrm{\ion{H}{i}}}$ and high $N_{\mathrm{metals}}$, these systems show a higher ionisation
and a higher metallicity
compared to a typical absorber with similar $N_{\mathrm{\ion{H}{i}}}$
\citep{Carswell:2002mb, Schaye:2007ss}. 
\citet{Schaye:2007ss} speculate that these systems 
could be responsible for transporting metals from galaxies to the surrounding IGM.
As the velocity difference between \ion{H}{i} and metal lines for these systems are 
usually very small,
they dominate the weaker \ion{C}{iv}-enriched forest at $\log N_{\mathrm{\ion{H}{i}}}\!<\!14$
for the $\Delta v_{\mathrm{\ion{C}{iv}}} = \pm 10 \textrm{ km s}^{-1}$.
In addition, the high-metallicity absorbers are more common at low redshift.

\begin{figure}
 \includegraphics[width=84mm]{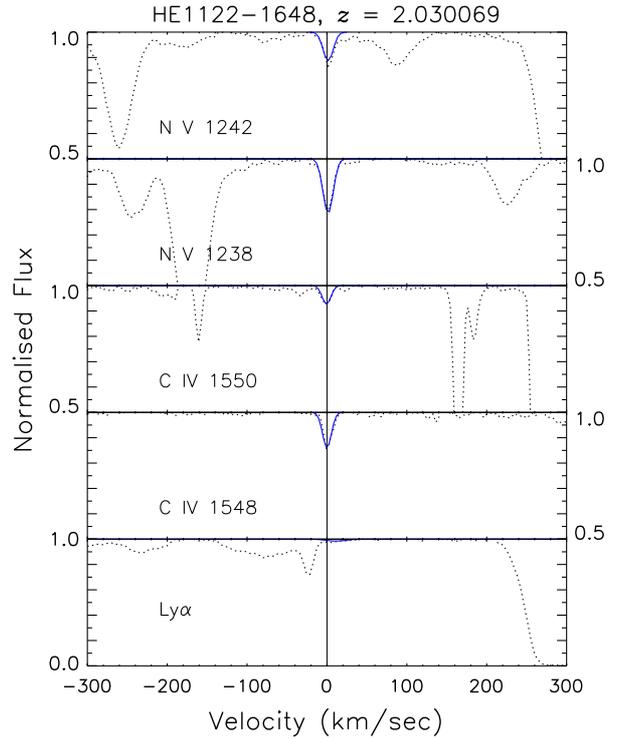}
 \vspace{-0.6cm}
 \caption{A velocity plot of a highly enriched \ion{C}{iv} absorber at $z = 2.030069$ toward HE1122$-$1648.
 The zero velocity is centered at $z = 2.030069$. Although there is no obvious Ly$\alpha$ absorption
 seen at the zero velocity, both \ion{C}{iv} and \ion{N}{v} doublets are present to secure
 the existence of this absorber.
 Note that the y-axis range for each ion is different: from the normalised flux 0 to 1
 for \ion{H}{i} and from 0.5 to 1 for the rest of the ions.}
 \label{fig:he1122_highly}
\end{figure}

\begin{figure}
 \includegraphics[width=84mm]{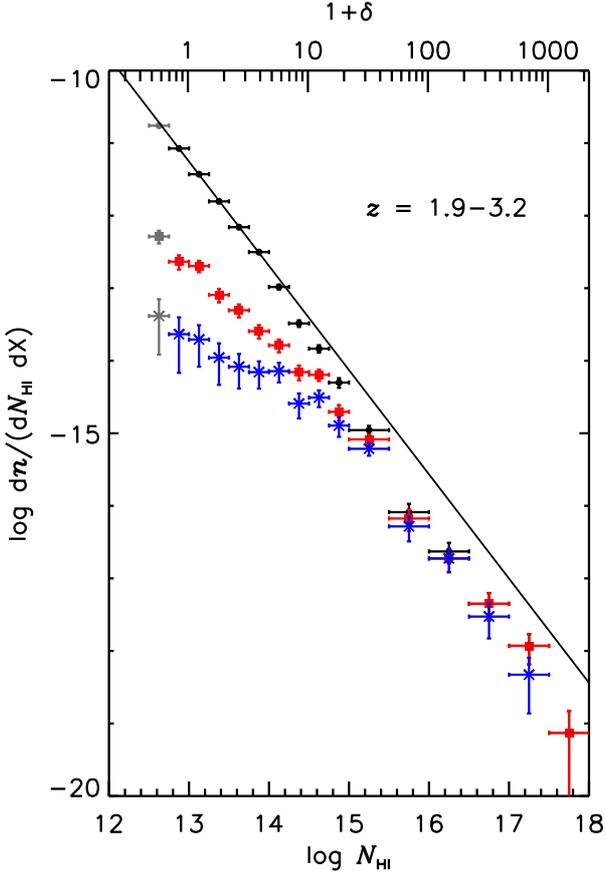}
 \caption{The distribution function
 for \ion{C}{iv}-enriched \ion{H}{i} lines for $\Delta v_{\mathrm{\ion{C}{iv}}}
 = \pm 100 \textrm{ km s}^{-1}$ (red filled squares) and for 
 $\Delta v_{\mathrm{\ion{C}{iv}}} = \pm 10 \textrm{ km s}^{-1}$ (blue stars)
 at $1.9\!<\!z\!<\!3.2$. Also shown is the 
 differential column density distribution function for all \ion{H}{i} Ly$\alpha$
 absorbers excluding Q0055$-$269 and J2233$-$606 (black filled circles)
 in the same redshift range analysed for 
 the \ion{C}{iv}-enriched forest.
 The solid line
 indicates the fit to filled circles for $\log N_\ion{H}{i} = [12.75, 14.0]:
 \log \dif N / (\dif N_{\mathrm{\ion{H}{i}}} \dif X) = (7.43\pm0.44)
 + (-1.44\pm0.03) \times \log N_{\mathrm{\ion{H}{i}}}$.
  The vertical errors indicate $1\sigma$ Poisson errors, while
the x-axis error bars show the $N_{\mathrm{\ion{H}{i}}}$ range covered by each data point.
All the grey data points indicate that the data are incomplete
 at $\log N_\ion{H}{i}\!<\!12.75$.}
 \label{fig:dNdNdXMetalsAll}
\end{figure}

The different characteristics of these two \ion{C}{iv} groups explains the different
$\dif n_\ion{C}{iv} / \dif z$ behaviour between the
$\Delta v_{\mathrm{\ion{C}{iv}}} = \pm 100 \textrm{ km s}^{-1}$
and $\Delta v_{\mathrm{\ion{C}{iv}}} = \pm 10 \textrm{ km s}^{-1}$ samples
at $\log N_{\mathrm{\ion{H}{i}}} = [12.75, 14.0]$.
With recent observational evidence that metals are only
found close to galaxies in the circum-galactic medium at $2\!<\!z\!<\!4$
and not far away from galaxies \citep{Adelberger:2005wa, Steidel:2010pd},
our results could provide further theoretical constraints for this interpretation.
It could well be that the high-metallicity forest population is completely different
from the typical, low-metallicity forest and resides in a different intergalactic
space. 
However, due to the low gas density and
high temperature, the Ly$\alpha$ forest does not have in-situ star formation.
Metals associated with the \ion{H}{i} forest should have been transported from
nearby galaxies. In other words, all the \ion{C}{iv}-enriched absorbers are close to
galaxies.

\subsubsection{Differential column density distribution function of the \ion{C}{iv}-enriched forest}

Fig.~\ref{fig:dNdNdXMetalsAll} shows the differential column
density distribution function for \ion{C}{iv}-enriched \ion{H}{i} absorbers for 
$1.9\!<\!z\!<\!3.2$. Red filled squares
and blue stars represent the
search velocity ranges of
$\Delta v_{\mathrm{metal}} = \pm 100 \textrm{ km s}^{-1}$ and
$\Delta v_{\mathrm{metal}} = \pm 10 \textrm{ km s}^{-1}$, respectively.
Black filled circles are for all \ion{H}{i} lines (excluding J2233$-$606 and Q0055$-$269),
regardless of their metal association. 
As in Figs~\ref{fig:figure7} and \ref{fig:dNdNdX},
the binsize of $\log N_{\mathrm{\ion{H}{i}}}\!=\!0.25$ is used
at $\log N_{\mathrm{\ion{H}{i}}}\!= [12, 15]$, then the binsize of 0.5 at
$\log N_{\mathrm{\ion{H}{i}}}\!= [15, 18]$.
The solid line
indicates the fit to filled circles for $\log N_\ion{H}{i} = [12.75, 14.0]:
\log \dif N / (\dif N_{\mathrm{\ion{H}{i}}} \dif X) = (7.43\pm0.44)
+ (-1.44\pm0.03) \times \log N_{\mathrm{\ion{H}{i}}}$.
The total absorption distance is $X(z) = 19.6652$ for the redshift ranges
analysed in this subsection.

For $\log N_{\ion{H}{i}}\!>\! 15$,  
the CDDF of the enriched forest is not sensitive to our choice of the search velocity
and 
the $\Delta v_{\mathrm{metal}} = \pm 100 \textrm{ km s}^{-1}$ CDDF becomes
almost identical with the CDDF of the entire \ion{H}{i} sample. 
For the column 
densities $\log N_{\mathrm{\ion{H}{i}}} = [14, 17]$, 
the CDDF functional form of the enriched forest shows a power-law with
a similar slope obtained for the entire \ion{H}{i} absorbers at
$\log N_{\mathrm{\ion{H}{i}}} = [12.75, 14.0]$, but with a smaller normalisation value. 

At $\log N_\ion{H}{i}\!<\! 15$, the distribution function of the \ion{C}{iv}-enriched
forest starts to deviate significantly from the CDDF of the entire \ion{H}{i} sample. 
The CDDF of the \ion{C}{iv}-enriched \ion{H}{i} forest starts to flatten out toward lower 
$N_{\mathrm{\ion{H}{i}}}$ at both search velocity ranges.
Furthermore the flattening of
the enriched forest depends strongly on the choice of $\Delta v_{\mathrm{\ion{C}{iv}}}$. The
large search velocity results in a steeper slope with a less fluctuation  
than the small one. This is due to 
the $\Delta v_{\mathrm{\ion{C}{iv}}} = \pm 10 \textrm{ km s}^{-1}$ sample being 
predominantly sensitive to highly enriched
absorbers at $\log N_\ion{H}{i}\!<\!14$ and 
less sensitive to mis-aligned broad \ion{C}{iv} lines with $b \ge 10$ km s$^{-1}$. Note that
our method to associate \ion{H}{i} with \ion{C}{iv} is only dependent on the relative
velocity difference between the line centers, but not the \ion{C}{iv} profile shape. 
The large velocity range includes broader \ion{C}{iv} lines up to $b\!\sim\!100$ 
km s$^{-1}$ as well as narrow, highly enriched absorbers. The 
$\Delta v_{\mathrm{\ion{C}{iv}}} = \pm 100 \textrm{ km s}^{-1}$ velocity
range is a better filter to associate \ion{H}{i} and \ion{C}{iv}.

The flattening of the distribution function seen at $\log N_\ion{H}{i}\!<\! 15$ 
by \ion{C}{iv}-enriched absorbers
cannot be caused by the incompleteness of the \ion{H}{i} sample. 
The \ion{H}{i} incompleteness
would result in a similar flattening
as is seen at $\log N_\ion{H}{i}\!<\!12.75$ for the entire sample (as seen in Fig.~\ref{fig:figure7}).
However, our sample of \ion{H}{i} absorbers is complete for column densities larger than 
$\log N_\ion{H}{i}\!>\!12.75$.

\begin{table}
\centering
\small
\caption{\label{tab07} Linear regression results for the differential column 
density distribution of the \ion{C}{iv}-enriched forest at $\log N_{\mathrm{\ion{H}{i}}}\!=\![14.5, 17.0]$.
Here the normalisation point $\log \left( \dif n / (\dif N_\ion{H}{i} \, \dif X) \right)_0$ is denoted by $B$.}
\begin{tabular}{@{}c@{~~~}c@{~~~}c@{~~~}c@{~~~}c@{}}
\hline
\noalign{\smallskip}

 & \multicolumn{2}{c}{$\Delta v_{\mathrm{\ion{C}{iv}}} = \pm 100 \textrm{ km s}^{-1}$} & 
   \multicolumn{2}{c}{$\Delta v_{\mathrm{\ion{C}{iv}}} = \pm 10 \textrm{ km s}^{-1}$}  \tabularnewline
$z$ & $B$ & $\beta$ & $B$ & $\beta$ \tabularnewline
\hline
\noalign{\smallskip}

$1.9-3.2$ & $6.93\pm1.17$ & $-1.45\pm0.08$ & $5.32\pm1.50$ & $-1.36\pm0.10$ \tabularnewline
$1.9-2.4$ & $5.94\pm2.51$ & $-1.40\pm0.17$ & $3.75\pm3.26$ & $-1.27\pm0.21$ \tabularnewline
$2.4-3.2$ & $6.90\pm1.45$ & $-1.44\pm0.10$ & $5.97\pm1.94$ & $-1.39\pm0.13$ \tabularnewline

\hline
\end{tabular}
\end{table}

\begin{figure*}
 \includegraphics[width=170mm]{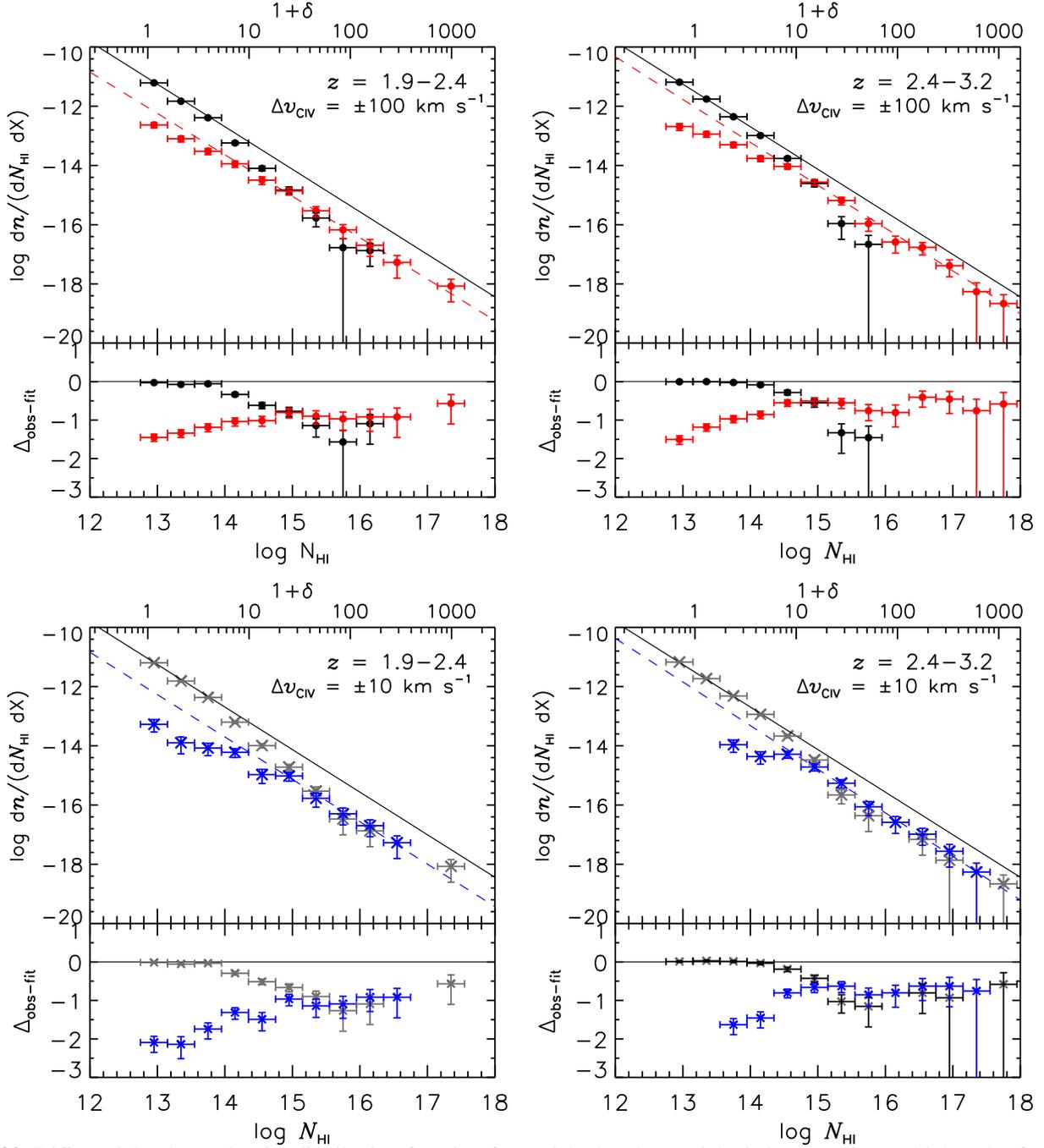}
 \vspace{-0.7cm}
 \caption{Differential column density distribution function for enriched and unenriched
  absorbers in our high-order fit sample using
  $\Delta v_{\mathrm{\ion{C}{iv}}}\!=\!\pm 100 \textrm{ km s}^{-1}$
  (upper panels) and $\Delta v_{\mathrm{\ion{C}{iv}}}\!=\!\pm 10 \textrm{ km s}^{-1}$ (lower panels)
  at the two different redshift ranges.
  Black filled circles and gray stars mark unenriched absorbers.
  Red filled squares and blue stars indicate the \ion{C}{iv}-enriched forest for
  the $\Delta v_{\mathrm{\ion{C}{iv}}}\!=\!\pm 100 \textrm{ km s}^{-1}$ sample
  and the $\Delta v_{\mathrm{\ion{C}{iv}}}\!=\!\pm 10 \textrm{ km s}^{-1}$ sample,
  respectively. Both Q0055$-$269 and J2233$-$606 are excluded in the analysis.
  The solid black line
  indicates the fit to the entire \ion{H}{i} sample and the whole redshift
  range at $\log N_{\mathrm{\ion{H}{i}}}\!=\![12.75, 14.0]$ as in Fig. \ref{fig:dNdNdXMetalsAll}.
  Red and blue dashed lines represent the fit to each \ion{C}{iv}-enriched forest
  sample at $\log N_{\mathrm{\ion{H}{i}}}\!=\![14.5, 17]$.
  The vertical errors indicate $1 \sigma$ Poisson errors, while
  the x-axis error bars show the $N_{\mathrm{\ion{H}{i}}}$ range covered by each data point.
  The lower parts of the panels show the difference between
  the observed CDDF and the expected CDDF from the power-law fit obtained for the 
  entire \ion{H}{i} samples
  (black solid lines).}
 \label{fig:dNdNdXMetals100}
\end{figure*}

As discussed in Section \ref{sec:Method}, the flattening of the   
\ion{C}{iv}-enriched forest 
at $\log N_\ion{H}{i}\!<\! 15$
could be in part caused by the missed weak, broad \ion{C}{iv} lines. 
However, Fig.~\ref{fig:litComp_KimFigNew} shows that the \ion{C}{iv} CDDF 
at $\log N_\ion{C}{iv}\!> 12.2$ is not strongly affected by the \ion{C}{iv}
incompleteness. The number ratio of the entire \ion{H}{i} forest lines and the
\ion{C}{iv}-enriched forest lines at $\log N_\ion{H}{i}\!\sim\!13$
is $\sim\!25$ for the 
$\Delta v_{\mathrm{\ion{C}{iv}}} = \pm 100 \textrm{ km s}^{-1}$ sample. 
This ratio increases to
$\sim 260$ for the $\Delta v_{\mathrm{\ion{C}{iv}}} = \pm 10 \textrm{ km s}^{-1}$ sample.
Even if we took a maximum correction for the \ion{C}{iv} incompleteness
of 50\%, roughly consistent with the results by \citet{Giallongo:1996lr} (see
their Section~2.3), 
the CDDF flattening of the enriched forest toward lower $N_\ion{H}{i}$ 
is still present. Therefore, this flattening is real and 
physically related to the characteristics of the \ion{C}{iv}-enriched absorbers only
with $N_{\mathrm{\ion{C}{iv}}}\!>\!12.2$.

The observation that the differential column density distribution
of the \ion{C}{iv}-enriched forest
flattens at low column densities can be easily explained by the fact that
the enrichment fraction with $\log N_{\mathrm{\ion{C}{iv}}} > 12.2$ becomes smaller as
$N_\ion{H}{i}$ decreases.

At $\log N_{\mathrm{\ion{H}{i}}} = [13.0, 13.5, 14.0, 14.5, 15.0, 15.5]$, the
fraction of the metal enriched forest 
for the $\Delta v_{\mathrm{\ion{C}{iv}}} = \pm 100 \textrm{ km s}^{-1}$ sample
is roughly $[4, 6, 11, 31, 49, 78]$\%, respectively.
This enrichment fraction can be roughly inferred from the difference between
the entire \ion{H}{i} CDDF and \ion{C}{iv}-enriched CDDF in Fig.~\ref{fig:dNdNdXMetalsAll}.

The different CDDF shape between the \ion{C}{iv}-enriched absorbers
and unenriched absorbers strongly supports that the \ion{C}{iv}-enriched
absorbers arise from the different physical environment, i.e. the
circum-galactic medium, while the unenriched forest has its origin as 
the intergalactic medium. The fact that
the number of \ion{C}{iv}-enriched absorbers decreases with decreasing $N_\ion{H}{i}$
is also consistent with the picture of IGM metal enrichment models by galactic winds
\citep{Aguirre:2001ud}.
The lower the \ion{H}{i} column density of absorbers is, the farther they are
from high-density gas concentrations where galaxies are formed. As galactic winds
have a limited life time and outflow velocity to transport metals in to the low-density
IGM, weaker absorbers will not be likely to be metal enriched.

The redshift evolution of the distribution function of 
\ion{C}{iv}-enriched and unenriched absorbers ({\it not} the entire \ion{H}{i} absorbers)
is shown in the upper panels of Fig.~\ref{fig:dNdNdXMetals100} for
the $\Delta v_{\mathrm{\ion{C}{iv}}}\!=\!\pm 100 \textrm{ km s}^{-1}$ sample
and in the lower panels for the
$\Delta v_{\mathrm{\ion{C}{iv}}}\!=\!\pm 10 \textrm{ km s}^{-1}$ sample
at the redshift ranges $z = [1.9, 2.4]$ and
[2.4, 3.2].
The total absorption distance is
$X(z) = 11.0940$ and 8.57061 at $z = [1.9, 2.4]$ and [2.4, 3.2], respectively,
excluding Q0055$-$269 and J2233$-$606.
To increase the absorber number for each $N_{\mathrm{\ion{H}{i}}}$ bin,
the binsize of $\log N_{\mathrm{\ion{H}{i}}}\!=\!0.4$ is used
at $\log N_{\mathrm{\ion{H}{i}}}\!= [12.75, 17.95]$.
Black filled circles and gray stars mark absorbers without \ion{C}{iv}.
Red filled squares and blue stars indicate the \ion{C}{iv}-enriched forest for
the $\Delta v_{\mathrm{\ion{C}{iv}}}\!=\!\pm 100 \textrm{ km s}^{-1}$ sample
and the $\Delta v_{\mathrm{\ion{C}{iv}}}\!=\!\pm 10 \textrm{ km s}^{-1}$ sample,
respectively. The solid black line
indicates the power-law fit to the entire \ion{H}{i} sample at $1.9\!<\!z\!<\!3.2$ 
and at $\log N_{\mathrm{\ion{H}{i}}}\!=\![12.75, 14.0]$ as in Fig. \ref{fig:dNdNdX}.
Red and blue dashed lines represent the power-law fit to each \ion{C}{iv}-enriched forest
sample at $\log N_{\mathrm{\ion{H}{i}}}\!=\![14, 17]$ (see Table~\ref{tab07}).

The upper panel of Fig.~\ref{fig:dNdNdXMetals100} suggests that the entire absorber population
can be considered as the combination of two populations of well-characterised absorbers, 
the enriched absorbers and the unenriched absorbers. 
The \ion{C}{iv}-enriched absorbers dominate at $\log N_{\mathrm{\ion{H}{i}}}\!>\!15$.
Their CDDF is well-described as a power law. The slope $\beta \sim -1.45$ obtained at 
$\log N_{\mathrm{\ion{H}{i}}} = [14.5, 17]$ (red dashed lines) is similar to the
slope $\beta \sim -1.44$ for the entire \ion{H}{i} sample 
at $\log N_{\mathrm{\ion{H}{i}}} = [12.75, 14.0]$ at
both redshifts. The normalisation value for the \ion{C}{iv}-enriched forest is smaller, with
about 10 times lower absorber numbers.
The enriched
absorbers do not show any strong redshift evolution at $\log N_{\mathrm{\ion{H}{i}}}\!>\!15$, 
while the CDDF flattening at $\log N_{\mathrm{\ion{H}{i}}}\!<\!15$
seems to be weaker at the low redshift.
The unenriched absorbers dominate at $\log N_{\mathrm{\ion{H}{i}}}\!<\!15$ with
a power-law CDDF. At higher $N_{\mathrm{\ion{H}{i}}}$, the unenriched
absorbers become significantly deviated from the extrapolated power law obtained at
lower $N_{\mathrm{\ion{H}{i}}}$. There are no unenriched absorbers at
$\log N_{\mathrm{\ion{H}{i}}}\!\ge\!16.0$.

The lower panels for the $\Delta v_{\mathrm{\ion{C}{iv}}}\!=\!\pm 10 \textrm{ km s}^{-1}$
sample show similar results. 
There is no strong redshift evolution
for the \ion{C}{iv}-enriched absorbers at $\log N_{\mathrm{\ion{H}{i}}}\!>\!15$.
Again, there is a suggestion that the flattening at $\log N_{\mathrm{\ion{H}{i}}}\!<\!15$
becomes less significant at lower redshifts. 
The unenriched forest starts to dominate at $\log N_{\mathrm{\ion{H}{i}}}\!<\!15$.
However, the $\Delta v_{\mathrm{\ion{C}{iv}}}\!=\!\pm 10 \textrm{ km s}^{-1}$ sample
shows a less-smooth CDDF.
The highest-$N_{\mathrm{\ion{H}{i}}}$ data point
at $z\!=\![2.4, 3.2]$ illustrates the inadequacy of
our assigning method of \ion{C}{iv} to \ion{H}{i} when a small search velocity was used. 
Absorbers contributing
this data point are part of multi-component high column density systems.
Their associated \ion{C}{iv} shows a rather simple, broad, but mis-aligned
profile from the \ion{H}{i} center of saturated Ly$\alpha$ profiles.
The \ion{H}{i} line center of some components resolved at high-order
Lyman lines is sometimes at $\ge\!10$ km s$^{-1}$ from the closest
\ion{C}{iv} component and thus they are flagged as the unenriched forest.

In both samples, the two populations overlap at $\log N_{\mathrm{\ion{H}{i}}}\!\sim\!15$.
The deviation from the power law starting
at $\log N_{\mathrm{\ion{H}{i}}}\!\sim\!14.0$--14.5 shown
in the CDDF for the entire \ion{H}{i} sample in Fig.~\ref{fig:dNdNdX}
is a result of combining two
different populations which show a different CDDF shape.

\begin{table}
\centering
\caption{\ion{H}{i} density relative to the critical density at
$\log N_{\mathrm{\ion{H}{i}}} = [12.75, 17.0]$: 
the entire \ion{H}{i} forest $\Omega_{\mathrm{\ion{H}{i}}}$, the
enriched forest $\Omega_{\mathrm{\ion{H}{i}+\ion{C}{iv}}}$ and the unenriched 
forest $\Omega_{\mathrm{\ion{H}{i}-\ion{C}{iv}}}$. Units are  
$10^{-6} h^{-1}$.}\label{tbl:massFracs}
\begin{tabular}{@{}c@{~~~}c@{~~~}c@{~~~}c@{~~~}c@{~~~}c@{}}
\hline
\noalign{\smallskip}

 &  & \multicolumn{2}{c}{$\Delta v_{\mathrm{\ion{C}{iv}}} = \pm 100 \textrm{ km s}^{-1}$} & 
\multicolumn{2}{c}{$\Delta v_{\mathrm{\ion{C}{iv}}} = \pm 10 \textrm{ km s}^{-1}$}\tabularnewline
$z$ & $\Omega_{\mathrm{\ion{H}{i}}}$ & $\Omega_{\mathrm{\ion{H}{i}+\ion{C}{iv}}}$ & 
 $\Omega_{\mathrm{\ion{H}{i}-\ion{C}{iv}}}$ & $\Omega_{\mathrm{\ion{H}{i}+\ion{C}{iv}}}$ & 
 $\Omega_{\mathrm{\ion{H}{i}-\ion{C}{iv}}}$ \tabularnewline
\hline
\noalign{\smallskip}

1.9--3.2 & 1.57 & 0.61 & 0.96 &      0.43 &  1.14 \tabularnewline
1.9--2.4 & 1.29 & 0.33 & 0.96 &      0.25 &  1.04 \tabularnewline
2.4--3.2 & 1.85 & 0.92 & 0.93 &      0.63 &  1.22 \tabularnewline
\hline
\end{tabular}
\end{table}

To obtain a rough idea on the \ion{H}{i} density relative
to the critical density of the entire forest 
($\Omega_{\mathrm{\ion{H}{i}}}$), 
the enriched forest ($\Omega_{\mathrm{\ion{H}{i}+\ion{C}{iv}}}$) and the unenriched 
forest ($\Omega_{\mathrm{\ion{H}{i}-\ion{C}{iv}}}$), we used Eq.~(A12) from
\citet{Schaye:2001yk} at $\log N_{\mathrm{\ion{H}{i}}}\!=\![12.75, 17.0]$,
by directly integrating the observed CDDF.
The ionisation fraction of \ion{H}{i} to obtain the total hydrogen
density $\Omega_{\mathrm{H}}$ 
was not corrected, since it is highly uncertain. 
The resulting mass fractions for $\log N_{\mathrm{\ion{H}{i}}} = [12.75, 17.0]$
are given in Table~\ref{tbl:massFracs}.
The derived CDDF is not well-constrained at 
$\log N_{\mathrm{\ion{H}{i}}}\!\ge\!16$ due to the low number statistics.
Therefore, the derived $\Omega$ values in Table~\ref{tbl:massFracs}
are only rough numbers.  
The ratio of $\Omega_{\mathrm{\ion{H}{i}+\ion{C}{iv}}}$ 
and $\Omega_{\mathrm{\ion{H}{i}}}$ might decrease by a factor
of 2 at low redshift for both
$\Delta v_{\mathrm{\ion{C}{iv}}}$ samples. However, a high uncertainty in
deriving $\Omega$ values does not allow to assure this decrease.
The \ion{C}{iv}-enriched forest accounts for $\sim 40$\% of the entire forest in mass
at $1.9\!<\!z\!<\!3.2$ for the $\Delta v_{\mathrm{\ion{C}{iv}}} = 100$ sample.

\section{Conclusions}
\label{Sec:Conclusions}

Based on an in-depth Voigt profile fitting analysis 
of 18 high-redshift quasars obtained from the ESO VLT/UVES
archive, we have studied $\sim 3100$ \ion{H}{i} absorbers to
investigate the number density evolution and the
differential column
density distribution function at $1.9\!<\!z\!<\!3.2$
and for $\log N_{\mathrm{\ion{H}{i}}} = [12.75, 17.0]$. 
Two methods of the Voigt profile fitting 
analysis have been applied, one 
by fitting absorption profiles only to the Ly$\alpha$ transition and another
by including higher order Lyman transitions such as Ly$\beta$ and Ly$\gamma$.
These higher order transitions provide 
a more reliable column density measurement of saturated absorption systems, since
saturated and blended lines 
often become unsaturated at higher order transitions.
This also enables us to resolve the structure of absorbers more reliably.
This study has increased the sample size by a factor of 3 from previous
similar studies at $z\!>\!2$.
In addition, we have investigated whether there exist any differences 
in the $N_{\mathrm{\ion{H}{i}}}$ evolution between the \ion{C}{iv}-enriched forest 
and the unenriched forest.

We have found that the results based on the Ly$\alpha$-only fit are in good
agreement with previous results on a quasar by quasar analysis.
For our data only (values in parenthesis indicate results including 
high-quality data from the literature at $z\!>\!1$), 
the number density
$\dif n/ \dif z$ is $\dif n / \dif z  = (1.46 \pm 0.11) \times (1+z)^{1.67\pm0.21}$ 
($(1.52\pm0.05) \times (1+z)^{1.51\pm0.09}$)
and $\dif n/ \dif z = (0.03 \pm 0.20) \times (1+z)^{3.40\pm0.36}$ 
($(0.72\pm0.08) \times (1+z)^{2.16\pm0.14}$) at
$\log N_{\mathrm{\ion{H}{i}}}\!=\![13.1$, 14.0] and [14, 17], respectively.
The noticeable difference in the exponent between our sample and the sample including 
the data from the literature for stronger absorbers 
is caused by the fact that our sample does not cover a large redshift
range and that the evolution of $\dif n / \dif z$ is more significant for stronger absorbers.
The scatter between different sightlines becomes larger at lower redshifts and
stronger absorbers due to the evolution of the large-scale structure. 

Combining our Ly$\alpha$-only fit analysis at $1.9\!<\!z\!<\!3.6$ with the high-quality
literature data at $0.0\!<\!z\!<\!4$, the {\it mean} number density evolution
is not well described by a single power law and strongly suggests
that its evolution slows down at $z \le 1.5$ at both high and low column density ranges.
Although a single power law does not give a good description, the number density is
$\dif n / \dif z \propto (1+z)^{0.89\pm 0.06}$ and $\dif n / \dif z \propto (1+z)^{1.61\pm 0.12}$
at $\log N_{\mathrm{\ion{H}{i}}} = [13.1, 14.0]$ and $[14, 17]$, respectively.

The differential column density distribution function (CDDF) from the Ly$\alpha$-only
fit analysis is also consistent with
previous results. The single power-law exponent is $-1.44 \pm 0.02$ at 
$1.9\!<\!z\!<\!3.2$ and $\log N_{\mathrm{\ion{H}{i}}}\!=\![12.75$--14.0], with
a deviation from the power law at $\log N_{\mathrm{\ion{H}{i}}} > 14.0-14.5$.

The high-order Lyman fits do not show any significantly different results from the ones
based on the Ly$\alpha$-only fits. The $\dif n / \dif z$ evolution based on a quasar by quasar 
analysis yields
a very similar result to the Ly$\alpha$-only fit. The {\it mean} $\dif n / \dif z$ based on
the combined sample from our quasars at $1.9\!<\!z\!<\!3.2$ is
$\dif n / \dif z  = (1.89\pm0.13) \, \, (1+z)^{1.28\pm0.24}$ 
and $\dif n/ \dif z = (-0.65\pm0.36) \, \, (1+z)^{4.65\pm0.66}$ at
$\log N_{\mathrm{\ion{H}{i}}}\!=\![12.75$, 14.0] and [14, 17], respectively.

Using the high-order fits, we have derived the differential column density distribution function at
$1.9\!<\!z\!<\!3.2$ and confirm the existence of
a dip at $\log N_{\mathrm{\ion{H}{i}}} = [14, 18]$
as seen in the Ly$\alpha$-only-fit CDDF analysis.
At $1.9\!<\!z\!<\!3.2$, the power-law exponent of the differential column density distribution
function is $-1.44\pm0.03$, $-1.67\pm0.09$ and $-1.55\pm0.08$ at
$\log N_{\mathrm{\ion{H}{i}}} = [12.75, 14.0]$, [14, 15] and [15, 18],
respectively. 

By obtaining the differential column density distribution 
function for two redshift bins $z = [1.9, 2.4],$ and [2.4, 3.2], we observe 
that a deviation from the expected power-law at $\log N_{\mathrm{\ion{H}{i}}}\!=\![14.0, 18.0]$ 
is more prominent at lower redshifts. In addition,
the power-law seems to be slightly steeper at the low redshift for the
column density range $\log N_{\mathrm{\ion{H}{i}}}\!=\![12.75, 14.0]$ in which
the distribution function follows a perfect single power law. However,
the CDDF at two redshift bins is consistent with no redshift evolution within 2$\sigma$.

Further, we have split the entire \ion{H}{i} absorbers excluding 2 quasars with
a lower S/N \ion{C}{iv} region into two samples:
absorbers associated with \ion{C}{iv} tracing
the metal enriched forest, and absorbers associated with no \ion{C}{iv} tracing the unenriched
forest.
A \ion{H}{i} absorber is considered \ion{C}{iv}-enriched, if a \ion{C}{iv} line
with $\log N_{\mathrm{\ion{C}{iv}}}$ greater than a threshold value
is found within a given search velocity interval
centered at each \ion{H}{i} absorption center.
The threshold $\log N_{\mathrm{\ion{C}{iv}}}\!=\!12.2$ was used
since the \ion{C}{iv} distribution function and the $N_{\mathrm{\ion{C}{iv}}}$--$b_{\mathrm{\ion{C}{iv}}}$ 
diagram show that the 
\ion{C}{iv} detection is reasonably complete down to $\log N_{\mathrm{\ion{C}{iv}}}\!=\!12.2$
for a typical $b_{\mathrm{\ion{C}{iv}}}$ value found at $\log N_{\mathrm{\ion{C}{iv}}}\!\ge\!12.2$
in our sample. We used two arbitrarily chosen search velocity 
intervals, 
$\Delta v_{\mathrm{\ion{C}{iv}}} = \pm 100 \textrm{ km s}^{-1}$ 
and $\Delta v_{\mathrm{\ion{C}{iv}}} = \pm 10 \textrm{ km s}^{-1}$. 

At $\log N_{\mathrm{\ion{H}{i}}}\!=\![14, 17]$, the $\dif n_{\ion{H}{i}+\ion{C}{iv}} / \dif z$ of 
the \ion{C}{iv}-enriched 
\ion{H}{i} absorbers show a similar evolution compared to the one of the
entire Ly$\alpha$ forest, with a power-law decrease in number density
with decreasing redshift.
The power-law slope is
$[0.78\!\pm\!0.92, 5.27\,\pm\,0.99]$ for 
$\log N_{\mathrm{\ion{H}{i}}} = [12.75, 14.0]$ and [14, 17] at $1.9 < z < 3.2$
for the $\Delta v_{\mathrm{\ion{C}{iv}}} = \pm\,100 \textrm{ km s}^{-1}$ sample.

The enriched fraction is fairly constant with redshift at $1.9\!<\!z\!<\!3.2$.  
About 5\% of all absorbers
show an association with \ion{C}{iv} at $\log N_{\mathrm{\ion{H}{i}}} = [12.75, 14]$,
while about 40\% are metal enriched at $\log N_{\mathrm{\ion{H}{i}}} = [14, 17]$
for the $\Delta v_{\mathrm{\ion{C}{iv}}} = \pm\,100 \textrm{ km s}^{-1}$ sample.

For $\Delta v_{\mathrm{\ion{C}{iv}}} = \pm\,10 \textrm{ km s}^{-1}$ sample,
the low column density enriched absorber suggests that $\dif n_{\ion{H}{i}+\ion{C}{iv}} / \dif z$
increases as redshift decreases, i.e. a negative slope of $-2.49 \pm 1.94$. Part of this
behaviour is caused by the fact that high-metallicity absorbers which are more
sensitive to the small search velocity become more
abundant at low redshift. However, this {\it negative} evolution should not be
taken literally since about a half of sightlines does not show enriched absorbers
at $\log N_{\mathrm{\ion{H}{i}}} = [12.75, 14.0]$. 

The differential column density distribution function for 
the enriched and unenriched
systems show a significant difference. However, each shows a well-characterised
CDDF.
At $\log N_{\mathrm{\ion{H}{i}}} \le 15.0$, the unenriched forest dominates
and its distribution shows a power law similar to the entire forest sample.
On the other hand, the \ion{C}{iv}-enriched forest
is found to flatten out at $\log N_{\mathrm{\ion{H}{i}}} \le 15$.
Depending on the search velocity interval, the number of enriched systems is a factor of
$25$ ($\Delta v_{\mathrm{\ion{C}{iv}}} = \pm 100 \textrm{ km s}^{-1}$) to 
$260$ ($\Delta v_{\mathrm{\ion{C}{iv}}} = \pm 10 \textrm{ km s}^{-1}$) lower than the one of the 
unenriched systems at $\log N_{\mathrm{\ion{H}{i}}}\!=\!13$.
This flattening is mainly caused by 
the fact that the enriched fraction
of the Ly$\alpha$ forest decreases as $\log N_\ion{H}{i}$ decreases.

At the higher $N_{\mathrm{\ion{H}{i}}}$ range, the \ion{C}{iv}-enriched forest
dominates. Its distribution function can be described as a power law with its slope of 
$-1.45\pm0.08$ similar
to the power-law slope ($-1.44\pm0.03$) of the entire \ion{H}{i} 
forest at $\log N_{\mathrm{\ion{H}{i}}} = [12.75, 14.0]$,
but a lower normalisation value, i.e. $\sim$ 10 times lower in the absorber number. 
The unenriched forest disappears very rapidly as 
$\log N_{\mathrm{\ion{H}{i}}}$ increases. 

The distribution function of the entire \ion{H}{i}
forest can be described as the combination of these two well-characterised populations,
overlapping at $\log N_{\mathrm{\ion{H}{i}}}\!\sim\!15$.
The deviation from the power law
at $\log N_{\mathrm{\ion{H}{i}}} = [14, 17]$ seen
in the CDDF for the entire \ion{H}{i} sample
is a result of combining two
different \ion{H}{i} populations with a different CDDF shape.
This result supports 
other observational evidence from absorber-galaxy studies at $z \sim 3$, namely that 
metals associated with the high-redshift Ly$\alpha$ forest are 
within $\sim\!100$~kpc of galaxies 
\citep{Adelberger:2005wa, Steidel:2010pd}. Absorber-galaxy studies suggest that
the \ion{C}{iv}-enriched and unenriched forest would arise from the different spatial
and physical locations, therefore having a different physical/evolutionary behaviour 
suggested by the different CDDF shape. Therefore, our results combined with 
absorber-galaxy studies
indicate that the \ion{C}{iv}-enriched forest is the circum-galactic medium,
while the unenriched forest has its origin as
the intergalactic medium.
 
At $1.9 < z < 3.2$, the \ion{C}{iv}-enriched forest contributes $\sim\!40$\% of 
the entire forest mass to the \ion{H}{i} density relative to the critical density for
the $\Delta v_{\mathrm{\ion{C}{iv}}} = \pm\,100 \textrm{ km s}^{-1}$ sample.

\section*{Acknowledgments}
We are grateful to Drs. Martin Haehnelt, Jamie Bolton
and Gerry Williger for insightful discussions and comments. We are also
grateful to our anonymous referee for very constructive comments.
A.P. acknowledges support in parts by the German Ministry
for Education and Research (BMBF) under grant FKZ 05 AC7BAA.

\bibliography{literature}

\appendix
\section{Data}

In this appendix we present all the data for the quasar by quasar number 
density evolution (Table~\ref{tbl:dndz_ap1}),
the {\it mean} number density evolution (Table~\ref{tbl:dndz_ap2}), 
the differential column density distribution (CDDF)
of the entire \ion{H}{i} sample (Table~\ref{tbl:dndz_ap3}), the CDDF of the \ion{C}{iv}-enriched forest
(Table~\ref{tbl:dndz_ap4}) and of the unenriched forest (Table~\ref{tbl:dndz_ap5})
for $\Delta v_{\mathrm{metal}} = \pm 100 \textrm{ km s}^{-1}$, and the CDDF of the
\ion{C}{iv}-enriched forest (Table~\ref{tbl:dndz_ap6}) and of the unenriched forest (Table~\ref{tbl:dndz_ap7}) 
for $\Delta v_{\mathrm{metal}} = \pm 10 \textrm{ km s}^{-1}$.

\begin{table*}
\centering 
\small
\caption{Number density evolution data for each quasar\label{tbl:dndz_ap1}}
\begin{tabular}{c c c c c c c c c}
\hline 
\noalign{\smallskip}
 &  \multicolumn{4}{c}{Ly$\alpha$-only fit} &  \multicolumn{4}{c}{high-order fit}\tabularnewline
 &  &  & $\log N_\ion{H}{i} = [12.75,14.0]$ & 
   $\log N_\ion{H}{i} = [14,17]$ &  &  & $\log N_\ion{H}{i} = [12.75,14.0]$ 
    & $\log N_\ion{H}{i} = [14, 17]$\tabularnewline
Quasar & $<z>$ & $\Delta z$ & $\log \dif n / \dif z$ & 
   $\log \dif n / \dif z$ & $<z>$ & $\Delta z$ & $\log \dif n / \dif z$ & 
   $\log \dif n / \dif z$ \tabularnewline
\noalign{\smallskip}

\hline 
\hline 
\noalign{\smallskip}

Q0055$-$269 & 3.270 & 0.669 & 2.672$\pm$0.024 & 2.157$\pm$0.042 & 3.071 & 0.269 & 2.600$\pm$0.040 & 2.269$\pm$0.057 \tabularnewline
PKS2126$-$158 & 3.010 & 0.390 & 2.642$\pm$0.032 & 1.886$\pm$0.073 & 3.010 & 0.390 & 2.642$\pm$0.032 & 1.965$\pm$0.067 \tabularnewline
Q0420$-$388 & 2.759 & 0.558 & 2.656$\pm$0.026 & 2.117$\pm$0.048   & 2.851 & 0.373 & $2.722\pm0.022$ & 2.160$\pm$0.055 \tabularnewline
HE0940$-$1050 & 2.729 & 0.554 & 2.628$\pm$0.028 & 1.997$\pm$0.055 & 2.729 & 0.554 & 2.635$\pm$0.027 & 2.070$\pm$0.051 \tabularnewline
HE2347$-$4342 & 2.577 & 0.484 & 2.602$\pm$0.030 & 1.918$\pm$0.064 & 2.577 & 0.484 & 2.600$\pm$0.030 & 1.918$\pm$0.064 \tabularnewline
Q0002$-$422 & 2.457 & 0.496 & 2.651$\pm$0.029 & 1.948$\pm$0.066   & 2.457 & 0.496 & 2.659$\pm$0.028 & 1.995$\pm$0.058 \tabularnewline
PKS0329$-$255 & 2.395 & 0.513 & 2.483$\pm$0.035 & 1.846$\pm$0.073 & 2.395 & 0.513 & 2.499$\pm$0.033 & 1.781$\pm$0.072 \tabularnewline
Q0453$-$423$^{a}$ & 2.340 & 0.354 & 2.665$\pm$0.033 & 1.812$\pm$0.082 & 2.340 & 0.354 & 2.663$\pm$0.033 & 1.913$\pm$0.074 \tabularnewline
HE1347$-$2457 & 2.300 & 0.505 & 2.506$\pm$0.033 & 1.658$\pm$0.082 & 2.300 & 0.505 & 2.525$\pm$0.032 & 1.695$\pm$0.079 \tabularnewline
Q0329$-$385 & 2.139 & 0.475 & 2.571$\pm$0.031 & 1.602$\pm$0.090   & 2.139 & 0.475 & 2.569$\pm$0.032 & 1.666$\pm$0.084 \tabularnewline
HE2217$-$2818 & 2.126 & 0.479 & 2.575$\pm$0.031 & 1.524$\pm$0.097 & 2.167 & 0.396 & 2.587$\pm$0.034 & 1.607$\pm$0.097 \tabularnewline
Q0109$-$3518 & 2.127 & 0.443 & 2.518$\pm$0.035 & 1.716$\pm$0.072  & 2.158 & 0.380 & 2.507$\pm$0.038 & 1.721$\pm$0.088 \tabularnewline
HE1122$-$1648 & 2.124 & 0.467 & 2.588$\pm$0.031 & 1.793$\pm$0.074 & 2.129 & 0.458 & 2.589$\pm$0.031 & 1.816$\pm$0.073 \tabularnewline
J2233$-$606 & 1.978 & 0.445 & 2.533$\pm$0.034 & 1.582$\pm$0.094   & 2.086 & 0.230 & 2.501$\pm$0.048 & 1.416$\pm$0.149 \tabularnewline
PKS0237$-$23 & 1.972 & 0.415 & 2.519$\pm$0.036 & 1.613$\pm$0.094  & 2.070 & 0.219 & 2.492$\pm$0.050 & 1.505$\pm$0.139 \tabularnewline
PKS1448$-$232 & 1.947 & 0.456 & 2.514$\pm$0.034 & 1.773$\pm$0.076 & 2.064 & 0.222 & 2.517$\pm$0.048 & 1.800$\pm$0.103 \tabularnewline
Q0122$-$380 & 1.921 & 0.442 & 2.434$\pm$0.038 & 1.735$\pm$0.081   & 2.059 & 0.165 & 2.456$\pm$0.059 & 1.687$\pm$0.131 \tabularnewline
Q1101$-$264 & 1.989 & 0.216 & 2.493$\pm$0.050 & 1.512$\pm$0.139  & 1.998 & 0.197 & 2.460$\pm$0.054 & 1.307$\pm$0.176 \tabularnewline
\hline
\end{tabular}
\begin{list}{}{}
\item[$^{\mathrm{a}}$]
It includes a sub-DLA, which introduces a gap in the
Ly$\alpha$ redshift range. The calculation of $\dif n / \dif z$ and the listed
$\Delta z$ take into account the
redshift gap.
\end{list}
\end{table*}

\begin{table*}
\centering 
\small
\caption{{\it Mean} number density evolution data for $\Delta z = 0.26$\label{tbl:dndz_ap2}}
\begin{tabular}{c c c c c}
\hline 
\noalign{\smallskip}
 &  \multicolumn{2}{c}{Ly$\alpha$-only fit} & \multicolumn{2}{c}{high-order fit}\tabularnewline
 &  $\log N_\ion{H}{i} = [12.75, 14.0]$ & 
 $\log N_\ion{H}{i} = [14,17]$ & $\log N_\ion{H}{i} = [12.75,14.0]$ 
 & $\log N_\ion{H}{i} = [14,17]$\tabularnewline
$<z>^{\mathrm{a}}$ & $\log \dif n / \dif z$ & $\log \dif n / \dif z$ &  
$\log \dif n / \dif z$ & 
$\log \dif n / \dif z$ \tabularnewline
\noalign{\smallskip}
\hline
\hline
\noalign{\smallskip}

2.067 &	2.543$\pm$0.016 & 1.630$\pm$0.044 & 2.540$\pm$0.0160 & 1.626$\pm$0.0442	\tabularnewline
2.263 &	2.516$\pm$0.017 & 1.680$\pm$0.044 & 2.542$\pm$0.0173 & 1.723$\pm$0.0422	\tabularnewline
2.552 &	2.542$\pm$0.020 & 1.874$\pm$0.042 & 2.545$\pm$0.0199 & 1.883$\pm$0.0416	\tabularnewline
2.813 &	2.647$\pm$0.022 & 2.024$\pm$0.045 & 2.664$\pm$0.0219 & 2.094$\pm$0.0412	\tabularnewline
3.027 &	2.640$\pm$0.024 & 2.078$\pm$0.045 & 2.650$\pm$0.0240 & 2.136$\pm$0.0424	\tabularnewline

\hline
\end{tabular}
\begin{list}{}{}
\item[$^{\mathrm{a}}$]
The exactly same redshift range was used for both fit samples.
\end{list}
\end{table*}

\begin{table*}
\centering 
\small
\caption{CDDF of the entire \ion{H}{i} forest 
$f = \log \left( \dif N / (\dif N_\ion{H}{i} \; \dif X) \right)$
for the high-order fit sample\label{tbl:dndz_ap3}}
\begin{tabular}{c c c c c c c c c c}
\hline 
\noalign{\smallskip}
 & \multicolumn{3}{c}{$z=1.9-3.2$} & \multicolumn{3}{c}{$z=1.9-2.4$} & \multicolumn{3}{c}{$z=2.4-3.2$} \tabularnewline
$\log N_\ion{H}{i}$ & $f$ & $+\Delta f$ & $-\Delta f$ & $f$ & $+\Delta f$ & $-\Delta f$ & $f$ & $+\Delta f$ & $-\Delta f$ \tabularnewline
\noalign{\smallskip}
\hline 
\hline 
\noalign{\smallskip}

  12.875 &  -11.077 &    0.015 &    0.016 &  -11.082 &    0.021 &    0.022 &  -11.071 &    0.022 &    0.023 \\ 
  13.125 &  -11.437 &    0.017 &    0.018 &  -11.441 &    0.023 &    0.024 &  -11.432 &    0.025 &    0.026 \\ 
  13.375 &  -11.795 &    0.019 &    0.020 &  -11.840 &    0.027 &    0.029 &  -11.747 &    0.027 &    0.028 \\ 
  13.625 &  -12.163 &    0.022 &    0.023 &  -12.206 &    0.031 &    0.034 &  -12.117 &    0.031 &    0.033 \\ 
  13.875 &  -12.507 &    0.024 &    0.026 &  -12.555 &    0.035 &    0.038 &  -12.457 &    0.034 &    0.037 \\ 
  14.125 &  -12.992 &    0.032 &    0.034 &  -13.118 &    0.049 &    0.055 &  -12.879 &    0.041 &    0.045 \\ 
  14.375 &  -13.459 &    0.040 &    0.045 &  -13.611 &    0.064 &    0.075 &  -13.329 &    0.051 &    0.058 \\ 
  14.625 &  -13.781 &    0.044 &    0.049 &  -14.083 &    0.081 &    0.099 &  -13.579 &    0.051 &    0.058 \\ 
  14.875 &  -14.290 &    0.058 &    0.067 &  -14.351 &    0.082 &    0.102 &  -14.227 &    0.078 &    0.095 \\ 
  15.250 &  -14.933 &    0.055 &    0.063 &  -15.203 &    0.097 &    0.125 &  -14.744 &    0.065 &    0.076 \\ 
  15.750 &  -15.970 &    0.097 &    0.125 &  -16.129 &    0.149 &    0.228 &  -15.835 &    0.119 &    0.165 \\ 
  16.250 &  -16.674 &    0.119 &    0.165 &  -16.629 &    0.149 &    0.228 &  -16.733 &    0.176 &    0.301 \\ 
  16.750 &  -17.219 &    0.125 &    0.176 &  -17.907 &    0.301 &          &  -16.932 &    0.131 &    0.189 \\ 
  17.250 &  -17.896 &    0.149 &    0.228 &  -18.106 &    0.232 &    0.533 &  -17.733 &    0.176 &    0.301 \\ 
  17.750 &  -19.174 &    0.301 &          &          &          &          &  -18.835 &    0.301 &          \\ 

\hline
\end{tabular}
\end{table*}

\begin{table*}
\centering 
\small
\caption{CDDF $f = \log \left( \dif N / (\dif N_\ion{H}{i} \; \dif X) \right)$ of 
the \ion{C}{iv}-enriched forest for $\Delta v_{\mathrm{metal}} = \pm 100 \textrm{ km s}^{-1}$\label{tbl:dndz_ap4}}
\begin{tabular}{c c c c c c c c c c}
\hline 
\noalign{\smallskip}

 & \multicolumn{3}{c}{$z=1.9-3.2$} & \multicolumn{3}{c}{$z=1.9-2.4$} & \multicolumn{3}{c}{$z=2.4-3.2$} \\
$\log N_\ion{H}{i}$ & $f$ & $+\Delta f$ & $-\Delta f$ & $f$ & $+\Delta f$ & $-\Delta f$ & $f$ & $+\Delta f$ & $-\Delta f$ \\
\noalign{\smallskip}

\hline 
\hline 
\noalign{\smallskip}

  12.950 &  -12.655 &    0.066 &    0.078 &  -12.632 &    0.084 &    0.104 &  -12.686 &    0.100 &    0.130 \\ 
  13.350 &  -13.021 &    0.064 &    0.075 &  -13.096 &    0.090 &    0.113 &  -12.940 &    0.086 &    0.107 \\ 
  13.750 &  -13.410 &    0.063 &    0.074 &  -13.519 &    0.092 &    0.117 &  -13.301 &    0.082 &    0.102 \\ 
  14.150 &  -13.855 &    0.066 &    0.078 &  -13.944 &    0.094 &    0.121 &  -13.762 &    0.088 &    0.110 \\ 
  14.550 &  -14.232 &    0.065 &    0.076 &  -14.495 &    0.110 &    0.148 &  -14.031 &    0.076 &    0.093 \\ 
  14.950 &  -14.705 &    0.070 &    0.083 &  -14.861 &    0.106 &    0.141 &  -14.562 &    0.088 &    0.110 \\ 
  15.350 &  -15.344 &    0.090 &    0.113 &  -15.530 &    0.139 &    0.206 &  -15.183 &    0.110 &    0.148 \\ 
  15.750 &  -16.069 &    0.125 &    0.176 &  -16.173 &    0.176 &    0.301 &  -15.964 &    0.161 &    0.257 \\ 
  16.150 &  -16.645 &    0.149 &    0.228 &  -16.697 &    0.198 &    0.374 &  -16.585 &    0.198 &    0.374 \\ 
  16.550 &  -16.978 &    0.139 &    0.206 &  -17.274 &    0.232 &    0.533 &  -16.764 &    0.161 &    0.257 \\ 
  16.950 &  -17.746 &    0.198 &    0.374 &               &              &              &  -17.385 &    0.198 &    0.374 \\ 
  17.350 &  -18.146 &    0.198 &    0.374 &  -18.074 &    0.232 &    0.533 &  -18.263 &    0.301 &  $\infty$ \\ 
  17.750 &  -19.023 &    0.301 &  $\infty$ &               &              &    $\infty$ &  -18.663 &    0.301 &  $\infty$ \\ 

\hline
\end{tabular}
\end{table*}

\begin{table*}
\centering
\small
\caption{CDDF $f = \log \left( \dif N / (\dif N_\ion{H}{i} \; \dif X) \right)$ of
the unenriched forest for $\Delta v_{\mathrm{metal}} = \pm 100 \textrm{ km s}^{-1}$\label{tbl:dndz_ap5}}
\begin{tabular}{c c c c c c c c c c}
\hline
\noalign{\smallskip}

 & \multicolumn{3}{c}{$z=1.9-3.2$} & \multicolumn{3}{c}{$z=1.9-2.4$} & \multicolumn{3}{c}{$z=2.4-3.2$} \\
$\log N_\ion{H}{i}$ & $f$ & $+\Delta f$ & $-\Delta f$ & $f$ & $+\Delta f$ & $-\Delta f$ & $f$ & $+\Delta f$ & $-\Delta f$ \\
\noalign{\smallskip}

\hline
\hline
\noalign{\smallskip}

  12.950 &  -11.200 &    0.013 &    0.014 &  -11.210 &    0.018 &    0.018 &  -11.187 &    0.020 &    0.020 \\ 
  13.350 &  -11.797 &    0.016 &    0.017 &  -11.831 &    0.023 &    0.024 &  -11.757 &    0.024 &    0.025 \\ 
  13.750 &  -12.374 &    0.020 &    0.021 &  -12.389 &    0.027 &    0.029 &  -12.355 &    0.029 &    0.032 \\ 
  14.150 &  -13.114 &    0.029 &    0.032 &  -13.240 &    0.044 &    0.050 &  -12.991 &    0.038 &    0.042 \\ 
  14.550 &  -13.920 &    0.046 &    0.051 &  -14.097 &    0.073 &    0.088 &  -13.764 &    0.057 &    0.066 \\ 
  14.950 &  -14.718 &    0.071 &    0.084 &  -14.828 &    0.103 &    0.135 &  -14.607 &    0.092 &    0.117 \\ 
  15.350 &  -15.845 &    0.149 &    0.228 &  -15.773 &    0.176 &    0.301 &  -15.962 &    0.232 &    0.533 \\ 
  15.750 &  -16.722 &    0.232 &    0.533 &  -16.775 &    0.301 &  $\infty$ &  -16.663 &    0.301 & $\infty$ \\ 
  16.150 &  -17.122 &    0.232 &    0.533 &  -16.874 &    0.232 &    0.533 &               &               &     \\ 
  
\hline
\end{tabular}
\end{table*}

\begin{table*}
\centering 
\small
\caption{CDDF $f = \log \left( \dif N / (\dif N_\ion{H}{i} \; \dif X) \right)$ of 
the \ion{C}{iv}-enriched forest for $\Delta v_{\mathrm{metal}} = \pm 10 \textrm{ km s}^{-1}$\label{tbl:dndz_ap6}}
\begin{tabular}{c c c c c c c c c c}
\hline 
\noalign{\smallskip}

 & \multicolumn{3}{c}{$z=1.9-3.2$} & \multicolumn{3}{c}{$z=1.9-2.4$} & \multicolumn{3}{c}{$z=2.4-3.2$} \\
$\log N_\ion{H}{i}$ & $f$ & $+\Delta f$ & $-\Delta f$ & $f$ & $+\Delta f$ & $-\Delta f$ & $f$ & $+\Delta f$ & $-\Delta f$ \\
\noalign{\smallskip}

\hline 
\hline 
\noalign{\smallskip}

  12.950 &  -13.524 &    0.161 &    0.257 &  -13.276 &    0.161 &    0.257 &      &      &      \\ 
  13.350 &  -14.146 &    0.198 &    0.374 &  -13.897 &    0.198 &    0.374 &      &      &      \\ 
  13.750 &  -14.023 &    0.119 &    0.165 &  -14.076 &    0.161 &    0.257 &  -13.964 &    0.161 &    0.257 \\ 
  14.150 &  -14.277 &    0.103 &    0.135 &  -14.220 &    0.125 &    0.176 &  -14.364 &    0.161 &    0.257 \\ 
  14.550 &  -14.544 &    0.090 &    0.113 &  -14.973 &    0.176 &    0.301 &  -14.286 &    0.100 &    0.130 \\ 
  14.950 &  -14.861 &    0.082 &    0.102 &  -15.020 &    0.125 &    0.176 &  -14.716 &    0.103 &    0.135 \\ 
  15.350 &  -15.477 &    0.103 &    0.135 &  -15.773 &    0.176 &    0.301 &  -15.263 &    0.119 &    0.165 \\ 
  15.750 &  -16.178 &    0.139 &    0.206 &  -16.297 &    0.198 &    0.374 &  -16.060 &    0.176 &    0.301 \\ 
  16.150 &  -16.645 &    0.149 &    0.228 &  -16.697 &    0.198 &    0.374 &  -16.585 &    0.198 &    0.374 \\ 
  16.550 &  -17.124 &    0.161 &    0.257 &  -17.274 &    0.232 &    0.533 &  -16.985 &    0.198 &    0.374 \\ 
  16.950 &  -17.922 &    0.232 &    0.533 &               &              &              &  -17.561 &    0.232 &    0.533 \\ 
  17.350 &  -18.623 &    0.301 & $\infty$ &               &              &              &  -18.263 &    0.301 &  $\infty$  \\

\hline
\end{tabular}
\end{table*}

\begin{table*}
\centering
\small
\caption{CDDF $f = \log \left( \dif N / (\dif N_\ion{H}{i} \; \dif X) \right)$ of
the unenriched forest for $\Delta v_{\mathrm{metal}} = \pm 10 \textrm{ km s}^{-1}$\label{tbl:dndz_ap7}}
\begin{tabular}{c c c c c c c c c c}
\hline
\noalign{\smallskip}

 & \multicolumn{3}{c}{$z=1.9-3.2$} & \multicolumn{3}{c}{$z=1.9-2.4$} & \multicolumn{3}{c}{$z=2.4-3.2$} \\
$\log N_\ion{H}{i}$ & $f$ & $+\Delta f$ & $-\Delta f$ & $f$ & $+\Delta f$ & $-\Delta f$ & $f$ & $+\Delta f$ & $-\Delta f$ \\
\noalign{\smallskip}

\hline
\hline
\noalign{\smallskip}

  12.950 &  -11.187 &    0.013 &    0.013 &  -11.197 &    0.017 &    0.018 &  -11.173 &    0.019 &    0.020 \\ 
  13.350 &  -11.774 &    0.016 &    0.017 &  -11.811 &    0.022 &    0.023 &  -11.730 &    0.023 &    0.024 \\ 
  13.750 &  -12.345 &    0.019 &    0.020 &  -12.366 &    0.026 &    0.028 &  -12.318 &    0.028 &    0.030 \\ 
  14.150 &  -13.067 &    0.028 &    0.030 &  -13.201 &    0.043 &    0.047 &  -12.939 &    0.036 &    0.039 \\ 
  14.550 &  -13.823 &    0.041 &    0.046 &  -13.995 &    0.065 &    0.077 &  -13.670 &    0.052 &    0.059 \\ 
  14.950 &  -14.600 &    0.062 &    0.073 &  -14.719 &    0.092 &    0.117 &  -14.482 &    0.081 &    0.099 \\ 
  15.350 &  -15.582 &    0.114 &    0.156 &  -15.530 &    0.139 &    0.206 &  -15.660 &    0.176 &    0.301 \\ 
  15.750 &  -16.421 &    0.176 &    0.301 &  -16.474 &    0.232 &    0.533 &  -16.361 &    0.232 &    0.533 \\ 
  16.150 &  -17.122 &    0.232 &    0.533 &  -16.874 &    0.232 &    0.533 &               &              &              \\ 
  16.550 &  -17.522 &    0.232 &    0.533 &               &              &               &  -17.161 &    0.232 &    0.533 \\ 
  16.950 &  -18.223 &    0.301 & $\infty$ &               &               &              &  -17.863 &    0.301 & $\infty$ \\ 
  17.350 &  -18.322 &    0.232 &    0.533 &  -18.074 &    0.232 &    0.533 &               &               &             \\ 
  17.750 &  -19.023 &    0.301 & $\infty$  &               &              &              &  -18.663 &    0.301 &   $\infty$   \\

\hline
\end{tabular}
\end{table*}

\end{document}